\documentclass[preprint,preprintnumbers,nofootinbib,amsmath,amssymb]{revtex4-2}

\usepackage[english]{babel}
\usepackage{color}
\usepackage{amsmath}
\usepackage{physics}
\usepackage[hidelinks]{hyperref}
\usepackage{diagbox}
\usepackage{multirow}
\usepackage{amsfonts}
\usepackage{amssymb}
\usepackage{latexsym}
\usepackage{graphicx}
\usepackage{adjustbox}
\usepackage{xr}
\usepackage{siunitx}
\usepackage{threeparttable}
\usepackage{braket}
\usepackage[justification=centering]{caption}
\usepackage{float}
\usepackage{caption}
\usepackage{subcaption}

\usepackage{xr-hyper}
\usepackage{enumitem}

\usepackage{ytableau}
\usepackage[nodisplayskipstretch]{setspace} %%%% Avoids big spaces between eqs

\newcommand{\al}{\alpha}

\newcommand{\pa}{\partial}

\newcommand{\veps}{\varepsilon}

\newcommand{\sig}{\sigma}

\newcommand{\la}{\lambda}

\newcommand{\ga}{\gamma}

\newcommand{\de}{\delta}

\newcommand{\De}{\Delta}

\newcommand{\tha}{\theta}

\newcommand{\rar}{\rightarrow}

\newcommand{\non}{\nonumber}

\newsavebox\CBox
\def\textBF#1{\sbox\CBox{#1}\resizebox{\wd\CBox}{\ht\CBox}{\textbf{#1}}}
\begin{document}

\title{\Large Two-body neutral Coulomb system in a magnetic field at rest: from Hydrogen atom to positronium}
\date{\today}
\author{J.C.~del~Valle}
\email{delvalle@correo.nucleares.unam.mx}

\author{A.V.~Turbiner}
\email{turbiner@nucleares.unam.mx}

\affiliation{Instituto de Ciencias Nucleares, Universidad Nacional Aut\'onoma de M\'exico,
A. Postal 70-543 C. P. 04510, Ciudad de M\'exico, M\'exico}

\author{Adrian M Escobar Ruiz}
\email{admau@xanum.uam.mx}
\affiliation{Departamento de Fisica, Universidad Aut\'onoma Metropolitana-Iztapalapa,
San Rafael Atlixco 186, C.P. 09340 Ciudad de M\'exico, M\'exico}

\begin{abstract}

A simple locally accurate uniform approximation for the nodeless wavefunction is constructed for a {\it neutral} system of two Coulomb charges of different masses $(-q,m_1)$ and $(q,m_2)$ at rest in a constant uniform magnetic field for the states of positive and negative parity, ${(1s_0)}$ and ${(2p_0)}$, respectively. It is shown that by keeping the mass and charge of one of the bodies fixed, all systems with different second body masses are related. This allows one to consider the second body as infinitely-massive and to take such a system as basic. Three physical systems are considered in details: the Hydrogen atom with (in)-finitely massive proton (deuteron, triton) and the positronium atom $(-e,e)$.
We derive the Riccati-Bloch and Generalized-Bloch equations, which describe the domains of small and large distances, respectively. Based on the interpolation of the small and large distance behavior of the logarithm of the wavefunction, a compact 10-parametric function is proposed. Taken as a variational trial function it provides accuracy of not less than 6 significant digits (s.d.) ($\lesssim 10^{-6}$ in relative deviation) for the total energy in the whole domain of considered magnetic fields $[0\,,\,10^4]$ a.u. and not less than 3 s.d. for the quadrupole moment $Q_{zz}$. In order to get reference points the Lagrange Mesh Method with 16K mesh points was used to get from 10 to 6 s.d. in energy from small to large magnetic fields. Based on the Riccati-Bloch equation the first 100 perturbative coefficients for the energy, in the form of rational numbers, are calculated and, using the Pad\'e-Borel re-summation procedure, the energy is found with not less than 10 s.d. at magnetic fields $\leq 1$\,a.u.

\end{abstract}

{\large \it
\centerline{Phys. Rev. {\bf A 103}, 032820 (2021)}

\centerline{Phys. Rev.  {\bf A 105}, 049901 (2022) (erratum)}

}

\maketitle

\section{Introduction}

Hydrogen atom in a constant uniform magnetic field is one of the first problems studied in quantum mechanics. It is stable at any magnetic field strength. Its importance is related to the fact that it arises in various domains of physics, in particular, in semiconductor physics \cite{Elliott:1960} and in astrophysics (e.g. the physics of strong surface magnetic fields of magnetic white dwarfs and neutron stars). In the former case, the excitons occur as hydrogen-like quasi-atoms with a small effective mass and a large dielectric constant. In the latter case, the atmosphere of white dwarfs and neutron stars can contain Hydrogen atoms subject to a strong magnetic field. Magnetic fields in Nature occur from a few Gauss (e.g. the Earth, Jupiter magnetic fields) up to about $10^{16}$\,Gauss corresponding to a surface magnetic field in a few explored magnetars. Hence, they range in 16 orders of magnitude! Hydrogen atom is the simplest Coulomb system in the sequence of one-electron hydrogenic atomic-molecular ions, both traditional and exotic, which may exist in a strong magnetic field \cite{Turbiner:2006}. All of that explains enormous amount of articles published on the subject.

A description of the problem with weak magnetic fields may be found in any textbook on quantum mechanics (see e.g. Landau and Lifshitz \cite{LL:1977}). Early attempts to explore the problem are summarized in the remarkable review paper by Garstang \cite{Garstang:1977}. In the overwhelming majority of considerations the proton is assumed explicitly to be infinitely-heavy, which implies that the atom is at rest, although the Hydrogen atom, since it is the neutral two body Coulomb system, can be at rest even for the case of finite proton mass - the case of zero pseudomomentum  \cite{Gorkov:1968}. Many years ago it was shown that the problems when the proton is infinitely-massive and finitely-massive but both at rest are connected via non-trivial scaling relation \cite{Pavlov-Verevkin:1980}. It is well known that in the finite mass case the center-of-mass motion is not separated unlike in the field-free case: it is replaced by the pseudo-separation. This does not lead to complexification in the case of zero pseudomomentum. In general, the specific coupling between the relative and the c.m. motion leads to the prediction of a giant-dipole moment \cite{Burkova:1976}.

By increasing the magnetic field the electronic density evolves from a spherical-symmetrical distribution at weak fields to a cigar-like one (elongated in the field direction) at atomic and larger fields, where the energy grows linearly with field strength. It was a challenge for many years to give a unified description of the evolution in the framework of the same approach with a sufficiently high, uniform in field strength accuracy. This would imply approximate solution of the problem.
It is worth mentioning three approaches, which treated the challenge: (i) power series expansion - the Method of Moments \cite{Kravchenko:1996}, (ii) a numerical approach - the Lagrange Mesh method \cite{Baye:2015} and (iii) the Variational Method with (a) multiconfigurational trial functions \cite{Stubbins:2004} and (b) with simple, few-parametric, single-configurational, physically-adequate trial functions \cite{Yafet:1956,Larsen:1968,Brandi:1975,Larsen:1982,Turbiner:1984,
Turbiner:1987,Potekhin:2001,Turbiner:2007}. As a result, in all three approaches the ground state energy (and the energies of a few excited states) were found with reasonably high (or excessively high) accuracy. Surprisingly, the quadrupole moment, which is one of the principal consequences of the presence of the magnetic field for the Hydrogen atom and which defines the van der Waals type constant for the repulsion at large distances of two H-atoms in H$_2$ molecule, was studied quantitatively in a reliable way recently \cite{Baye:2008} only. The situation with finite-mass effects \cite{Gorkov:1968,Burkova:1976,Schmelcher:1988}, relativistic and QED corrections is far to be complete, see e.g. \cite{Pop-Kar:2014}. { In general, the finite nuclear mass effects do not change 4-5 significant digits, and following Salpeter at al. estimates the leading relativistic and QED effects leave unchanged 3-4 significant digits in the ground state energy.}

It should be mentioned that the neutral system can move across magnetic field, see e.g. \cite{Potekhin:1998} and references therein. The two-dimensional case, where the neutral atom moves on a plane
subject to a magnetic field perpendicular to it, has been analyzed in detail (see \cite{Escobar-Turbiner:2014} and references therein). In particular, a simple physically-adequate trial function with the property of being a uniform local approximation of the exact eigenfunction { in any point of coordinate space} was constructed for the lowest states and {\it any} constant uniform magnetic field.
Remarkably, when the system possesses azimuthal symmetry, the hidden $sl(2)$ algebra occurs and there exists an infinite number of exact analytic eigenfunctions. These eigenfunctions occur for specific values of the magnetic field only. The existence of exact eigenfunctions for the three-dimensional two body neutral Coulomb system is still an interesting open problem.
It should be noted that usually studies in a magnetic field are characterized by a high degree of  technicality. In order to get reliable numerical results two (or more) independent calculations have to be carried out.

The aim of this paper is twofold: (i) using perturbation theory and semiclassical consideration to construct a compact function as a uniform local approximation of the exact (unknown) eigenfunction in the whole range of magnetic fields and (ii) to study finite mass effects for the neutral atom at rest. Main emphasis will be given to the ground state - the state of lowest energy. It will be revisited and then will be provided the high accuracy estimates of the quadrupole moment for the Hydrogen atom.

The paper consists of two large parts. Part I is about the infinite (proton) mass case where the Riccati-Bloch and generalized Bloch equations are derived, the perturbation theory in powers of the magnetic field strength is constructed for both equations and the approximate expression (the Approximant) for the ground state eigenfunction of positive parity and for another one of negative parity is introduced. In Part II the case of the two-body neutral Coulomb system of finite masses is studied. The hydrogen atom $(p,e)$ and Positronium $(e^+,e^-)$ are considered.

Atomic units will be used through out the paper, the energy will be measured in Rydbergs.
\numberwithin{equation}{section}
\section{Infinite Mass Case}
\begin{center}
\begin{figure}[h]
\includegraphics[scale=1.0]{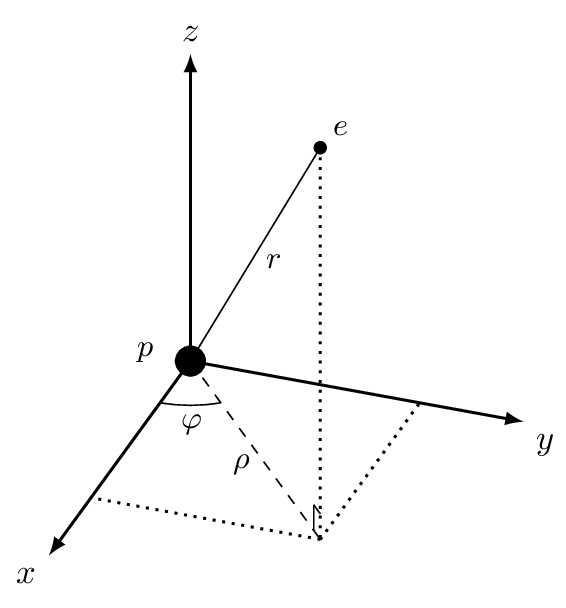}
\caption{Coordinates $(\rho,r,\varphi)$ at half-space $z \geq 0$. The infinitely heavy proton is located at the   origin.}
\label{fig:coordinates}
\end{figure}
\end{center}

The Hamiltonian
\begin{equation}
\label{H}
\hat{H}^{(\infty)}\  = \ \frac{1}{2\,m_e}\left(\hat{{\bf p}}\,+\,\frac{e}{c}\bf{A}\right)^2\  -\ \frac{e^2}{r}\  \ ,\qquad r = \sqrt{x^2+y^2+z^2}\ ,
\end{equation}
describes a hydrogen atom in presence of a constant uniform magnetic field $\textbf{B}=\gamma\,\hat{\textbf{z}}$, directed along $z$-axis, in the static approximation, when the mass of the proton is infinite, $m_p=\infty$ and $c$ is the speed of light. Here $m_e$ and $(-e)$ are the mass and charge of the electron, respectively, $\hat{\bf p}$ is its momentum, $r$ is its distance from the origin. The infinitely heavy proton of charge $e > 0$ is situated at the origin. In symmetric gauge
\begin{equation}
 \textbf{A}\ =\frac{1}{2}\textbf{B}\times\textbf{r}\ ,
\end{equation}
the Hamiltonian (\ref{H}) takes the form
\begin{equation}
\label{Hamiltonian}
   \hat{H}^{(\infty)}\ =\ -\frac{\hbar^2}{2m_e}\Delta\ +\ \frac{e\gamma}{2m_e c}\hat{L}_z\
   +\ V\ ,\qquad   \De\ =\ \pa_x^2\ +\ \pa_y^2\ +\ \pa_z^2\ ,
\end{equation}
where the potential
\begin{equation}
\label{potential}
   V\ =\ -\frac{e^2}{r}\ +\ \frac{e^2\gamma^2}{8m_ec^2}\,\rho^2\ ,\qquad \rho=\sqrt{x^2+y^2}\ ,
\end{equation}
depends on two variables $\rho$ and $r$, and $\hat{L}_z$ is the projection of the angular momentum operator in the direction of the magnetic field,
\begin{equation}
\hat{L}_z \ =\ -i\hbar\,(x\,\pa_y\ -\ y\,\pa_x)\ ,
\end{equation}
which is conserved, $[\hat{H}^{(\infty)}, \hat{L}_z]=0$. The parity operator ${\hat\Pi} \Psi(x,y,z)= \Psi(x,y,-z)$ also commutes with $\hat{H}^{(\infty)}$.
The  Schr\"odinger equation associated to (\ref{Hamiltonian}) is
\begin{equation}
 \label{schroedinger}
 \hat{H}^{(\infty)}\psi\ =\  E^{(\infty)}\psi\ ,\qquad E^{(\infty)}=E^{(\infty)}(\gamma,e,m_e)\ ,
\end{equation}
with boundary conditions imposed in such a way that the wave function is normalizable,
\begin{equation}
\label{normalizabilty}
      \int |\psi|^2\,d\,{\bf r} \ <\ \infty\ .
\end{equation}
The dependence of the potential (\ref{potential}) on the variables $\rho,r$ hints (see for discussion \cite{twe}) that it might be convenient to write the Schr\"odinger equation (\ref{schroedinger}) in the non-orthogonal system of coordinates $(\rho,r,\varphi)$, see Fig. \ref{fig:coordinates} and make a search for subfamily of eigenfunctions with $(\rho,r)$ dependence alone. In these coordinates we get
\[
 -\frac{\hbar^2}{2m_e}\left[\pa_\rho^2\ +\ \frac{2\rho}{r}\pa_{\rho r}\ +\ \pa_r^2\ +\ \frac{1}{\rho}\pa_{\rho}\ +\ \frac{2}{r}\pa_r\ +\ \frac{1}{\rho^2}\pa_\varphi^2\right]\psi\ -\ \frac{i e \hbar \gamma}{2m_ec}\pa_\varphi\psi \ +\
\]
\begin{equation}
\label{Sch-rho-r-phi}
 \left[\ -\frac{e^2}{r}\ +\ \frac{\gamma^2e^2}{8m_ec^2}\rho^2\right]\psi\ =\
       E^{(\infty)}\,\psi\ .
\end{equation}
Needless to say, this equation is non-solvable for $\gamma \neq 0$: the energy $E^{(\infty)}$ and
wave function $\psi$ can not be written in terms of known elementary and transcendental functions, they can be found approximately {\it only} \footnote{It should be emphasized that the title {\it Exact solution for a hydrogen atom in a magnetic field of arbitrary strength} in the article \cite{Kravchenko:1996} is misleading.}.

Due to cylindrical symmetry any state is characterized by two quantum numbers: the magnetic quantum number $m$ ($\hbar m$ the eigenvalue of the operator $\hat{L}_z$) and the parity $\nu=\pm$ with respect to a reflection $z \rar -z$. It suggests to represent a wavefunction in the form
\begin{equation}
\label{representation}
  \psi(\rho,r,\varphi)\ =\ \rho^{|m|}\,z^p\,\Psi(\rho,r)\, e^{im\varphi}\ ,
  \qquad m\ =\ 0,\pm 1,\pm2,\, \ldots \ ,\ p=0,1\ ,
\end{equation}
for the states with magnetic quantum number $m$ and of positive/negative parity $\nu=+/-$, hence, $p=0,1$ and $\nu=(-1)^p$, respectively, here $z=\sqrt{r^2-\rho^2}$, see Fig.\ref{fig:coordinates}. The problem is reduced to find the function $\Psi(\rho,r)$, which satisfies the (gauge-rotated) Schr\"odinger equation
\[
  -\frac{\hbar^2}{2m_e}\left[\pa_\rho^2\ +\ \frac{2\rho}{r}\pa_{\rho r}\ +\ \pa_r^2\ +\ \frac{2|m|+1}{\rho}\pa_{\rho}\ +\ \frac{2(|m|+p+1)}{r}\pa_r\right]\Psi
\]
\begin{equation}
\label{schroedinger_new}
  \ +\ \left[\ -\frac{e^2}{r}\ +\ \frac{\gamma^2 e^2}{8 m_e c^2}\rho^2\right]\Psi\
  =\ \mathcal{E}_{m,p}^{(\infty)}(\gamma^2,e^2,m_e)\,\Psi\ ,
\end{equation}
where
\begin{equation*}
\label{E --> eps}
  \mathcal{E}_{m,p}^{(\infty)}\ =\ E_{m,p}^{(\infty)}\ -\frac{|e|\hbar \gamma\,m}{2m_ec}\ ,
\end{equation*}
and the magnetic quantum number $m$ plays a role of parameter.
From equation (\ref{schroedinger_new}) one can explicitly see that $\mathcal{E}^{(\infty)}_{m,p}$, which is equal to the energy with linear Zeeman term subtracted, should be even function with respect to the magnetic quantum number $m$,
\begin{equation}
\label{Eparity}
  \mathcal{E}_{m,p}^{(\infty)}\ =\ \mathcal{E}_{-m,p}^{(\infty)}\ ,
\end{equation}
for both positive and negative parity states, hence, $\mathcal{E}_{m,p}^{(\infty)}$ depends on $m^2$, therefore,
\begin{equation}
  E_{m,p}^{(\infty)}\ =\ E_{-m,p}^{(\infty)}\ +\ \frac{e\,\hbar \gamma\,m}{m_e c}\ .
\end{equation}
In one-dimensional quantum mechanics this phenomenon was called the {\it Energy Reflection Symmetry}
\cite{Sh-Tur:1999}.

The spectra of the Schr\"odinger equation (\ref{schroedinger_new}) consists of the infinite families characterized by different magnetic quantum numbers $m$, each family splits into two subfamilies of different parities. For fixed $m$ and $\nu$ the energy levels form infinitely-sheeted Riemann surface
in space of magnetic field $\gamma$ with square-root branch points, hence, there are quasi-crossings at real $\gamma \geq 0$ and two-level crossings at complex $\gamma$'s with vanishing discriminant (the Landau-Zener phenomenon, see e.g. \cite{LL:1977}). Levels with different $m$'s and/or different $\nu$'s can intersect without forming square-root branch points (true crossings). Interestingly, at large $\gamma$ the zone structure occurs, see e.g. \cite{Garstang:1977}. In particular, the lowest energy states with non-positive $m=0, -1, -2, \ldots$ and of positive parity ($1s_0, 2p_{-1}, 3d_{-2}, \ldots$) form zeroth (lowest) Landau zone, while in the case of negative parity ($2p_0, 3d_{-1}, 4f_{-2}, \ldots$) the first Landau zone occurs. Inside of these zones for sufficiently large $\gamma$ the energy levels can be ordered following the decrease of $m$, these levels never have quasi-crossings.
Higher Landau zones can be obtained through analytic continuation in $\gamma$ from the first two ones. Levels with $m=0$ define the lower edges of zones.

\subsection{Riccati Equation: Ground State}

\begin{center}
\begin{figure}[h]
\includegraphics[scale=1.0]{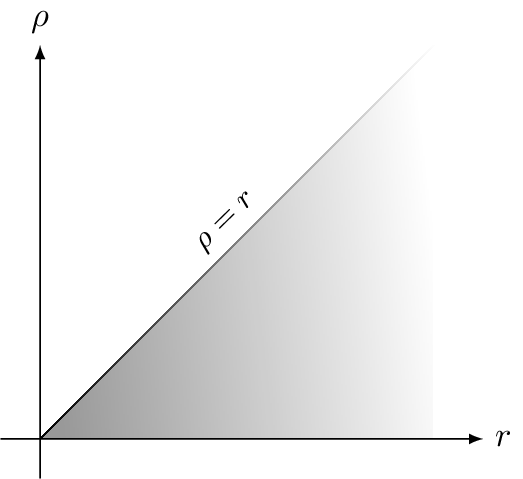}
\caption{Domain (shaded in gray) for the equation (\ref{Riccati}) in  $(\rho,r)$ variables.}
\label{fig:domain}
\end{figure}
\end{center}

For any magnetic field $\gamma$ the global ground state is non-degenerate and is characterized by the quantum numbers $m=0$ and $\nu=+ (p=0)$. It depends on two variables $(\rho,r)$ only. At $\gamma=0$ it corresponds to $1s_0$ state of the Hydrogen atom. At large $\gamma$ this state defines the lower edge of the lowest (zero) Landau zone.  From now on we write the ground wave function and its energy dropping labels corresponding to quantum numbers, presenting them as $\Psi$ and $\mathcal{E}^{(\infty)}$, respectively. Sometimes, this state is denoted as $1s_0$ even for $\gamma \neq 0$. Following (\ref{schroedinger_new}), the  equation that determines $\Psi$ and $\mathcal{E}$ reads
\begin{equation}
\label{Sch-rho-r-phi-0}
  -\frac{\hbar^2}{2m_e}\left[\pa_\rho^2\ +\ \frac{2\rho}{r}\pa_{\rho r}\ +\ \pa_r^2\ +\ \frac{1}{\rho}\pa_{\rho}\ +\ \frac{2}{r}\pa_r\right]\Psi\ +\ \left[\ -\frac{e^2}{r}\ +\ \frac{\gamma^2e^2}{8m_ec^2}\rho^2\right]\Psi\ =\
  \mathcal{E}^{(\infty)}\,\Psi\ .
\end{equation}
If $\Psi(\rho,r)$ is written in exponential representation,
\begin{equation}
\label{Phase}
  \Psi(\rho,r)\ =\ e^{-\Phi(\rho,r)}\ ,
\end{equation}
the \textit{phase} $\Phi(\rho,r)$ satisfies a non-linear partial differential equation of second order,
\[
  \pa_\rho^2\Phi\ +\ \frac{2\rho}{r}\pa_{\rho r}\Phi\ +\ \pa_r^2\Phi\ +\ \frac{1}{\rho}\pa_{\rho}\Phi\ +\    \frac{2}{r}\pa_r\Phi\ -\ (\pa_\rho\Phi)^2\ -\ \frac{2\rho}{r}(\pa_\rho\Phi)(\pa_r\Phi)\ -\ (\pa_r\Phi)^2
\]
\begin{equation}
\label{Riccati}
  \ =\ \frac{2m_e}{\hbar^2}\left[\mathcal{E}^{(\infty)}\ +\ \frac{e^2}{r}\ -\ \frac{\gamma^2e^2}{8m_ec^2}\rho^2\right]\ ,
\end{equation}
where $E^{(\infty)}=\mathcal{E}^{(\infty)}$, see (\ref{Eparity}). This equation is defined in the domain $0 \leq \rho \leq r$ and $0 \leq r < \infty$, see Fig. \ref{fig:domain}. At $\gamma=0$ this equation can be solved exactly: the solution corresponds to lowest Coulomb orbital,
\begin{equation}
\label{Coulomb-1s0}
  \Phi_0\ =\ \al r\ ,\ \mathcal{E}_0^{(\infty)}\ =\ -\frac{\hbar^2}{2m_e} \al^2  \ ,\ \al\ =\ \frac{m_e\,e^2}{\hbar^2}\ .
\end{equation}
Note that (\ref{Riccati}) can be regarded as a generalization to two dimensions of the well-known one-dimensional Riccati equation. We will call it {\it the (two-dimensional) Riccati equation}. The equation (\ref{Riccati}) is the key equation of the present work.

{ In Section III, we will rewrite the fundamental non-linear equation (II.15) in two forms introducing two sets of dimensionless variables but with the same effective dimensionless magnetic field $\lambda$ instead of the original magnetic field $|{\bf B}|=\gamma$. One equation is suitable to develop perturbation theory in powers of $\lambda$ and study the domain of small distances. Another equation can be used to
study the domain of large distances and develop semiclassical expansion. This information
is important to design our ground state trial function.
}

\vspace{-14mm}

{ \section{From Riccati Equation to Riccati-Bloch and to Generalized Bloch Equations}
\label{RBGB} }

\subsection{Riccati-Bloch Equation}

Let us introduce the dimensionless variables
\begin{equation}
 \label{changesRB}
 s\ =\ \frac{\rho}{a_0}\ ,\qquad t\ =\ \frac{r}{a_0}\ ,
\end{equation}
where
\begin{equation}
\label{r0}
 a_0=\frac{\hbar^2}{m_e\, e^2}\ \simeq\ 5.29 \times 10^{-9}\text{cm}\ ,
\end{equation}
is the Bohr radius. In new variables (\ref{changesRB}) the Riccati equation (\ref{Riccati}) appears without explicit dependence on parameters $c$, $e$, $\hbar$, and $m_e$,
\begin{equation}
\label{riccatiad}
 \pa_{s}^2\Phi\ +\ \frac{2s}{t}\pa_{s t}\Phi\ +\ \pa_{t}^2\Phi\ + \frac{1}{s}\pa_s\Phi\ +\ \frac{2}{t}\pa_t\Phi\ -\ (\pa_s\Phi)^2\ -\  \frac{2s}{t}(\pa_s\Phi)(\pa_t\Phi)\ -\ (\pa_t\Phi)^2\ =\ \veps\ +\ \frac{2}{t}\ -\ \frac{\la^2s^2}{4}\ ,
\end{equation}
where
\begin{equation}
\label{effenergy}
  \veps\ =\ \frac{\mathcal{E}^{(\infty)}}{\mathcal{E}^{(\infty)}_0}\ ,\qquad \mathcal{E}^{(\infty)}_0\ =\ \frac{m_e\, e^4}{2\hbar ^2} \ ,
\end{equation}
and
\begin{equation}
\label{effmag}
    \la\ = \frac{ \gamma}{\gamma_0}\ ,\qquad  \gamma_0\ =\ \frac{c\,e^3\, m_e^2}{\hbar^3}\ .
\end{equation}
Note that $\mathcal{E}^{(\infty)}_0$ is the Rydberg constant - the unit of energy, while $\gamma_0$ is the atomic unit of magnetic field (the magnetic field generated by the electron on the Bohr orbit), respectively,
\begin{align}
\label{au_energy}
   \mathcal{E}^{(\infty)}_0\ &\ \approx \ 2.18 \times 10^{-18} \text{J}\ =\ 13.6\,\text{eV}\ ,\\
\label{au_magnetic}
   \gamma_0\ &\ \approx\ 2.35 \times 10^5\, \text{T}\ =\ 2.35 \times10^9\,\text{G}\ .
\end{align}
Expressions (\ref{au_energy}) and (\ref{au_magnetic}) suggest that $\la$ is the
magnetic field measured in atomic units $\gamma_0$ (a.u.) \footnote{Sometimes, a.u. is defined with
$\gamma_0 = 2.3505 \times 10^5\, \text{T}$}, which occur instead of $\gamma$,
while $\veps$ plays the role of energy measured in Rydbergs (Ry).

Equation (\ref{riccatiad}) is nothing but the dimensionless version of the Riccati equation, we call it \textit{Riccati-Bloch (RB) equation} as in \cite{DELVALLE:2019-20} for the case of radial anharmonic oscillator. It is evident that both equations coincide when we set parameters $\hbar=1$, $m_e=1$ and $(-e)=1$. The Riccati-Bloch equation governs the dynamics via the phase $\Phi$ in the $(s,t)$-space. At zero magnetic field $\la=0$ the exact solution of (\ref{riccatiad}) reads,
\[
   \Phi_0\ =\ t\ ,\ \veps_0\ =\ -1\ ,
\]
see (\ref{Coulomb-1s0}). At $\la \rar \infty$ the leading behavior of the phase is given by
\[
   \Phi = \frac{\la}{4} s^2\ +\ \ldots \ ,
\]
it corresponds to Landau orbital, it is consistent with $\veps=\la$.

\subsection{Generalized Bloch Equation}

Let us introduce in (\ref{Riccati}) different dimensionless variable $u$ instead of $s$ keeping $t$ unchanged,
\begin{equation}
\label{changesGB}
u\ =\ \frac{\rho}{\rho_0}\ ,\qquad v\ =\ \frac{r}{a_0}\ =\ t\ ,
\end{equation}
where
\begin{equation}
\label{rho0}
    \rho_0\ =\ \frac{m_e\,c\,e}{\hbar\,\gamma}\ =\ \frac{l_{\gamma}^2}{a_0},
\end{equation}
cf. (\ref{changesRB}), $l_{\gamma}=\sqrt {\frac{\hbar c}{e\,\gamma}}$ is the magnetic length. Note that $\rho_0$ has explicit dependence on $\gamma$, unlike the Bohr radius $a_0$, being singular at $\gamma=0$, in turn, the variable $v$ coincides with $t$, see (\ref{changesRB}) and (\ref{changesGB}). Introducing (\ref{changesGB}) into the Riccati equation (\ref{Riccati}), we obtain two-dimensional \textit{Generalized Bloch (GB) equation},
\[
      \la^2\,\pa_u^2\Phi\ +\ \frac{2u}{v}\pa_{uv}\Phi\ +\ \pa_v^2\Phi\ +\ \frac{\la^2}{u}\pa_u\Phi\ +\ \frac{2}{v}\pa_v\Phi\ -\ \la^2(\pa_u\Phi)^2\ - \frac{2u}{v}(\pa_u\Phi)(\pa_v\Phi)\ -\ (\pa_v\Phi)^2
\]
\begin{equation}
\label{Bloch}
  \ =\ \veps\ +\ \frac{2}{v}\ -\ \frac{u^2}{4}\ ,
\end{equation}
cf. \cite{DELVALLE:2019-20,ESCOBAR:2016,ESCOBAR:2017}, where the potential in the rhs does not have any explicit dependence on the parameters of the problem including the magnetic field.
The definitions of $\varepsilon$ and $\la$ are given in (\ref{effenergy}) and (\ref{effmag}), respectively. Just as it occurred  for the Riccati-Bloch equation, all variables/quantities involved in (\ref{Bloch}) are dimensionless. The generalized Bloch equation (\ref{Bloch}) governs the dynamics via the phase $\Phi$ in the $(u,v)$-space.
Let us note that  variables $u$ and $s$ are related via remarkably easy relation
\begin{equation}
\label{relation}
u\ =\ \la\,s\ ,
\end{equation}
which allows to connect RB and GB equations, see (\ref{changesRB}), (\ref{effmag}) and (\ref{changesGB}). This relation is $\hbar$-independent: it holds for {\it any} value of $\hbar$. The variable $u$ looks similar to the classical coordinate introduced in \cite{ESCOBAR:2016}, see also \cite{DELVALLE:2019-20}.

{ \section{Perturbation theory and Asymptotic analysis}

In this Section we will obtain the basic building blocks to construct the ground state trial function. First of all, from the Riccati-Bloch equation we will determine the perturbative expansion of the energy $\varepsilon$ (\ref{effenergy}) in powers of $\la^2$ (\ref{effmag}). Second of all, we will derive from the Riccati-Bloch and Generalized Bloch equations the exact asymptotic behavior of the phase $\Phi$ in (\ref{Phase}) at small and large distances, respectively. An interpolation of all this information between two limits will lead to our trial ground state function. The connection between the RB and GB equations will be explained as well.
}

\vspace{-13mm}
\subsection{Weak Magnetic Field Expansion from the Riccati-Bloch Equation}
\vspace{-8mm}

One of the ways to study the RB equation (\ref{riccatiad}) in weak magnetic field regime  is to develop perturbation theory in powers of $\la^2$,
\begin{equation}
\label{ptrb}
     \Phi(s,t; \la^{2})\ =\ \sum_{n=0}^{\infty}\la^{2n}\Phi_n(s,t)\quad ,
     \quad \veps(\la^{2})\ =\ \sum_{n=0}^{\infty}\la^{2n}\veps_n\ .
\end{equation}
where the zero order approximation
\begin{equation}
\label{RB-0}
     \Phi_0(s,t)\ =\ t \quad , \quad \veps_0\ =\ -1\ ,
\end{equation}
see (\ref{Coulomb-1s0}), corresponds to the phase and energy of the $1s_0$ ground state of the hydrogen atom at $\gamma=0$, respectively. It must be emphasized that if $c=1$, $m_e=1$ and $e=1$ the coupling constant is equal to $\la=\gamma \hbar^3$ and the perturbation theory for energy $\veps$ in powers of $\gamma$ coincides (!) with semi-classical expansion in powers of $\hbar^3$.
Corrections $\veps_n$ are numbers. This statement is not valid for the perturbation expansion of phase $\Phi$: the $n$th correction depends explicitly on the Planck constant, $\Phi_n=\Phi_n(\rho \hbar^{-2},r \hbar^{-2})$.

The $n$th order corrections $\Phi_n$ and $\veps_n$ at $n \geq 1$ are determined by a linear partial differential equation,
\begin{equation}
\label{Phi_n-eq}
    \pa_{ss}\Phi_n\ +\ \frac{2s}{t}\pa_{st}\Phi_n\ +\ \pa_{tt}\Phi_n\ +\ \left(\frac{1}{s}-\frac{2s}{t}\right)\pa_s\Phi_n\ +\  2\,\left(\frac{1}{t}-1\right)\pa_t\Phi_n\ =\ \veps_n\ -\ Q_n\ ,
\end{equation}
where in the r.h.s.
\[
      Q_1\ =\ \frac{s^2}{4}\ ,
\]
plays the role of perturbation and at $n > 1$
\begin{equation*}
    Q_n\ =\ -\sum_{k=1}^{n-1} \left[\pa_s\Phi_k\pa_s\Phi_{n-k}\ +\ \pa_t\Phi_k\pa_t\Phi_{n-k}\ +\ \frac{s}{t}\left(\pa_s\Phi_k\pa_t\Phi_{n-k}\ +\ \pa_t\Phi_k\pa_s\Phi_{n-k}\right)\right]\ ,
\end{equation*}
is defined by the previous corrections as well as the energy correction,
\[
\veps_n=\frac{\int Q_n e^{-2\Phi_0} dV}{\int e^{-2\Phi_0} dV}\ .
\]
Eventually, the scheme leads to iterative procedure. The first order correction is
\begin{equation}
\label{Phi_1}
      \Phi_1(s,t)\ =\ \frac{1}{24}s^2t\ +\ \frac{1}{16}s^2\ +\frac{1}{24}t^2 \quad , \quad \veps_1\ =\ \frac{1}{2}\ .
\end{equation}
In general, the  $n$th correction $\Phi_n$ has the form of polynomial in variables $(s,t)$ of the following structure
\begin{equation}
\label{structure}
        \Phi_n(s,t)\ =\ \sum_{j=0}^{n-1}\sum_{k=j}^{n}\left(a_{j,k}^{(n)}\,t\ +\ b_{j,k}^{(n)}\right)\,(s^2)^{(n-k)}\,(t^2)^{(k-j)}\quad , \quad a_{0,n}^{(n)}\
        =\ 0\ .
\end{equation}
By substituting (\ref{structure}) in the equation (\ref{Phi_n-eq}) we arrive at the system of recurrence equations. Energy corrections $\veps_n$ are found following the consistency of the procedure and related to lowest order coefficients of $\Phi_n$.
Interestingly, for any $n>1$ the relation
\begin{equation}
\label{energyconnection}
         \veps_n\ =\ 4\,{b}_{n-1,n-1}^{(n)}\ +\ 6\,{b}_{n-1,n}^{(n)} \ ,
\end{equation}
holds. It is clear that for any integer $n$ the coefficients $a_{j,k}^{(n)}$ and $b_{j,k}^{(n)}$ are rational numbers. Hence, the energy correction $\veps_n$ is also rational number. Several corrections can be easily computed in this framework as a consequence of the polynomial nature of $\Phi_n$ in $(s,t)$ variables. The coefficients $a$ and $b$ in (\ref{structure}) are determined by solving recurrence relations by algebraic means. The construction of perturbation theory is ultimately an algebraic procedure. A finite number of terms in expansions (\ref{ptrb}) can be calculated explicitly.  In Appendix \ref{appendixA} the first three corrections in the expansion of the phase $\Phi$ for $n=2,3,4$ are presented. In turn, the expansion of $\veps$ in powers of $\la$ is easily computed up to 100th order in the form of rational numbers using MATHEMATICA 12, see Tables \ref{Table:upto10}, \ref{Table:upto100}. Following the Dyson instability argument  \cite{Dyson:1952}, in $(s,t)$-space (\ref{changesRB}) both series (\ref{ptrb}) should be divergent. Using multidimensional semi-classical analysis, the asymptotic behavior of $\veps_n$ at large order was found \cite{Avron:1979,Avron:1981} in the form of $1/n$ expansion
\begin{equation}
\label{Avron}
     \veps_n\ =\
       64\,\frac{(-1)^{n+1}}{\pi^{\frac{5}{2}+2n}}\,\Gamma\left(2n+\frac{3}{2}\right)\,
       \left(1\ -\ \frac{A}{n}\ +\ \mathcal{O}(\frac{1}{n^2})\right)
       \quad , \quad n\rar\infty\ ,
\end{equation}
with $A > 0$.
Thus, it has the index of divergence equals to 2: $\veps_n \sim (n!)^2$. The $1/n$-expansion
demonstrates a convergence, for example, at $n=70-100$ the leading contribution in (\ref{Avron}) agrees with exact $\veps_{70-100}$ in about 2 s.d. (depending on rounding) \footnote{It contradicts to the statement in \cite{Avron:1981} about the agreement in 9 s.d.}\,, see Table \ref{Table:upto100}. Next-to-leading term in (\ref{Avron}) with $A=2.61$ improves
the agreement to 4-5 s.d. In spite of the fact that PT for $\veps$ is Borel-summable, one of the best known summation procedure for asymptotic series - the Pad\'e-Borel procedure, see e.g. \cite{Bender-O}  - does not provide accurate results for large $\gamma \gtrsim 10$\,a.u. even taking into account the first hundred terms $\veps_{1-100}$. Although at small $\gamma \lesssim 1$\,a.u. it leads to accurate results providing 11 d.d. correct (or more), see below Table II for $\gamma=1$\,a.u. as the example.

Following the equation (\ref{energyconnection}) the coefficients $b_{{n-1},n-1}^{(n)}$ and $b_{{n-1},n}^{(n)}$ grow factorially at large $n$.
It is worth mentioning that the perturbative approach used to solve the Riccati-Bloch equation is nothing but an application of the so-called Non-Linearization Procedure \cite{Turbiner:1984} (sometimes referred for the ground state case as the Logarithmic Perturbation Theory). The general description can be found in \cite{Turbiner:1981}.

%\vskip -2.9cm

\subsubsection*{Behavior of phase $\Phi$ at Small Distances}

The structure of the Taylor series of the phase $\Phi$ at small $s$ and $t$ can be obtain from the polynomial form of the correction $\Phi_n$, see (\ref{structure}). Collecting the same degrees in
$s$ and $t$ coming from different corrections $\Phi_n$, their formal sums result in expansion
\begin{equation}
 \label{taylor}
  \Phi(s,t;\la^{2})\ =\ t\ +\ \sig_1(\la^2)\,s^2\ +\ \sig_2(\la^2)\,t^2\ +\ \sig_3(\la^2)\,s^2t\ +\
  \ldots \quad ,\quad (s,t) \rar 0 \ ,
\end{equation}
where the first functions $\sig_1$, $\sig_2$ and $\sig_3$ are given by
\begin{equation}
   \sig_1(\la^2)\ =\ \sum_{n=1}^{\infty}b_{n-1,n-1}^{(n)}\la^{2n}\ , \quad
   \sig_2(\la^2)\ =\ \sum_{n=1}^{\infty}b_{n-1,n}^{(n)}\la^{2n}\ ,\quad
   \sig_3(\la^2)\ =\ \sum_{n=1}^{\infty}a_{n-1,n-1}^{(n)}\la^{2n}\ .
\end{equation}
From equation (\ref{energyconnection}), it is clear that
\begin{equation}
\varepsilon(\lambda^2)\ =\ -1\ +\ 4\,\sig_1(\la^2)\ +\ 6\,\sig_2(\la^2)\ .
\end{equation}
From Taylor series (\ref{taylor}), once variables $\rho$ and $r$ are restored, one can immediately conclude that the presence of a magnetic field does not break the \textit{cusp condition} for the exact ground state  function
\begin{equation}
\label{cuspcondition}
   C\ \equiv \ -\frac{\Braket{\psi|\de(\vec{r})\frac{\pa}
   {\pa r}|\psi}}{\Braket{\psi|\de(\vec{r})|\psi}}\ =\ \frac{1}{a_0}\ .
\end{equation}
In atomic physics the parameter $C$ is known as the \textit{cusp parameter}.
For non-exactly-solvable Coulomb systems which consist of electrons and (infinitely massive) positive charged nuclei this parameter is used to \textit{measure} the local quality of the approximate wave function near the Coulomb singularities. This parameter has a meaning of residue at the Coulomb singularity. The easiest way to find the cusp parameter in approximate trial function is to calculate the coefficient (the slope) standing in front of linear in $r$ term at small $r$ behavior of the phase.

\subsection{The Weak Magnetic Field Expansion from the Generalized Bloch Equation}

One of the ways to solve the generalized Bloch equation (\ref{Bloch}) is to develop perturbation theory in powers of $\la^2$,
\begin{equation}
   \Phi(u,v;\la^{2})\ =\ \sum_{n=0}^{\infty}\la^{2n}\phi_n(u,v)\quad ,\quad
   \veps(\la^{2})\ =\ \sum_{n=0}^{\infty}\la^{2n}\veps_n\ .
\label{ptgb}
 \end{equation}
%Following normalizability (\ref{normalizabilty}), boundary condition for $\phi_n(u,v)$ reads
%\begin{equation}
%\phi_n(u,v)\ =\ 0,\qquad u,v\ \rar\ 0\ .
%\end{equation}
where the expansion for $\veps$ coincides with one presented in (\ref{ptrb}).
%Before proceed with the discussion on the calculation of correction $\phi_n$, let us note the following. Keeping all \textit{classical quantities} ($c,e,m_e,\gamma$) fixed, perturbation series (\ref{ptgb}) are expansions in powers of $\hbar^6$, see (\ref{effmag}). Therefore, in principle they can be regarded as semi-classical expansions of the phase and the energy, respectively. This statement was explicitly demonstrated for a special type of anharmonic oscillators in a path-integral formalism \cite{ESCOBAR:2016,ESCOBAR:2017} and also using standard WKB in quantum mechanics \cite{DELVALLE:2019-20}.
%
The zero order approximation $\phi_0(u,v)$ is determined by the non-linear partial differential equation of the second order
\begin{equation}
\label{GB-zero_order}
 \frac{2u}{v}\pa_{uv}\phi_0\ +\ \pa_v^2\phi_0\ +\ \frac{2}{v}\pa_v\phi_0\ - \frac{2u}{v}(\pa_u\phi_0)(\pa_v\phi_0)\ -\ (\pa_v\phi_0)^2\ =\
 \veps_0\ +\ \frac{2}{v}\ -\ \frac{u^2}{4} \ ,
\end{equation}
at $\veps_0=-1$.
Surprisingly, it can be solved explicitly in closed analytic form
\begin{equation}
\label{gen0}
 \phi_0(u,v)\ =\ A_0^{(0)}(u)\,v\ +\ B_0^{(0)}(u)\ ,
\end{equation}
where
\begin{equation*}
%\label{first-gen}
  A_0^{(0)}(u)\ =\ \sqrt{1+\frac{u^2}{12}}\ ,\qquad B_0^{(0)}(u)\ =\ \frac{1}{2}\log(1+\frac{u^2}{12})\ +\  \log(1+\sqrt{1+\frac{u^2}{12}})\ .
\end{equation*}
It is convenient to introduce a new variable
\begin{equation}
\label{w-variable}
     w\ =\ \sqrt{1+\frac{u^2}{12}}\ \geq \ 1\ ,
\end{equation}
it allows to represent the zero order approximation $\phi_0$ in compact form,
\begin{equation}
\label{firstgen-w}
  \phi_0(u,v)\ =\ w\,v \ +\ \log [w\,(1 + w)]\ .
\end{equation}
The $\phi_0(u,v)$ plays a role of classical action, although the classical trajectory is unknown.
Evidently, the function $\Psi_0=e^{-\phi_0(u,v)}$ is square integrable, it can be taken as variational trial function to study $(1s_0)$ state, see below.
The correction $\phi_n(u,v), n=1,2, \ldots $ obeys a linear partial differential equation,
\[
 \frac{2u}{v}\pa_{u,v}\phi_n\ +\ \pa_v^2\phi_n\ +\ \frac{2}{v}\pa_v\phi_n\ -\ \frac{2u}{v}\left(\pa_u\phi_0\,\pa_v\phi_n\ +\ \pa_u\phi_n\,\pa_v\phi_0\right)\ -\ 2(\pa_v\phi_n\,\pa_v\phi_0)
\]
 \begin{equation}
 \ =\ \varepsilon_n\ -\ q_n
 \end{equation}
where
\begin{align}
  q_n\ = &\ \pa_u^2\phi_{n-1}\ +\ \left(\frac{1}{u}-\pa_u\phi_0\right)\pa_u\phi_{n-1}\non
\\
  &\ -\sum_{k=1}^{n-1}\left\{ \pa_u\phi_{n-k-1}\pa_u\phi_{k}\ +\ \frac{2u}{v}\pa_u\phi_{n-k}\pa_v\phi_{k}\ +\ \pa_v\phi_{n-k}\pa_v\phi_{k}\right\}\ .
\end{align}
It can be shown that the correction $\phi_n(u,v)$ is a polynomial in $v$ of degree $(2n+1)$ with $u$-dependent coefficients,
\begin{equation}
\label{phigen}
 \phi_n(u,v)\ =\  \sum_{k=0}^{n}\left\{A_k^{(n)}(u)\,v\ +\ B_{k}^{(n)}(u)\right\}v^{2(n-k)}\ .
\end{equation}
Functions $A_k^{(n)}(u)$ and $B_k^{(n)}(u)$ are determined by solving (ordinary) linear differential  equations of first degree. In particular,
\begin{equation}
 -\phi_1(u,v)\ =\ A_{0}^{(1)}(u)\,v^3\ +\ B_0^{(1)}(u)\,v^2\ +\ A_{1}^{(1)}(u)\,v \ +\ B_{1}^{(1)}(u)
\end{equation}
where the coefficients can be written conveniently in variable $w$ (\ref{w-variable}) as follows
\begin{eqnarray*}
 A_{0}^{(1)}&\ =\ & \frac{(w-1) (w+1)}{120\,w^3}\ ,
\\
 B_0^{(1)}  &\ =\ & \frac{6 w^3-w^2-9 w-6}{120 \,(w+1)\,w^4}\ ,
\\
 A_{1}^{(1)}&\ =\ & \frac{(w-1) \left(30 w^4+52 w^3+54 w^2+42 w+15\right)}{120 \,(w+1)\,w^5}\ ,
\\
 B_{1}^{(1)}&\ =\ &\frac{(w-1) \left(9 w^6+18 w^5+38 w^4+46 w^3+42 w^2+30 w+10\right)}
    {80\,(w+1)\,w^6 }\ .
\end{eqnarray*}

Several next order corrections $\phi_2,\phi_3,\ldots$ can also be calculated explicitly, for all of them the coefficients $A$ and $B$ appear usually as rational functions in $w$.

In the next sub-Section, some properties of functions $A_k^{(n)}(u)$ and $B_{k}^{(n)}(u)$ related with their asymptotic behavior at large $u$ and $v$ are presented.

\subsubsection*{Asymptotic Behavior of $\Phi$ for Large Distances}

Using the perturbation theory corrections obtained from the Generalized Bloch equation (\ref{Bloch}), asymptotic expansions can be calculated along particular directions in the plane $(s,t)$. Let us consider the line
\begin{equation}
s\ =\ \al\,t \ ,
\end{equation}
where $\al \in(0,1]$ is a parameter. One can see that along this line (by keeping the value of $\al$ fixed), the dominant asymptotic behavior of the $n$th correction to the phase, see (\ref{ptgb}),
\begin{equation}
\label{lineasymp}
   \phi_n(\la s,t)\vert_{s=\al t}\ ,
\end{equation}
where the relation (\ref{relation}) is taken into account,
at large $t$ comes from the term $A_0^{(n)}(\la(\al t))t^{2n+1}$ in (\ref{phigen}) leading to
\begin{equation}
\label{asymptotic}
   \phi_n(\la\,s,t)_{s=\al t}\ \sim\ \frac{\mathcal{A}_n}{(\al\la)^{2n-1}}\,t^2\ +\ \mathcal{O}\left(t^{0}\right)\quad ,\quad t\rar\infty \quad ,
   \quad n\ =\ 0,1,2,\ldots \ ,
\end{equation}
where in the dominant term the coefficients $\mathcal{A}_n$ for the first three corrections $n=0,1,2$ are
\begin{equation*}
    \mathcal{A}_0\ =\ \frac{1}{2\sqrt{3}}\quad ,\quad \mathcal{A}_1\ =\ -\frac{1}{20\sqrt{3}}\quad ,\quad\mathcal{A}_2\ =\ -\frac{23}{2800\sqrt{3}}\ .
\end{equation*}
Finally, from (\ref{asymptotic}) one can find the behavior of $\Phi(\la\,s,t)\vert_{s=\al t}$\,,
\begin{equation}
\label{explicitasymp}
    \Phi(\la\,s,t)_{s=\al t}\ =\ \la \al\left(\sum_{n=0}^\infty \mathcal{A}_n\al^{-2n}\right)t^2\ +\ 2\log(t)\ +\ \ldots \quad ,\quad t\ \rar\ \infty\ ,
\end{equation}
where logarithmic term comes from the $B_0^{(0)}(u)$, see (\ref{gen0}). In order to determine the sum in ( \ref{explicitasymp}), we can define the generating function
\begin{equation}
 \mathcal{A}(\al)\ =\ \al\sum_{n=0}^\infty \mathcal{A}_n \al^{-2n}\ ,
\end{equation}
which satisfies the equation,
\begin{equation}
  (1-\al^2)\,(\mathcal{A}')^2\ +\ 4\,\mathcal{A}^2\ -\ \frac{1}{4}\,\al^2\ =\ 0 \quad ,\quad \mathcal{A}(1)\ =\ \frac{1}{4}\ .
\end{equation}
This equation can be solved in closed analytic form
\begin{equation}
\mathcal{A}(\al)\ =\ \frac{1}{4}\,\al^2\ .
\end{equation}
Hence, the behavior of the phase (\ref{lineasymp}) at $t\rar\infty$ results in
\begin{equation}
\label{arho}
\Phi(\la\,s,t)\vert_{s=\al t}\ =\ \frac{1}{4}\,\al^2\,\la\,t^2\ +\ 2\log(t)\ +\ O(t^{0})\ ,\qquad\qquad t \rar \infty\ .
\end{equation}
%There is another relevant limit, it corresponds to the behavior of $\Phi(\rho,r)$ at large distances when $z$ is fixed, say $z=z_0$. Following (\ref{potential}), it constraints the coordinates $(\rho,r)$ to lay in the hyperbola defined by the equation
%\begin{equation}
%\left(\frac{r}{z_0}\right)^2\ -\ \left(\frac{\rho}{z_0}\right)^2\ =\ 1\ .
%\end{equation}
%If $r$ is large, $\rho\sim r$, hence the hyperbola is very close to its asymptote ($\rho=r$). It implies that we return to case discussed above.

Some other limits in different directions can be also studied. For example, if $s$ is fixed, $s=s_0$, the asymptotic behavior at large $t$ is of the form
\begin{equation}
\label{s0const}
   \Phi(s_0,t)\ =\ C_0(s_0,\la)\,t\ +\ C_1(s_0,\la)\,\log t\ +\ O(t^{0}) \quad , \quad
   t\rar\infty\ .
\end{equation}
where constants $C_1$ and $C_2$ are unknown generally.

\subsection{Connection between Riccati-Bloch and Generalized Bloch Equations}

We have constructed two different representations for the phase $\Phi$ (\ref{Phase}) of the ground state wave function. From RB equation it is obtained
\begin{equation}
\label{expst}
   \Phi(s,t;\la^2)\ =\ \sum_{n=0}^{\infty}\la^{2n}\Phi_n(s,t)\ ,\qquad s\ =\ \frac{\rho}{a_0}\ ,\quad t\ =\ \frac{r}{a_0}\ ,
\end{equation}
while from the GB equation
\begin{equation}
\label{expuv}
    \Phi(u,v;\la^2)\ =\ \sum_{n=0}^{\infty}\la^{2n}\phi_n(u,v)\ ,\qquad u\ =\ \frac{\rho}{\rho_0}\ ,\quad v\ =\ \frac{r}{a_0}\ ,
\end{equation}
where $a_0$ and $\rho_0$ are defined in (\ref{r0}), (\ref{rho0}), respectively.

It is clear that there must exist a connection between corrections $\Phi_n(s,t)$ and $\phi_n(u,v)$. It order to establish it we use the polynomial representation of corrections (\ref{structure}) in  (\ref{expst})
\begin{equation}
\label{sumphin}
   \Phi\ =\ \sum_{n=0}^{\infty}\la^{2n}\sum_{j=0}^{n}\sum_{k=j}^{n}\left(a_{j,k}^{(n)}\,t\ +\ b_{j,k}^{(n)}\right)s^{2(n-k)}t^{2(k-j)}\ ,\ a_{0,0}^{(0)}=1\ ,\  b_{0,0}^{(0)}=0\ ,
\end{equation}
then change the order of summation and use the relation (\ref{relation}) between variables $s$ and $u$. As the result we arrive at
\begin{equation}
   \Phi \ =\ \sum_{n=0}^{\infty}\la^{2n}\sum_{k=0}^{n}\sum_{j=n}^{\infty}\left(a_{k,n}^{(j)}\,v\ +\ b_{k,n}^{(j)}\right)u^{2(j-n)}v^{2(n-k)}\ .
\label{sumphin*}
\end{equation}
Comparing (\ref{expuv}) and (\ref{sumphin*}) one can conclude that
\begin{equation}
\label{phin}
    \phi_n(u,v)\ =\ \sum_{k=0}^{n}\sum_{j=n}^{\infty}\left(a_{k,n}^{(j)}\,v\ +\ b_{k,n}^{(j)}\right)u^{2(j-n)}v^{2(n-k)}\ .
\end{equation}
Eventually, after simple manipulations the equation (\ref{phin}) can be written as follows
\begin{equation}
\label{phigen*}
    \phi_n(u,v)\ =\ \sum_{k=0}^{n}v^{2(n-k)}\sum_{j=0}^{\infty}\left(a_{k,n}^{(n+j)}\,v\ +\ b_{k,n}^{(n+j)}\right)u^{2j}\ .
\end{equation}
Making a comparison (\ref{phigen}) and (\ref{phigen*}) we see explicitly that
\begin{equation}
      A_k^{(n)}(u)\ =\ \sum_{j=0}^{\infty}a_{k,n}^{(n+j)}u^{2j}\ ,
\end{equation}
and
\begin{equation}
      B_k^{(n)}(u)\ =\ \sum_{j=0}^{\infty}b_{k,n}^{(n+j)}u^{2j}\ .
\end{equation}
Therefore, the meaning of the connection between the expansion of phase in generalized Bloch an Riccati-Bloch equations is the following: the coefficient functions $A_k^{(n)}(u)$ and $B_k^{(n)}(u)$ are nothing but the generating functions of the coefficients $a_{k,n}^{(n+j)}$ and $b_{k,n}^{(n+j)}$, respectively. Similar connection exists for anharmonic oscillators \cite{Turbiner:1984,DELVALLE:2019-20}.

{ \section{The Approximant}}

\subsection{Approximant of the Ground State}

The analytical information for the phase (\ref{Phase}), obtained from the Riccati-Bloch  and  Generalized Bloch equations, Taylor expansion at small distances and asymptotic series for large distances, will be now used to design the \textit{Approximant}: an approximation of the exact (unknown) ground state wave function $(1 s_0)$, denoted by $\Psi^{(t)}$, in the form of interpolations of different expansions.
To do so, we follow the prescription proposed in \cite{Turbiner:1981,Turbiner:1984}, further
developed and applied in \cite{TURBINER:2005,TURBINER:2010,DELVALLE:2019-20}, where it was successfully constructed the Approximant for 1D anharmonic oscillator and double-well potential, for the $D$-dimensional radial polynomial anharmonic oscillator and some other potentials.

We assume the exponential representation (\ref{Phase}) for the Approximant in coordinates
$(\rho,r)$,
\begin{equation}
\label{Psit}
  \Psi^{(t)}(\rho,r)\ =\ e^{-\Phi_t(\rho,r)}\ ,
\end{equation}
and focus on the construction of $\Phi_{t}(\rho,r)$. According to the prescription,
the approximate phase has to interpolate the expansions at small and large distances, see (\ref{taylor}), (\ref{arho}), and (\ref{s0const}).
%One can intermediately check that the simplest interpolation with such properties is given by
%\begin{equation}
%\Phi(\rho,r)\ =\ a\,r\ +\ b\,\gamma\,\rho^2\ ,
%\label{simplesttrial}
%\end{equation}
%where $a$ and $b$ are free parameters. In fact, this \textit{simplest and  physically adequate} trial function was introduced  in \cite{Turbiner:1984} and leads to estimates of the energy via the Variational Method with sufficiently high accuracy for small magnetic fields.
%
In addition, the zero order approximation $\phi_0$ in GB equation, see (\ref{gen0}), should be reproduced for particular values of parameters as well as the first terms in weak magnetic field expansions. Following the reflection symmetry $(\rho \rar -\rho)$, which holds in RB and GB equations, $\Phi_t$ has to be function of  $\rho^2$. One of the simplest interpolations, which accomplishes the prescription given above, is of the form,
\begin{equation}
\label{approximatephase}
    \Phi_t(\rho,r)\ = \ \frac{\al_0\,+\,\al_1\,r\,+\,\al_2\,r^2\,+\,\al_3\,\gamma\,\rho^2+\al_4\,\gamma\,\rho^2\,r}
    {\sqrt{1 + \beta_0 w + \beta_1\,r + \beta_2\,r^2+\beta_3\,\rho^2}}
    \ +\
    q\,\log(1 + \beta_0 w + \beta_1\,r + \beta_2\,r^2 + \beta_3\,\rho^2)\ ,
\end{equation}
where $\{\al_0,\al_1,\al_2,\al_3,\al_4,\beta_0,\beta_1,\beta_2,\beta_3;q\}$ are ten free parameters that later will be fixed in variational calculation, $w=\sqrt{1+\frac{\gamma ^2 \rho ^2}{12}}$, see (\ref{w-variable}). We call it the {\it Phase Approximant}. Making straightforward preliminary minimization we found that for all studied magnetic fields up to 10,000\,a.u. the parameter $\beta_0$ is extremely small and parameter $q$ is invariably close to 1 \footnote{In detailed minimization it was found that at large magnetic fields $\gamma > 100$\,a.u. the parameter $q$ jumps sharply down to zero, while the variational energies at $q=1$ and $q=0$ are very close. Present authors have no explanation of this phenomenon.}. Thus, without loosing much in accuracy in variational energy we put $\beta_0=0$ and $q=1$ from the very beginning in trial function (\ref{approximant1s0}), see below, which becomes 8-parametric in minimization procedure. Interestingly, the deviation of $q$ from 1
and $\beta_0$ from 0 influences far distant decimal digits in energy, although it significantly improves the cusp parameter $C^{(t)}$ (\ref{cuspcondition}) making it closer to the exact value $C^{(exact)}=1$. Releasing those parameters makes minimization procedure much more complicated and slow.
It will be checked for two magnetic fields, 1\,a.u. and 10,000\,a.u. only.

The logarithmic term (\ref{approximatephase}) is included to mimic the appearance of logarithmic terms in the exact wave function treated in the GB equation at $\la \rar 0$,  see (\ref{gen0}). It will generate a pre-factor in the approximate wave function, which ultimately is given by
\[
   \Psi_{1s_0}^{(t)}(\rho,r)\ =\
\]
\begin{equation}
\label{approximant1s0}
    \frac{1}{(1+ \beta_0 w+\beta_1\,r+ \beta_2\,r^2+\beta_3\,\rho^2)^q}\,
    \exp\left(-\frac{\al_0\,+\,\al_1\,r\,+\,\al_2\,r^2\,+\,\al_3\,\gamma\,\rho^2\,+
    \,\al_4\,\gamma\,\rho^2\,r}
    {\sqrt{1+ \beta_0 w\,+\,\beta_1\,r\, +\,\beta_2\,r^2\,+\,\beta_3\,\rho^2}}\right)\ .
\end{equation}
This is the key expression for the trial function which is going to be used through all this paper. We have labeled the Approximant (\ref{approximant1s0}) with $(1s_0)$, using the standard notation for the ground state of hydrogen atom in absence of magnetic field. Calculations are made mainly with 8-parametric trial function (\ref{approximant1s0}) with $\beta_0=0$ and $q=1$. The 10-parametric trial function (\ref{approximant1s0}) with both $\beta_0, q$ as variational parameters is used for two magnetic fields $\gamma=1$ and $10^4$\,a.u. only in order to estimate the maximal accuracy which can be reached with it, see discussion above.

Note that the reduced and modified versions of the trial function (\ref{approximant1s0}) (and the trial phase (\ref{approximatephase})) had appeared in the previous investigations:

(i) If $\beta_0=\beta_1=\beta_2=\beta_3=0$, thus, the pre-factor is reduced to one, if $\al_0=\al_2=\al_4=0$, this function becomes the product of the ground state Coulomb orbital and the ground state Landau orbital, see \cite{Turbiner:1984}; if $\al_0=\al_1=\al_4=0$, the function becomes
one which was originally proposed by Yafet et al, \cite{Yafet:1956};

(ii) If $\beta_1=\beta_2=\beta_3=0$, and $\al_0=0$ \cite{Larsen:1982}; in spite of giving wrong asymptotic behavior at large distance, this function  provided sufficiently high accuracy at small magnetic fields which then deteriorated with magnetic field increase: at $\gamma=100$\, a.u. it reproduces in the ground state energy 3 s.d. only;

(iii) For $\beta_1=\beta_2=\beta_3=0$, there were other attempts to modify the numerator in the first term in (\ref{approximatephase}), see e.g. \cite{Turbiner:1987,Potekhin:2001}, to keep functionally correct asymptotic behavior at large distances and adding as many as possible free parameters; none of these attempts allowed to get accurate results for energy beyond 2-3 s.d. for magnetic fields larger than 10\,a.u.

(iv) In 2007 one of the present authors (AVT, see \cite{Turbiner:2007}) demonstrated that even keeping the pre-factor in (\ref{approximant1s0}) {\it equal to one}, $q=0$ (thus, no logarithmic term in (\ref{approximatephase})) and taking $\al_0=\beta_1=0$ in exponent (thus, having in total six free parameters) it allows for $\gamma=10\,000$\,a.u. to obtain the variational energy 9972.05\,a.u., which differs from the exact value in the 5th s.d.; many years later, recently, another one among the present authors (JCdV, see \cite{delValle:2016}) carrying out more accurate minimization procedure was able to improve the above result up to 9971.95\,a.u. and then performed  variational studies in the whole domain $\gamma \in [0, 10\,000]$\ a.u. reaching the accuracy of 3-4 s.d. (which was even higher for weak magnetic fields).

(v) It is worth noting that if $\beta_1=\beta_2=0$\,,\, $\al_0=\al_2=\al_3=0$ and $q=1$,
then for
$\al_1=1$ and $\gamma \,\al_4 =\beta_3= \frac{1}{12\,\rho_0^2}$ the non-logarithmic term in (\ref{approximatephase}) coincides to the non-logarithmic term of zero-order approximation $A_0^{(0)}(u)\,v$ (\ref{gen0}) to GB equation while the logarithmic terms in $B_0^{(0)}(u)$
coincide with logarithmic term in (\ref{approximatephase}) at $\beta_0=1$. Hence, the trial function (\ref{approximant1s0}) is reduced to
\begin{equation}
\label{approximant0}
    \Psi_0\ =\ \frac{1}{\left(1+\sqrt{1+\frac{\gamma^2 \rho ^2}{12}}+\frac{\gamma^2 \rho ^2}{12}\right)}\ e^{-r \sqrt{1+\frac{\gamma^2 \rho ^2}{12}}}\ ,
\end{equation}
which contains no free parameters. This function leads to very accurate energies at
$\gamma \leq 1$\,a.u., see Table VIII, cf. Table I, but it fails for larger $\gamma$. Surprisingly, if pre-factor in 
(\ref{approximant0}) is dropped by putting $q=0$ in (\ref{approximatephase}), (\ref{approximant1s0}),
then the amazingly simple, parameter-free function
\begin{equation}
\label{approximant1}
    \Psi_1\ = \ e^{-r \sqrt{1+\frac{\gamma^2 \rho ^2}{12}}}\ ,
\end{equation}
leads to reasonably accurate energies for $\gamma \geq 10$\,a.u. \cite{Turbiner:1987}. For both formulas (\ref{approximant0})-(\ref{approximant1}) the value of cusp parameter is exact being equal to 1 for all $\gamma$.

\vskip -0.3cm

As was indicated before the variational method allows us to fix the values of the free parameters of the Approximant (\ref{approximant1s0}) by minimizing variational energy $E_{var}$: it is calculated the parameter-dependent, expectation value of the Hamiltonian (\ref{Hamiltonian}) over trial function (\ref{approximant1s0}), which then it is minimized. The variational principle guarantees that the variational energy gives an upper bound of the exact energy, $E_{var} \geq E_{exact}$. However, it is still open question how close the variational energy is to the exact one. Hence, the quality of the variational results should be checked by making comparison
with reliable established data obtained independently. Such data are supplied by the Lagrange
Mesh Method (LMM) \cite{Baye:2015}, which is proved to be among the most reliable numerical methods, where convergence can be easily established.

\subsubsection{Lagrange Mesh Method: Results}

In this Section we will consider the hydrogen atom in a constant magnetic field in the Lagrange Mesh Method (LMM), for review see \cite{Baye:2015}, which is alternative to the variational method in order to establish benchmark results.

It has been known for quite some time that LMM is a highly accurate method leading to benchmark results and also simple to implement in order to solve the Schr\"odinger equation, see \cite{Baye:2008}. Using the formulation of the method in spherical coordinates $(r,\tha,\phi)$ presented in \cite{Baye:2008},  Section 2.5, we calculated the ground state energy and its quadrupole moment for magnetic fields in the range $\gamma \in [0.01,10\,000]$ a.u. Since the Schr\"odinger equation for the ground state is essentially two-dimensional: $\phi$ dependence is absent, the mesh is realized on the plane parametrized by $r$ and $u=\cos \tha$. We implemented the LMM in MATHEMATICA~12.
For the whole range of studied magnetic fields the mesh was kept unchanged and consisted of $N_r=80$ radial functions and $N_u=200$ angular ones. Hence, the approximate ground state function is represented by the expansion in terms of $80 \times 200=16\,000$ functions. With this enormous mesh we are able to reproduce and confirm the results obtained for all magnetic fields studied in previous LMM calculation \cite{Baye:2008}: $1, 10, 100\ \mbox{and}\ 1\,000$ a.u. and furthermore improve them, see Table \ref{Table:Energy}. It is shown in Table \ref{Table:Energy} the maximal number of figures obtained by other methods which are in agreement
with the LMM results. It ranges from 11 figures for $\gamma=0.01$\,a.u. up to 6 figures for $10\,000$\,a.u.
From the point of physics such accuracies are excessive since finite mass contribution changes usually the 4th figure, see a discussion in the Part II.

In general, for $\gamma \lesssim 1\,000$ a.u. the LMM allows to reach the accuracy in energy of 10 figures (or more) giving benchmark results except for the outstanding results by Stubbins et al, \cite{Stubbins:2004} which are checked and confirmed with high accuracy, see e.g. Table \ref{historic}. The maximum accuracy in energy - 19 figures - is reached for magnetic fields $\gamma \lesssim 1$ a.u., see e.g. Table \ref{historic} as for $\gamma=1$\,a.u. However, for large magnetic fields $\gamma \gtrsim 1\,000$ a.u. the accuracy begins to reduce dramatically. For example, at $\gamma=10\,000$ a.u. it reaches only 6 figures.
The results for energy and quadrupole moment are shown in Table \ref{Table:Energy}. It must be emphasized that for quadrupole moment the LMM - the present calculation and one performed in \cite{Baye:2008} lead to benchmark results.

In order to check the local accuracy of the solution for the eigenfunction of the Schr\"odinger equation obtained in the LMM one can calculate the cusp parameter $C$ (\ref{cuspcondition}). As the result, the calculated cusp parameter $C$ deviates from the exact value, $C^{(exact)}=1$, in the 6th d.d. in the whole range of magnetic fields considered.

%\vskip -3.5cm

\subsubsection{Variational Results}

For simplicity we set $c=\hbar=e=m_e=1$ in numerical computations, see (\ref{effenergy}).
In order to calculate the expectation value of the Hamiltonian over the trial function (\ref{approximant1s0}) and minimize it with respect to free parameters to get the optimal variational energy we need to perform the numerical integration and then minimization.
The computer code was written in FORTRAN 77 with use of the integration routine D01FCF from NAG-LIB employing the minimization routine MINUIT from CERN-LIB. Resulting variational parameters
are presented in Appendix B. Variational energies are shown in Table \ref{Table:Energy} for different magnetic fields in the range $\gamma \in [0.01,10\,000]$ a.u.

Our variational calculations with trial function (\ref{approximant1s0}) are compared with accurate results known in the past, in particular, with those obtained in the LMM  \cite{Baye:2015} in the paper \cite{Baye:2008}, which are extended and improved in the present paper, see Section G.1. It was paid a special attention to the {\it critical magnetic field} $\gamma_c$, for which the ground state energy (II.6) vanishes,
\begin{equation*}
    E^{(\infty)}(\gamma_c)\ =\ 0\ .
\end{equation*}
The value of $\gamma_c$, obtained in LMM, results in
\begin{equation}
\label{gamma_0}
 \gamma_c^{(LMM)}\ =\ 2.065\,211\,858\,a.u.\ ,
\end{equation}
with $E^{(\infty)}_{LMM}(\gamma_c)\ \sim\ 10^{-10}$\,Ry, while with variational trial function (\ref{approximant1s0}) with 8 parameters it gives $E_{var}^{(\infty)}(\gamma_c)\sim 10^{-6}$\,Ry,
see Table \ref{Table:Energy}. Note that for the magnetic field $\gamma_c$ the Hamiltonian (\ref{Hamiltonian}) has the normalizable zero mode. Surprisingly, for this magnetic field the value of quadrupole moment $(-Q_{zz})$ appears to be close to its maximal value
\footnote{In LMM the maximum of the quadrupole moment is reached at $\gamma=2.968 69$\,a.u.: $\max{(-Q_{zz})} = 0.524 52$\,(a.u.)${}^2$.}.

For $\gamma=10,000$\,a.u., see Table \ref{Table:Energy}, the LMM allows to reach 6 figures in energy {\it only} even taking 16K mesh points basis, while in the calculation by Wang-Hsue \cite{Wang:1995}, based on use of splines, it was reached 10 figures.
Taking the 8-parametric function (\ref{approximant1s0}) with $(q=1,\beta_0=0)$ the energy differs from established value in 6th figure in two units, while taking $(q=0,\beta_0=0)$ the 6 figures in energy are reproduced exactly. The 10-parametric function (\ref{approximant1s0}), where the parameters $(q,c)$ are released, allows us to reproduced 7 figures with difference in 2 units in 8th figure in comparison with results obtained in \cite{Wang:1995}.

In general, the relative deviation of the variational energy from the exact one is small
in the whole domain of considered magnetic fields,
\begin{equation}
 \left| \frac{E^{(\infty)}_{var}-E^{(\infty)}_{exact}}{E^{(\infty)}_{exact}} \right|
 \lesssim 10^{-6}\quad ,\quad \gamma \in [0.01,10\,000]\ .
\end{equation}

Due to Howard-Hasegawa \cite{Hasegawa-Howard:1961}, see also \cite{Avron:1981}, the ground state energy at large $\gamma$ behaves like
\begin{equation}
\label{HH}
   E^{(\infty)}\ =\ \gamma \ -\ \log^2 \gamma + O(\log \gamma)\ .
\end{equation}
Here the second term defines the asymptotic behavior at large $\gamma$ of the binding energy, $E_{binding}^{(\infty)}=\gamma - E^{(\infty)}$. It can be immediately seen that the asymptotic expansion (\ref{HH}) is slow convergent: even at $\gamma=10^4$\,a.u. the binding energy is equal to $\sim 28.25$\,Ry, see Table I, it differs from $\log^2 \gamma = 16$\,Ry in $\sim 50$\%\,. This magnetic field is close to the Schwinger limit $\gamma_{schwinger} \sim 2 \times 10^4$\,a.u., which limits the domain of applicability of non-relativistic quantum mechanics. It implies that the expansion (\ref{HH}) does not seem relevant to study the non-relativistic domain
$\gamma \lesssim \gamma_{schwinger}$.

\begin{table}[ht]
\centering
\caption{Energies $E^{(\infty)}$ (\ref{schroedinger}) and $E^{(m_p)}$ (\ref{spsi}) in Ry and
    quadrupole moment $Q_{zz}^{(\infty)}$ in $(a.u.)^2$
	for the ground state $1s_0$ of the static Hydrogen atom in magnetic field
   $\gamma \in [0.01, 10 000]$\,in a.u. found in Variational Method with 8-parametric trial function (\ref{approximant1s0}) $(q=1, \beta_0=0)$ and comparison with results of other calculations (rounded),
   confirmed and established in LMM.}
\label{Table:Energy}
\begin{threeparttable}
{\setlength{\tabcolsep}{0.3cm}
\begin{tabular}{clll}
\hline
\hline
$\gamma$\,(a.u.) & \hskip1.0cm$ E^{(\infty)}$     & $E^{(m_p)}$&-$Q^{(\infty)}_{zz}$  \\
\hline
\rule{0pt}{4ex}0&-1.000\,000\,000\,000&-0.999\,455\,679\,426 &0.000\,000
\\[5pt]
0.01           & -0.999\,950\,005\,51           & -0.999\,405\,603\,19 & 0.000\,248                 \\
       	       & -0.999\,950\,005\,52 $^{a,e}$&  & 0.000\,249 $^{e}$
\\[5pt]
0.1            & -0.995\,052\,960\,5            & -0.994\,500\,663\,9  & 0.023\,270                 \\
		       & -0.995\,052\,960\,8 $^{a,e}$    &                      & 0.023\,2712 $^{e}$     \\[5pt]
0.5            & -0.894\,421\,065               & -0.893\,731\,173     & 0.256\,143                 \\
     	       & -0.894\,421\,075 $^{a,e}$       &                      & 0.256\,156\,21 $^{e}$
\\[5pt]
1.0            & -0.662\,337\,66                & -0.661\,393\,27      & 0.417\,618\                 \\
               & -0.662\,337\,70                & -0.661\,393\,31      & 0.417\,635\ \ ($\star$)                 \\
   		       & -0.662\,337\,79 $^{a,d}$        &                      & 0.417\,654 $^{d}$
\\[5pt]
2.0            & -0.044\,426\,7                 & -0.042\,924\,9       & 0.511\,354                 \\
		       & -0.044\,427\,8 $^{a,e}$         &                      & 0.511\,432 $^{e}$       \\[5pt]
$\gamma_c$     & 0.000\,001                     & 0.001\,540           & 0.513\,561                 \\
   		       & 0.000\,000 $^{e}$               &                      & 0.513\,537 $^{e}$
\\[5pt]
5.0            & 2.239\,209                     & 2.242\,422           & 0.506\,493
\\
    	       & 2.239\,202 $^{a,e}$             &                      & 0.506\,331 $^{e}$
\\[5pt]			
10.0           & 6.504\,427                     & 6.510\,476           & 0.445\,22                   \\
   		       & 6.504\,405 $^{a,d}$            &                      & 0.445\,09 $^{d}$
\\[5pt]
100.0          & 92.420\,7                      & 92.476\,6            & 0.217\,5                     \\
   		       & 92.420\,4 $^{a,d}$             &                      & 0.216\,8 $^{d}$
\\[5pt]
500.0          & 487.487\,31                    & 487.762              & 0.125\,1                     \\
               & 487.485\,9477                  &                      &0.123\,87 ($\dagger$)
\\
   		       & 487.485\,82 $^{a,e}$           &                      & 0.123\,87 $^{e}$
\\[5pt]
1\,000.0       & 984.678                        & 985.226              & 0.099\,4                     \\
   		       & 984.675 $^{a,d}$               &                      & 0.098\,2 $^{d}$
\\[5pt]
10\,000.0      & 9\,971.74                      & 9\,977.22            & 0.049\,3 $(q=1,\beta_0=0)$                    \\
		       & 9\,971.72                      & 9\,977.18            & 0.049\,1 $(q=0,\beta_0=0)$
\\
               & 9\,971.718\,490                & 9\,977.173           & 0.048\,5\ ($\star \star$)
\\
		       & 9\,971.72 $^{e}$               &                      & 0.047\,9 $^{e}$
\\
		       & 9\,971.718 316 $^{f}$          &                      &           \\[5pt]   		                   		
\hline
\hline
\end{tabular}}
\begin{tablenotes}
\small
\item ($\star$)\ Eq.(I.65), 10 parameters, $q\, =\, 0.994 509,\ \beta_0\, =\,  0.000 000 003 6$,
      see~Table~II
\item ($\dagger$)  Eq.(I.65), 10 parameters, $q\, =\, -0.130\,064\ , \beta_0\, =\,  0.003\,583$
\item ($\star \star$)\ Eq.(I.65), 10 parameters, $q\, =\, -0.089 408,\ \beta_0\, =\,  0.000 029$
\item $\gamma_c=2.065\,211\,858$
\item $^{a}$\,Power series - method of moments \cite{Kravchenko:1996},
%      $^{{c}}$\,Variational Method \cite{Potekhin:2001},
      $^{{d}}$\,Lagrange Mesh \cite{Baye:2008},
      $^{{e}}$\,Lagrange Mesh (present work, 16K mesh points),
      $^{{f}}$\,Basis of Splines \cite{Wang:1995}.
\end{tablenotes}
\end{threeparttable}
\end{table}

\begin{table}[ht]
\caption{The ground state energy $E^{(\infty)}$ in Ry for the static Hydrogen atom at magnetic field $\gamma=1$\,a.u. obtained by different methods. The results ranked by accuracy. Rounding up to digits relevant for comparison performed, excessive digits not confirmed by the most accurate calculations not shown. Digits beyond 12th decimal having no chance to be verified experimentally at present times (see text) shown by italics. Mass effects change 4th~d.d., see Table I}
\label{historic}
{\setlength{\tabcolsep}{0.9cm}
\resizebox{\textwidth}{!}{
\begin{tabular}{|lll|}
\hline
\hline
\rule{0pt}{4ex}		
               Reference     &       $E^{(\infty)}$               & Method \\[5pt]
\hline
\rule{-4pt}{6ex}
\cite{Yafet:1956}
Yafet et al., 1956           &-0.523                                    & Variational	
\\[3pt]
\cite{Turbiner:1984}
Turbiner, 1984               &-0.61                                     & Variational
\\[3pt]
\cite{Larsen:1968}
Larsen, 1968                 &-0.661                                    & Variational
\\[3pt]
\cite{Praddaude:1972}
Praddaude, 1972              &-0.662\,33                                & Power Series
\\[3pt]
\cite{Potekhin:2001}
Potekhin \& Turbiner, 2001   &-0.662\,332                               & Variational
\\[3pt]
\textBF{Present Work}, see
(\ref{approximant1s0})       &\textBF{-0.662\,337\,66}                  & \textBF{Variational}
(8 parameters)\ ${}^a$
\\[3pt]
\textBF{Present Work}, see
(\ref{approximant1s0})       &\textBF{-0.662\,337\,70}                  & \textBF{Variational}
(10 parameters)
\\[3pt]
\cite{Wang:1995}
Wang and Hsue, 1995          & -0.662\,337\,785                         & B splines
\\[3pt]
\textBF{Present Work}        & \textBF{-0.662\,337\,793\,46}            &\textBF{Pad\'e-Borel} (100 coeffs)
\\[3pt]
\cite{Kravchenko:1996}
Kravchenko et al., 1996      & -0.662\,337\,793\,466                    & Method of Moments
\\[3pt]
\cite{Baye:2008}
Baye et al., 2008            & -0.662\,337\,793\,466\,{\it 315\,9}      & Lagrange Mesh
\\[3pt]
\textBF{Present Work}        & \textBF{-0.662\,337\,793\,466\,\it 316\,071\,2} & \textBF{Lagrange Mesh, 16K mesh points}      \\[3pt]
\cite{Stubbins:2004}
Stubbins et al., 2004        & -0.662\,337\,793\,466\,\it 316\,6        & Variational (multiconfiguration)
\\[10pt]
\hline
\hline
\end{tabular}}}
\begin{tablenotes}
\small
\item $^{a}$\,$q=1, \beta_0=0$
\end{tablenotes}
\end{table}

In Table \ref{historic} different estimates  of the ground state energy are presented following the order of reached accuracy for $\gamma=1$ a.u. So far the most accurate energy is found in \cite{Stubbins:2004} via multiconfigurational trial function, in the LMM
we are able to confirm 19 d.d., while our 8-parametric compact trial function (\ref{approximant1s0}) at $(q=1,\beta_0=0)$ gives 6 d.d. correctly with difference
in 7th d.d. in 1 unit.  10-parametric compact trial function (\ref{approximant1s0}) with released
parameters $(q,\beta_0)$ gives 7 d.d. correctly with difference in 8th d.d. in 9 units.

{ As for the binding energy $E_{binding}^{(\infty)}=\gamma - E^{(\infty)}$ it follows from Table \ref{Table:Energy} that the variational calculations with 8-parametric trial function provide not less than $6$ significant digits in the domain $\gamma \leq 100$ a.u. . This accuracy drops to $5$ and $4$ s.d. at $\gamma \approx 100$ a.u. and $\gamma \approx 10000$ a.u., respectively. Similar accuracies are inherited by the energy gap.
}

It is well-known that the Hydrogen atom acquires a quadrupole moment in magnetic field $\gamma > 0$, see e.g. \cite{Potekhin:2001}.  Due to azimuthal symmetry of the system, the quadrupole moment tensor is diagonal and is characterized by a single independent element only, i.e.
\begin{equation}
\label{qzz}
   Q_{zz}=\braket{r^2}-3\braket{z^2}\ .
\end{equation}
Confident results for $Q_{zz}$ are established in \cite{Baye:2008} for the first time, see also \cite{Potekhin:2001}, they are improved in the present recalculation in the LMM
in 3 - 8 s.d. depending on the magnetic field strength, see Table \ref{Table:Energy}. Expectation value $Q_{zz}$ (\ref{qzz}), found with compact variational trial function (\ref{approximant1s0}) with parameters from Appendix B, agrees with the LMM result with high accuracy for all studied magnetic fields.

Local deviation of the Approximant from the exact wave function can be estimated by studying the vicinity near the Coulomb singularity - located at the origin, $r=0$ - it can be \textit{measured} via the cusp parameter (\ref{cuspcondition}). A straightforward calculation shows that the cusp parameter $C^{(t)}$ derived from the Approximant (\ref{approximant1s0}) is given
\begin{equation}
\label{cusp-t}
       C^{(t)}\ =\ \al_1\,\kappa \ +\ \kappa^2 \left(q - \frac{\al_0\,\kappa}{2}\right)\beta_1\ ,\quad \kappa=(1+\beta_0)^{-1/2}\ ,
\end{equation}
which is the coefficient in front of the linear in $r$ term in the expansion of (\ref{approximant1s0}) at small distances \footnote{Since $\beta_0$ is always small being $\lesssim 10^{-3}$, one can place $\kappa = 1$}. Results are presented in Table \ref{Table:cusp} where it can be seen that $C^{(t)}$, calculated  with 8 optimal parameters and $(q=1,\beta_0=0)$ as entry, satisfies the cusp condition accurately with error $ \lesssim 1\%$ for $\gamma \lesssim 10$\,a.u. then it begin to grow reaching $\sim 16\%$ at $500$\,a.u. Note that in spite of so large deviation of $C^{(t)}$ from $C^{(exact)}=1$ the variational energy obtained is highly accurate.
It implies that the vicinity around Coulomb singularity gives very small contribution to the energy integrals. As for larger magnetic fields $\gamma \gtrsim 500$\,a.u. the deviation continues to grow and at $\gamma=10\,000$\, a.u. the deviation reaches $70\%$. Situation changes dramatically when the trial function (\ref{approximant1s0}) becomes 10-parametric, upon releasing
the parameters $(q,\beta_0)$. Although the energy improves in 1-2 far distant digits, see Table \ref{Table:Energy}, the cusp parameter gets smaller than 1, it deviates from the exact value $C=1$ in 3-4-5 d.d. for $\gamma \leq 1000$\,a.u., then it starts to grow and reaches its maximal deviation at $\gamma=10\,000$\,a.u. being $\sim 6\%$. It reflects sensitivity of the cusp parameter to values of parameters $(q,\beta_0)$, see (\ref{cusp-t}). Note that the cusp parameter calculated in the LMM provides the value of $C$ with not less than 6 d.d. in the whole range of studied magnetic fields $\gamma \leq 10\,000$\, a.u.

\begin{table}[ht]
\centering
\caption{Nuclear cusp parameter $C^{(t)}$ (\ref{cusp-t}) for the ground state $(1s_0)$ for
         different magnetic fields calculated with 8-parametric Approximant
         (\ref{approximant1s0}) with $(q=1,\beta_0=0)$.}
\label{Table:cusp}
		{\setlength{\tabcolsep}{0.4cm}	
\begin{tabular}{llll}
\hline \hline
		$\gamma$\,(a.u.) & $C^{(t)}$ & $\gamma$\,(a.u.) & $C^{(t)}$ \\
\hline
		\rule{0pt}{4ex}
	   0.01        & 1.000\,002         & 5.0          &   0.997  \\[4pt]
		0.1        & 0.999\,97          & 10.0         &   1.002  \\[4pt]
		0.5        & 0.999\,7           & 100.0        &   1.065  \\[4pt]
		1.0        & 0.999\,30          & 500.0        &   1.159  \\[4pt]
		1.0        & 0.999\,34 $^{{a}}$ & 500.0        &   0.977\,767 $^{a}$
\\[4pt]
		2.0        & 0.996\,5           & 1\,000.0     &  1.23    \\[4pt]
		$\gamma_c$ & 0.996\,7           & 10\,000.0    &  1.7 (I.65),  $(q=1,\beta_0=0)$  \\[4pt]
                   &                    & 10\,000.0    &  1.104 (I.65), $(q=0,\beta_0=0)$ \\[4pt]
		           &                    & 10\,000.0    &  0.939 $^{{a}}$ \\[4pt]
\hline \hline
\end{tabular}}
\begin{tablenotes}
\small
%\item $\gamma_c=2.065\,211\,858$
\item $^{{a}}$\,Variational Method:\ (I.65) with 10 parameters
\end{tablenotes}
\end{table}

Above-presented formalism developed for the ground state of positive parity $(1s_0)$ can be easily generalized for the family of excited states with $m=0$ and $\nu=+$.
At $\gamma=0$ the excited states with $\ell=m=0$ and $\nu=+$ are the $S$-states of the Hydrogen atom, $(n\,s_0)$ states with principal quantum number $n=2,3,4,\ldots$ and radial quantum number $n_r=n-1$. Its spectra is of the form
\[
    \Psi_{(n\,s_0)}\ =\ P_{n-1}(r) e^{-\frac{r}{n}}\ ,\ E_{(n\,s_0)}\ =\ - \frac{1}{2 n^2}\ ,
\]
where $P_{n-1}(r)$ is the Laguerre polynomial of degree $(n-1)$. Taking the exponential representation of the wave function,
\begin{equation}
\label{psi-(0+)}
  \Psi_{(m=0,+)}\ =\ P(\rho,r)\,e^{-\Phi(\rho,r)} \ ,
\end{equation}
cf.(\ref{representation}), (\ref{Phase}), and making substitution to (\ref{Sch-rho-r-phi-0}), we arrive at a generalized Riccati equation, see e.g. \cite{Turbiner:1984}, which later on can be transformed into the generalized RB equation and/or the GB equation. Similar analysis of these equations to that one made for the ground state $(1\,s_0)$ can be performed. It leads
for $\gamma \neq 0$ to the conclusion that an excited state at $m=0$ and $\nu=+$ can be studied using the trial function (\ref{psi-(0+)}) in the form of polynomial in $(r, \rho)$ variables multiplied by the Approximant (\ref{approximant1s0}), see discussion in Conclusiones. It can be done elsewhere.

\vspace{-8mm}
\subsection{Lowest energy state of negative parity}

The lowest energy state of negative parity is described by quantum numbers $m=0$ and $\nu=-$. For Hydrogen atom (at $\gamma=0$) it is $(2p_0)$ excited state. Sometimes it is called the ground state of  negative parity. Its eigenfunction can be written as the product of factor $z$ and nodeless function $\Psi^{(-)}(\rho, r)$, see (\ref{representation}). The Schr\"odinger equation that determines $\Psi^{(-)}$ and $\mathcal{E}^{(\infty)}$ reads, see (\ref{schroedinger_new}) at $m=0,p=1$,
\begin{equation}
\label{Sch-rho-r-phi-0-}
  -\frac{\hbar^2}{2m_e}\left[\pa_\rho^2\ +\ \frac{2\rho}{r}\pa_{\rho r}\ +\ \pa_r^2\ +\ \frac{1}{\rho}\pa_{\rho}\ +\ \frac{4}{r}\pa_r\right]\Psi\ +\ \left[\ -\frac{e^2}{r}\ +\ \frac{\gamma^2e^2}{8m_ec^2}\rho^2\right]\Psi^{(-)}\ =\
  \mathcal{E}^{(\infty)}\,\Psi^{(-)}\ .
\end{equation}
cf.(\ref{Sch-rho-r-phi-0}). Taking $\Psi^{(-)}(\rho,r)$ in exponential form,
\[
  \Psi^{(-)}(\rho,r)\ =\ e^{-\Phi^{(-)}(\rho,r)}\ ,
\]
one can see that \textit{phase} $\Phi^{(-)}(\rho,r)$ satisfies a non-linear partial differential equation of second order,
\[
  \pa_\rho^2\Phi\ +\ \frac{2\rho}{r}\pa_{\rho r}\Phi\ +\ \pa_r^2\Phi\ +\ \frac{1}{\rho}\pa_{\rho}\Phi\ +\    \frac{4}{r}\pa_r\Phi\ -\ (\pa_\rho\Phi)^2\ -\ \frac{2\rho}{r}(\pa_\rho\Phi)(\pa_r\Phi)\ -\ (\pa_r\Phi)^2
\]
\begin{equation}
\label{Riccati-}
  \ =\ \frac{2m_e}{\hbar^2}\left[ \mathcal{E}^{(\infty)}\ +\ \frac{e^2}{r}\ -\ \frac{\gamma^2e^2}{8m_ec^2}\rho^2\right]\ ,
\end{equation}
cf. (\ref{Riccati}), where for simplicity we dropped the superindex ${}^{(-)}$. We continue to call it the Riccati equation. By introducing new variables $(s,t)$, see (\ref{changesRB}), (\ref{r0}), the equation (\ref{Riccati-}) is transformed into the Riccati-Bloch equation (\ref{riccatiad}) (with term $\frac{2}{t}\pa_t\Phi$ replaced by $\frac{4}{t}\pa_t\Phi$) for the energy $\veps=\frac{\mathcal{E}^{(\infty)}}{\mathcal{E}^{(\infty)}_0}$, cf.(\ref{effenergy}) and magnetic field $\la$, see (\ref{effmag}). It is easy to see that the perturbation theory in powers of $\la^2$ (\ref{ptrb}) remains algebraic, its zero order correction
\begin{equation}
\label{RB-1}
     \Phi_0(s,t)\ =\ \frac{t}{2} \quad , \quad \veps_0\ =\ -\frac{1}{4}\ ,
\end{equation}
cf.(\ref{RB-0}), corresponds to $(2p_0)$ state of the Hydrogen atom, the structure of $n$th correction remains the same as for the ground state, see (\ref{structure}), and the expansion of $\veps$ in powers of $\la$ coincides with semi-classical expansion in powers of $\hbar^3$. It can be easily calculated any finite number of corrections to energy and phase like it was done for the ground state of positive parity, see App.A.

By introducing new variables $(u,v)$, see (\ref{changesGB}), (\ref{rho0}), the equation (\ref{Riccati-}) is transformed into the generalized Bloch equation (\ref{Bloch}) (with term $\frac{2}{v}\pa_v\Phi$ replaced by $\frac{4}{v}\pa_v\Phi$) for the same energy $\veps$ and magnetic field $\la$. Surprisingly, at $\la=0$ this equation can be solved exactly in the same form
\begin{equation}
\label{gen0-}
  \phi_0^{(-)}(u,v)\ =\ A_0^{(0,-)}(u)\,v\ +\ B_0^{(0,-)}(u)\ ,
\end{equation}
as in (\ref{gen0}), but with $\veps_0=-\frac{1}{4}$, see (\ref{RB-1}), where
\begin{equation*}
  A_0^{(0,-)}(u)\ =\ \sqrt{\frac{1}{4}+\frac{u^2}{12}}\ ,\qquad B_0^{(0,-)}(u)\ =\ \frac{1}{2}\log(\frac{1}{4}+\frac{u^2}{12})\ +\  2\log(\frac{1}{4}+\frac{1}{2}\sqrt{\frac{1}{4}+\frac{u^2}{12}})\ .
\end{equation*}
In new variable
\begin{equation}
\label{w-variable-}
     w_-\ =\ \sqrt{\frac{1}{4}+\frac{u^2}{12}}\ ,
\end{equation}
cf.(\ref{w-variable}) the zero order approximation is in the form,
\begin{equation}
\label{firstgen-w-1}
  \phi_0^{(-)}(u,v)\ =\  w_- v\ +\ \log w_-\ +\  2\log(\frac{1}{4} + \frac{w_-}{2})\ .
\end{equation}
Similar to the ground state the $\phi_0^{(-)}(u,v)$ plays a role of classical action, although the classical trajectory seems unknown.
Evidently, the function $\Psi_0=ze^{-\phi_0^{(-)}(u,v)}$ is square integrable, it can be taken as variational trial function to study $(2p_0)$ state in a way similar to what was done for the ground state, cf. (\ref{approximant0}).

Similar consideration, which had led to the Approximant (\ref{approximatephase}), can be repeated and we eventually arrive at
\begin{equation}
\label{approximatephase-}
    \Phi_t^{(-)}(\rho,r)\ = \ \frac{\al_0\,+\,\al_1\,r\,+\,\al_2\,r^2\,+\,\al_3\,\gamma\,\rho^2+\al_4\,\gamma\,\rho^2\,r}
    {\sqrt{1 + \beta_0 w_- + \beta_1\,r + \beta_2\,r^2+\beta_3\,\rho^2}}\ +\
    q\,\log(1 + \beta_0 w_- + \beta_1\,r + \beta_2\,r^2 + \beta_3\,\rho^2)\ ,
\end{equation}
where $\{\al_0,\al_1,\al_2,\al_3,\al_4,\beta_0,\beta_1,\beta_2,\beta_3;q\}$ are ten free parameters that later will be fixed in variational calculation, $w_-=\sqrt{\frac{1}{4}+\frac{\gamma ^2 \rho ^2}{12}}$, see (\ref{w-variable-}). We call it the {\it Phase Approximant} for the ground state of negative parity. Based on (\ref{approximatephase-}) one can build the 10-parametric trial function
\begin{equation}
\label{approximant2p0}
   \Psi_{2p_0}\ =\ z \, e^{-\Phi_t^{(-)}}\ .
\end{equation}

The results of variational calculations are presented in Table \ref{Table:Energy-} and compared with results by \cite{Wang:1995}
\footnote{In paper \cite{Wang:1995} there were presented the binding energies contrary to what was named as the energies} and \cite{Kravchenko:1996}. For all studied magnetic fields the variational results based on the 8-parametric variational function (\ref{approximant2p0}) with $(q=1,\beta_0=0)$ reproduce 5-6 s.d. (or more) in the energy $E^{(\infty)}$ of a static Hydrogen atom
\footnote{
It is worth noting that the results for energies by Wang-Hsue \cite{Wang:1995} and Kravchenko et al, \cite{Kravchenko:1996} presented in Table \ref{Table:Energy-} were recalculated and confirmed in LMM with 16K mesh points (it is not printed in Table).}
. Needless to say that by taking 10-parametric variational function (\ref{approximant2p0}) with
released parameters $(q,\beta_0)$ allows to increase accuracy similar to what happened for the ground state $(1s_0)$.

Binding energy $(\gamma-E^{(\infty)})$ grows in a very slow pace with magnetic field increase from
0.25\,Ry at $\gamma=0$ reaching $\sim 1$\,Ry at $10^4$\,a.u. Note that the critical magnetic
field found in LMM, when $E^{(\infty)}(\gamma_c=0)=0$, hence the Schr\"odinger operator in (\ref{Sch-rho-r-phi-0-}) has the zero mode, drops dramatically to $\gamma_c=0.436\,663\,244$ in comparison
with the ground state $(1s_0)$, see (\ref{gamma_0}).

\begin{table}[]
\centering
\caption{Excited $(2p_0)$ state (ground state of negative parity): energies $E^{(\infty)}$ and
        $E^{(m_p)}$ in Ry and quadrupole moment $Q_{zz}^{(\infty)}$ in $(a.u.)^2$ for Hydrogen atom in magnetic field for $\gamma \in [0.01, 10 000]$ in Variational Method
        with 8-parametric trial function (\ref{approximant2p0}) with $(q=1,\beta_0=0)$, comparison with calculations \cite{Wang:1995}, \cite{Kravchenko:1996} (rounded) made.}
\label{Table:Energy-}
\begin{threeparttable}
{\setlength{\tabcolsep}{0.35cm}
\begin{tabular}{clll}
\hline
\hline
$\gamma$ (a.u) & \hskip1.0cm $E^{(\infty)}$ & \hskip1.0cm  $E^{(m_p)}$ & -$Q^{(\infty)}_{zz}$  \\
\hline
\rule{0pt}{4ex}
0 & -0.250\,000\,000\,00  & -0.249\,863\,919\,86 & 24.000
\\[5pt]
   0.01     & -0.249\,700\,831\,66         & -0.249\,564\,266\,90 & 23.990     \\
       	    & -0.249\,700\,83  $^{a}$    \\
            & -0.249\,700\,831\,67 $^{f}$  &    & \\[5pt]
   0.1      & -0.224\,820\,1               & -0.224\,649\,6       & 23.064     \\
		    & -0.224\,820\,15 $^{a,f}$     &    &    \\[5pt]
$\gamma_c$  & 0.000\,001                   & 0.000\,374           & 18.781  \\[5pt]
   0.5      & 0.050\,480\,0                & 0.050\,892\,4        & 18.260     \\
     	    & 0.050\,479\,3 $^{f}$         &    &            \\[5pt]
   1.0      & 0.479\,989                   & 0.480\,701           & 15.585     \\
   		    & 0.479\,987 $^{a,f}$            &    &            \\[5pt]
   2.0      & 1.404\,583                   & 1.405\,877           & 13.177     \\
	        & 1.404\,578 $^{f}$            &    &            \\[5pt]
   5.0      & 4.304\,78                    & 4.307\,77            & 10.663     \\
   		    & 4.304\,76  $^{f}$            &    &            \\[5pt]			
  10.0      & 9.234\,73                    & 9.240\,49            & 9.267\,7   \\
   	        & 9.234\,70  $^{a,f}$          &    &            \\[5pt]
 100.0      & 99.072\,95                   & 99.127\,89           & 6.865\,7   \\
            & 99.072\,801                  &                  & 6.830\,912  ($\dagger$)\\
            & 99.072\,774  &                  & 6.824\,619  ($\dagger\dagger$)\\
   		    & 99.072\,76 $^{a,f}$            &    &            \\[5pt]
 500.0      & 499.025\,2                   & 499.298\,1           & 6.301\,2   \\
   		    & 499.025\,0 $^{a}$            &    &            \\[5pt]
1\,000.0    & 999.015\,2                   & 999.560\,4           & 6.194\,9   \\
   		    & 999.015\,0 $^{a,f}$            &    &            \\[5pt]
10\,000.0   & 9\,999.003                   & 10004.450            & 6.093\,2   \\  		                   		
\hline
\hline
\end{tabular}}
\begin{tablenotes}
	\small
	\item
		$^{{a}}$\,Basis of Splines \cite{Wang:1995}, $^{f}$\,Power series - method of moments
         \cite{Kravchenko:1996} \\
    \item
        $\gamma_c=0.436\,663\,244$ - found in LMM with 16K points (see text), $E^{(\infty)}\sim 10^{-10}$
    \item
        ($\dagger$)  Eq.(I.81), 8 parameters, $q\, =\,0\ , \beta_0\, =\,0$
    \item
        ($\dagger\dagger$)  Eq.(I.81), 10 parameters, $q\, =\,-0.078\,589\ , \beta_0\, =\,0.000\,46$
\end{tablenotes}
\end{threeparttable}
\end{table}

%\newpage

\section{Finite Mass Case}

\begin{center}
	\begin{figure}[h]
		\includegraphics[scale=1.0]{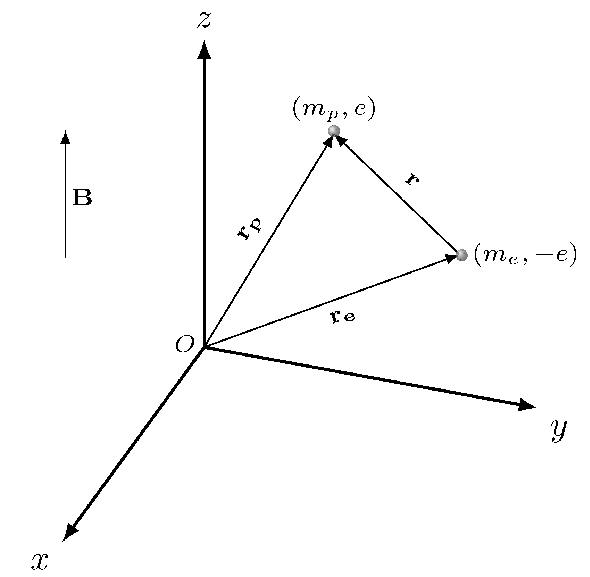}
		\caption{The three-dimensional two-body neutral system. }
		\label{fig:coordinatesfm}
	\end{figure}
\end{center}

We now investigate the effects which occur when finite mass ($m_p$) of the proton is taken into account. In this case, the Hamiltonian which describes the system is of the form
\begin{equation}
\label{Hamiltonian-fm}
    {\hat H}\ = \ \frac{1}{2\,m_p}\left(\hat{{\bf p}}_p-\frac{e}{c}{\bf A}_p\right)^2\ + \ \frac{1}{2\,m_e}\left(\hat{{\bf p}}_e+\frac{e}{c}{\bf A}_e\right)^2\ - \ \frac{e^2}{r}\ ,
\end{equation}
where
\begin{equation}
    \hat{{\bf p}}_{p,e}\ =\ \left({\hat p}_{x_{p,e}},\hat{p}_{y_{p,e}},{\hat p}_{z_{p,e}}\right)\ ,\qquad\qquad {\bf r}_{p,e}\ =\ (x_{p,e},y_{p,e},z_{p,e})\ ,
\end{equation}
are the momentum operator and vector position of the proton and electron, respectively.
Here $r \ = \ | {\bf r}_p-{\bf r}_e |$ is the relative distance between the charges.
Now, the configuration space is $6$-dimensional. For the geometrical setting of the system,
see Fig. \ref{fig:coordinatesfm}.

Just like in the infinite mass case, the {\it symmetric gauge}
\begin{equation}
{\bf A}_{p,e}\ = \ \frac{1}{2}{\bf B}\times{\bf r}_{p,e}\ ,
\end{equation}
is assumed for both vector potentials.

\subsection{Integrals of Motion}

The total pseudomomentum \cite{Gorkov:1968}
\begin{equation}
\label{pseudomomentum}
 \hat{{\bf K}}\ =\ \hat{{\bf p}}_p\ +\ \hat{{\bf p}}_e\ +\ \frac{e}{c}({\bf A}_p-{\bf A}_e) \ ,
\end{equation}

is an integral of motion,
\begin{equation}
 [\hat{{\bf K}},\hat{H}]\ = 0\ .
\end{equation}
Explicitly, the Cartesian components of $\hat{{\bf K}}$ are given by
\begin{equation}
\begin{aligned}
 \hat{K}_x \ &=\ \hat{p}_{x_p}\ +\ \hat{p}_{x_e}\ +\frac{e\,\gamma}{2c}(y_e-y_p)\ ,\\
 \hat{K}_y \ &=\ \hat{p}_{y_p}\ +\ \hat{p}_{y_e}\ +\frac{e\,\gamma}{2c}(x_p-x_e)\ ,\\
 \hat{K}_z \ &=\ \hat{p}_{z_p}\ +\ \hat{p}_{z_e}\ ,
\end{aligned}
\end{equation}
and obey the following commutation relations
\begin{equation}
 [\hat{K}_x,\hat{K}_y]\ =\ \hat{[K}_x,\hat{K}_z]\ =\ [\hat{K}_y,\hat{K}_z]\ =\ 0\ .
\end{equation}
Thus, they span 3-dimensional Abelian Lie algebra. The $z$-component of the total angular momentum
\begin{equation}\label{}
\begin{aligned}
\hat{L}_z\ & =\ ({\bf r}_p\times \hat{{\bf p}}_p)_z\ +\ ({\bf r}_e\times \hat{{\bf p}}_e)_z\ \nonumber \\
%&=\ (x_1\,p_{y_1}\ -\ y_1\,p_{x_1})\ +\ (x_2\,p_{y_2}\ -\ y_2\,p_{x_2}) \ ,
\end{aligned}
\end{equation}
is also conserved, $[\hat{L}_z,\hat{H}] = 0$. Hence, the total number of integrals is five $\{\hat{H},\hat{K}_x,\hat{K}_y,\hat{K}_z,\hat{L}_z\}$; the system is not (completely)-integrable: the sixth integral is missing and five known integrals do not form commutative algebra. It can be checked that the components of $\hat{{\bf K}}$ and $\hat{L}_z$ do not commute,
\begin{equation}
[\hat{K}_x,\hat{L}_z]\ =\ -i\hbar\,\hat{K}_y\ ,\qquad[\hat{K}_y,\hat{L}_z]\ =\ i\hbar\,\hat{K}_x\ ,\qquad[\hat{K}_{z},\hat{L}_z]\ =\ 0\ .
\end{equation}
%Note that the Lie subalgebra  spanned by $\{\hat{\bf K},\hat{L}_z\}$ is closed.
The second-order Casimir operator of the subalgebra $\{\hat{\bf K},\hat{L}_z\}$ is given by
\begin{equation}
\hat{C}\ =\ \hat{K}_x^2\ +\ \hat{K}_y^2\ .
\end{equation}
Indeed, for the case of a single particle in a constant magnetic field, the Casimir operator of the algebra of the corresponding integrals of motions is nothing but the Hamiltonian.

\subsection{Pseudo-separation of the Center of Mass Variables}

It presence of a magnetic field the center-of-mass motion can not be separated out, in this case it occurs the so-called {\it pseudo-separation}. Pseudo-separation of variables as introduced in \cite{Gorkov:1968} is achieved via three steps:

\noindent
{\bf (i)}\ We introduce the center-of-mass vectorial variables

\begin{equation}
\label{CM}
{\bf R}\ = \ \mu_p\,{\bf r}_p\ +\ \mu_e\,{\bf r}_e\ ,\qquad{\bf r}\ =\ {\bf r}_p\ -  {\bf r}_e\ ,
\end{equation}
\begin{equation}
\label{CMconjugated}
    \hat{{\bf P}}\ = \ \hat{{\bf p}}_p\ +\ \hat{{\bf p}}_e\ ,\qquad\hat{{\bf p}}\ =\
    \mu_e\,\hat{{\bf p}}_p\ -\ \mu_p\,\hat{{\bf p}}_e \non \ ,
\end{equation}
\begin{equation}
  M\ =\ m_p\ +\ m_e \ ,\qquad \mu \ = \ \frac{m_p\,m_e}{M}\ ,  \qquad \mu_{p,e} \ = \ \frac{m_{p,e}}{M} \non \ ,
\end{equation}
where $\hat {\bf P}$ and $\hat {\bf p}$ are the canonical conjugate momenta of ${\bf R}$ and ${\bf r}$, respectively, $\mu$ is the reduced mass of the system. In the variables (\ref{CM}), the Hamiltonian (\ref{Hamiltonian-fm}) takes the form
\begin{equation*}
\label{HamiltonianCM}
\begin{aligned}
   \hat{H}\ =\ &\ \frac{\hat{{\bf P}}^2}{2\,M}\ +\ \frac{\hat{{\bf p}}^2}{2\mu}\ -\frac{e}{2\,M\,c} ({\bf B}\times {\bf r})\cdot\hat{{\bf P}}\ -\ \frac{e}{2\,\mu\, c}\cdot({\bf B}\times {\bf R})\cdot\hat{{\bf p}}\ -\ \frac{e(\mu_e-\mu_p)}{2\,\mu\, c} ({\bf B}\times{\bf r})\cdot \hat{{\bf p}} \\
   &+\frac{e^2}{8\,\mu\, c^2}({\bf B}\times{\bf R})^2\ +\ \frac{e^2(\mu_e-\mu_p)}{4\mu c^2}({\bf B}\times {\bf R})\cdot({\bf B}\times {\bf r})\ +\ \frac{e^2}{8\,c^2}\left(\frac{\mu_e^2}{m_p}\ +\ \frac{\mu_p^2}{m_e}\right)({\bf B}\times {\bf r})^2
\end{aligned}
\end{equation*}
\begin{equation}
%\label{HamiltonianCM}
  -\ \frac{e^2}{r} \ .
\end{equation}

\noindent
{\bf (ii)}\ Also, in the coordinates (\ref{CM}) the conserved pseudomomentum (\ref{pseudomomentum}) takes the form
\begin{equation}
\label{KCMS}
\hat{{\bf K}} \ = \ \hat{{\bf P}}\ + \ \frac{e}{2\,c}{\bf B}\times{\bf r}\ ,
\end{equation}
see (\ref{CM}). Substituting (\ref{KCMS}) into the Hamiltonian (\ref{HamiltonianCM}) we obtain
\begin{equation*}
\begin{aligned}
%\label{HamiltonianK}
   \hat{H}\ =&\ \frac{(\hat{{\bf K}}-\frac{e}{2\,c}{\bf B}\times{\bf r})^2}{2\,M}\ +\
   \frac{\hat{{\bf p}}^2}{2\mu}\ -\frac{e}{2\,M\,c} ({\bf B}\times {\bf r})\cdot
   \left(\hat{{\bf K}}-\frac{e}{2\,c}{\bf B}\times{\bf r}\right)\ -\ \frac{e}{2\,\mu\, c}
   \cdot({\bf B}\times {\bf R})\cdot\hat{{\bf p}} \\
   &-\ \frac{e(\mu_e-\mu_p)}{2\mu c} ({\bf B}\times{\bf r})\cdot \hat{{\bf p}}\
   +\ \frac{e^2}{8\,\mu\, c^2}({\bf B}\times{\bf R})^2\ +\
   \frac{e^2(\mu_e-\mu_p)}{4\,\mu\, c^2}({\bf B}\times {\bf R})\cdot({\bf B}\times {\bf r}) \\
\end{aligned}
\end{equation*}
\begin{equation}
\label{HamiltonianK}
  +\ \frac{e^2}{8c^2}\left(\frac{\mu_e^2}{m_p}\ +\ \frac{\mu_p^2}{m_e}\right)({\bf B}\times {\bf r})^2\ -\ \frac{e^2}{r}\ .
\end{equation}

\noindent
{\bf (iii)}\ Since (\ref{HamiltonianK}) describes a neutral system, which can move across magnetic field, it is natural to look for a unitary-equivalent Hamiltonian $\hat{\mathcal{H}}$ such that $[\hat{\mathcal{H}},\,\hat{{\bf P}}]=0$, for which cms momentum $\hat{{\bf P}}$ is conserved, being the integral of motion. To this end, the operator (\ref{HamiltonianK}) is transformed via the gauge rotation
\begin{equation}
\label{Hamiltonianrotated}
\hat{\mathcal{H}}\ \equiv\ U^{-1} \hat{H}\,U\ \ ,
\end{equation}
with the gauge factor
\begin{equation}
U\ =\ \exp\left(\frac{i}{\hbar}\left[{\bf P}\ -\ \frac{ e}{2c}({\bf B}\times {\bf r})\right]\cdot {\bf R}\right)\ ,
\end{equation}
here $\bf{P}$ denotes the eigenvalue of the total (cms) momentum operator $\hat{\bf{P}}$.
For the gauge-rotated Hamiltonian $\hat{\mathcal{H}}$ defined in (6.14), we look for eigenfunctions with ${\bf r}$-dependence (i.e., ${\bf R}$-independent) only.
 The action of the gauge rotation\footnote{It is worth mentioning that the operator $\hat{L}_z$ is gauge invariant with respect to $U$, i.e. $U^{-1}\hat{L}_z\,U\ =\ \hat{L}_z$.} to $\hat{\bf K}$ and $\hat{\bf p}$ reads
\begin{equation}
U^{-1}\hat{{\bf K}}\,U\ =\ {\bf P}\ ,\qquad U^{-1}\hat{{\bf p}}\,U\ =\ \hat{{\bf p}}\ +\ \frac{e}{2c}({\bf B}\times {\bf R})\ ,
\end{equation}
%\begin{equation}
%U^{-1}\hat{{\bf P}}\,U\ =\ {\bf P}\ -\ \frac{q}{2c}({\bf B}\times {\bf r})\ ,\qquad U^{-1}\hat{{\bf p}}\,U\ =\ \hat{{\bf p}}\ +\ \frac{q}{2c}({\bf %B}\times {\bf R})\ ,
%\end{equation}
whereas $\bf R$ and $\bf r$ remain unaffected. Eventually, the gauge rotated Hamiltonian (\ref{Hamiltonianrotated}) takes the form
\begin{align}
\label{schroedingerpsi}
  \hat{\mathcal{H}}\ =\  \frac{1}{2\,M}\left({\bf P}-\frac{e}{c}{\bf B}\times{\bf r}\right)^2\ +\ \frac{1}{2\mu}\left(\hat{{\bf p}}-\frac{e_{\rm eff}}{2\,c}{\bf B}\times{\bf r}\right)^2\  -\ \frac{e^2}{r}\ ,
\end{align}
here
\[
e_{\rm eff} = e\,(\mu_e-\mu_p)\ ,
\]
is an effective charge,
it vanishes for the case of equal masses (like for positronium) and becomes $-e$ for $m_p=\infty$. It can be checked that the first and third terms in (\ref{schroedingerpsi}) are gauge invariant.

\vspace{-8mm}
\subsection{ Case ${\bf P}={\bf 0}$: Atom at Rest}

At zero momentum ${\bf P}={\bf 0}$ (atom at rest) the Schr\"odinger equation for the Hamiltonian (\ref{schroedingerpsi}) takes the form
\begin{equation}
\begin{aligned}
\label{spsi}
     \hat{\mathcal{H}}_{0}\,\psi({\bf r})\ \equiv \ \left[\frac{1}{2\mu}\hat{{\bf p}}^2\ -\ \frac{e_{\rm eff}}{2\,\mu \,c}({\bf B}\times{\bf r})\cdot \hat{{\bf p}}\  - \
     \frac{e^2}{r}\ +\ \frac{e^2}{8\,\mu\, c^2}({\bf B}\times{\bf r})^2\right]\psi({\bf r})\ =\ E^{(m_p)}\,\psi({\bf r})\ ,
\end{aligned}
\end{equation}
with energy denoted $E^{(m_p)}=E^{(m_p)}(\gamma, e, \mu)$. In the coordinates $\{\rho,r,\varphi\}$, see Fig.1, the eigenvalue problem (\ref{spsi}) reads
\begin{equation*}
\label{Sch-rho-r-phi-m}
  -\bigg[\frac{\hbar^2}{2\,\mu}\bigg(\pa_\rho^2\ +\ \frac{2\,\rho}{r}\pa^2_{\rho, r}\ + \
  \pa_r^2\ +\ \frac{1}{\rho}\pa_{\rho}\ +\ \frac{2}{r}\pa_r\,\bigg)\ - \
  \frac{{\hat \ell}_z^2}{2\,\mu\,\rho^2}\ + \ \frac{e_{\rm eff}\, \gamma}{2\,\mu\, c}{\hat \ell}_z
\end{equation*}
\begin{equation}
\label{eqpzero}
  +\ \frac{e^2}{r}\ -\ \frac{\gamma^2\,e^2}{8\,\mu\, c^2}\rho^2\,\bigg]\,\psi \ = \ \ E^{(m_p)}\,\psi\ ,
\end{equation}
cf.(\ref{Sch-rho-r-phi}), with ${\hat \ell}_z \equiv ({\bf r} \times \hat{{\bf p}})_z = -i\,\hbar\,\pa_\varphi$ being the $z$-component of the relative angular momentum. It is evident that the operator ${\hat \ell}_z$ is an integral, $[{\hat \ell}_z,\,\hat{\mathcal{H}}_{0}]=0$.
The Hamiltonian $\hat{\mathcal{H}}_{0}$ is $z$-reflection invariant, $\hat{\mathcal{H}}_{0}(-z)=\hat{\mathcal{H}}_{0}(z)$, hence, the eigenfunctions are characterized by parity, $\psi(-z)=\pm\psi(z)$. In the variables $(\rho,\,r,\,\varphi)$ the eigenfunctions have a factorized form
\begin{equation}
\label{representation-m}
  \psi(\rho,r,\varphi)\ =\ \rho^{|m|}\,z^p\,\chi(\rho,r)\, e^{im\varphi}\ ,
  \qquad m\ =\ 0,\pm 1,\pm 2,\, \ldots \ ,\ p=0,1\ ,\ z=\sqrt{r^2-\rho^2}\ ,
\end{equation}
similar to (\ref{representation}), where $m$ is the magnetic quantum number corresponding to the relative motion and $p$ is parity. Substituting (\ref{representation-m}) into (\ref{eqpzero}) we arrive at the two-dimensional Schr\"odinger equation
\begin{equation*}
  -\bigg[\frac{\hbar^2}{2\,\mu}\bigg(\pa_\rho^2\ +\ \frac{2\,\rho}{r}\pa^2_{\rho, r}\ + \
  \pa_r^2\ +\ \frac{2|m|+1}{\rho}\pa_{\rho}\ +\ \frac{2(|m|+p+1)}{r}\pa_r\bigg)\ + \
\end{equation*}
\begin{equation}
\label{eqpzeros}
   \frac{e^2}{r}\ -\ \frac{\gamma^2\,e^2}{8\,\mu\, c^2}\,\rho^2\,\bigg]\,\chi^{(m)}(\rho,r) \ = \  \mathcal{E}_{m,p}^{(m_p)}\,\chi^{(m)}(\rho,r)\ ,
\end{equation}
with eigenvalue
\begin{equation*}
\label{Em --> epsm}
  \mathcal{E}_{m,p}^{(m_p)}\ =\ E_{m,p}^{(m_p)}\ +\ \frac{e_{\rm eff}\,\hbar\, \gamma}{2\,\mu\, c}\,m\ ,
\end{equation*}
cf.(\ref{schroedinger_new}): they coincide if the replacement
\[
m_e \rar \mu\ ,\quad e \rar (-e_{eff})\ ,\quad \mathcal{E}_{m,p}^{(\infty)} \rar \mathcal{E}_{m,p}^{(m_p)}\ ,
\]
is made in (\ref{schroedinger_new}); the wavefunctions are related
\begin{equation}
\label{wavef-related}
	\Psi^{(\infty)}\left(\frac{\mu}{m_e} {\bf r}\, ;\ \frac{m_e^2}{\mu^2}\gamma,e,m_e\right)\ = \ 	\Psi^{(m_p)}\left({\bf r} \, ;\ \gamma, e, \mu\right)\ .
\end{equation}
It should be mentioned that the definition of Rydberg constant and atomic unit for magnetic field
is changed as well
\begin{equation}
\label{effenergy-fm}
  \veps\ =\ \frac{\mathcal{E}^{(m_p)}}{\mathcal{E}_0(\mu)}\ ,\qquad \mathcal{E}_0(\mu)\ =\ \frac{\mu\, e^4}{2\hbar ^2}\ ,
\end{equation}
cf.(\ref{effenergy}), and
\begin{equation}
\label{effmag-fm}
    \la\ = \frac{ \gamma}{\gamma_0(\mu)}\ ,\qquad  \gamma_0(\mu)\ =\ \frac{c\,e^3\, \mu^2}{\hbar^3}\ .
\end{equation}
cf.(\ref{effmag}).

As for the ground state energy, $m=0$,
\begin{equation}
\label{energypzero}
    \mathcal{E}_{0,p}^{(m_p)}\ =\ E_{0,p}^{(m_p)}\ ,
\end{equation}
while in general,
\begin{equation}
\label{Eparity-m}
  \mathcal{E}_{m,p}^{(m_p)}\ =\ \mathcal{E}_{-m,p}^{(m_p)} \ ,
\end{equation}
cf.(\ref{Eparity}).

%\subsection{ Atomic Case: $m_p \rightarrow \infty$}

We consider two special cases. One of them is when in (\ref{spsi}) the proton mass $m_p \rightarrow \infty$ while the electron mass $m_e$ is kept finite and another one when both masses are equal. In former case $\mu \rightarrow m_e,\,e_{\rm eff}\rightarrow -e $. The Hamiltonian (\ref{schroedingerpsi}) takes the form
\begin{equation}
\begin{aligned}
\label{Hatomic}
\hat{\mathcal{H}}\ \equiv \ \frac{1}{2\,m_e}\left(\hat{{\bf p}}\,+\,\frac{e}{2\,c}{\bf B}\times{\bf r}\right)^2\  -\ \frac{e^2}{r}\ ,
\end{aligned}
\end{equation}
where dependence on $\mathbf{P}$ disappears, it coincides with (\ref{H}). In general, the limit $m_p \rightarrow \infty$ corresponds to the atomic system where one mass is much heavier than other (for instance, as in the hydrogen atom). We call this case {\it atomic}.
Latter case corresponds to positronium Ps, when $m_p=m_e$ and $e_{\rm eff}=0$. Linear Zeeman effect is absent in this case, the Schr\"odinger equation is of the form (\ref{eqpzeros}) with $\mu$ replaced by $m_e/2$ and
\[
\mathcal{E}_{m,p}^{(m_e)}\ =\ E_{m,p}^{(m_e)}\ .
\]

%\vspace{-25mm}
\subsection{Scaling Relations}
%\vspace{-6mm}

In general, for non-moving neutral system, one can relate the atomic case $m_p \rightarrow \infty$ with the finite-mass case of the system at rest ${\bf P}=0$. In order to do that we have to make a scale transformation $r \rightarrow \frac{\mu}{m_e}\,r$ and $\gamma \rightarrow \frac{m_e}{\mu}\,\gamma$. Then the following remarkable scaling relation between the corresponding ground state energies (\ref{schroedinger}) and (\ref{energypzero}), respectively, emerges \cite{Pavlov-Verevkin:1980},
\begin{equation}
\label{PV-relation}
   \frac{\mu}{m_e}\,E_{0,0}^{(\infty)}\bigg(\frac{m_e^2}{\mu^2}\,\gamma,\,e,\,m_e\bigg) \ = \
         E_{0,0}^{(m_p)}(\gamma,\,e,\,\mu) \ ,
\end{equation}
where for the case of Hydrogen atom, taking $m_p/m_e=1836.152 673$\,(from NIST data), the mass ratio takes the value
\begin{equation}
    \frac{m_e}{\mu}\ \approx\ 1.000\,545\ .
\end{equation}
Note that the critical magnetic field (\ref{gamma_0}) effectively decreases,
\begin{equation}
\label{gamma_0m}
  \gamma_0^{(m_p)}\ =\ \frac{\mu^2}{m_e^2}\,\gamma_0\ ,
\end{equation}
while for the case of positronium it becomes
\[
\gamma_0^{(m_e)}\ =\ \frac{\gamma_0}{4}\ .
\]
It is evident that the relation (\ref{PV-relation}) holds for excited states
\begin{equation}
\label{PV-relation-excited}
   \frac{\mu}{m_e}\,\mathcal{E}_{m,p}^{(\infty)}\bigg(\frac{m_e^2}{\mu^2}\,\gamma,\,e,\,m_e\bigg)\ =\  \mathcal{E}_{m,p}^{(m_p)}(\gamma,\,e,\,\mu) \ ,
\end{equation}
and also for quadrupole momenta
\begin{equation}
 \frac{m_e^2}{\mu^2}Q_{zz}^{(\infty)}\left(\frac{m_e^2}{\mu^2}\gamma,e,m_e\right)\ = \ Q_{zz}^{(m_p)}(\gamma,e,\mu)\ .
\end{equation}

\subsection{Energy}

Since the equation (\ref{eqpzeros}) coincides with (\ref{schroedinger_new}) once $\mu$ is identified with $m_e$, both the LMM and the variational method with the
8-parametric Approximant (\ref{approximant1s0}) at $(q=1,\beta_0=0)$ with parameters presented in App.B and with the 8-parametric Approximant (\ref{approximant2p0}) at $(q=1,\beta_0=0)$ with parameters presented in App.C can be applied. In Table \ref{Table:Energy} the ground state energies for the Hydrogen atom with finite proton mass are presented for different magnetic fields, all printed digits correspond to the situation when the results obtained in both methods coincide. In a similar way in Table \ref{Table:Energy-} the variational energies for $(2p_0)$ excited state for the Hydrogen atom with finite proton mass are presented for magnetic fields ranging from 0 to $10^{4}$\,a.u.

Making comparison of the energies for infinite and finite mass cases in both Tables one can see that, in general, for fixed magnetic field the finite mass effects increase the ground state energy and the energy of $(2p_0)$ excited state by changing the 4th s.d. (and subsequent ones) independently on the magnetic field, they are of the order of $(m_e/m_p)\,\gamma$. This result is checked separately via the scaling relation (\ref{PV-relation}).

Positronium atom Ps is much less studied Coulomb system than the Hydrogen atom, see e.g. \cite{PsShertzer,PsWunner}. In Table \ref{Table:Energy-Ps} the results of independent calculations
of the ground state energy performed in variational method with the 8-parametric Approximant (\ref{approximant1s0}) at $(q=1,\beta_0=0)$ with parameters taken from App.B - and in LMM with 16,000 basic functions for different magnetic fields are presented. The obtained energies in both methods coincide systematically in 10 s.d. for weak magnetic fields and up to 6 s.d. for strong magnetic fields being far superior than previous results. Quadrupole moment versus magnetic field is calculated for the first time in two independent methods, see Table \ref{Table:Energy-Ps}. There is a good coincidence for all studied magnetic fields. The validity of the scaling relations for energy and quadrupole moment was checked in a separate calculation for different masses $m_p$ and magnetic fields.

\begin{table}[ht]
	\centering
	\caption{Energy $E^{(Ps)}$ in Ry and quadrupole moment $Q_{zz}^{(Ps)}$ in $(a.u)^2$ of the
     ground state of positronium Ps in magnetic field calculated in Variational Method with 8-parametric trial function (\ref{approximant1s0}) with $(q=1,\beta_0=0)$ for $\gamma \in [0.01,10\,000]$ (first lines) and in LMM with 16,000 basic functions marked by ${}^a$. Comparison with available results presented.}
	\label{Table:Energy-Ps}
	\begin{threeparttable}
{\setlength{\tabcolsep}{0.4cm}
\begin{tabular}{cllcll}
\hline
\hline
 $\gamma$  & \hskip 1.0cm $E^{(m_e)}$ & $-Q_{zz}^{(m_e)}$ & $\gamma$ & \hskip 0.5cm $E^{(m_e)}$
           & $-Q_{zz}^{(m_e)}$
\\
\hline
\rule{0pt}{6ex}
 0.01  & -0.499\,600\,701\,76       & 0.015\,805\,4         & 5.0  & 7.784\,63   & 1.478
\\
       & -0.499\,600\,701\,77 $^{a}$ & 0.015\,806\,8 $^{{a}}$ &  & 7.784\,60 $^{a}$ & 1.477 $^{{a}}$
\\	
       & -0.499\,6 $^{b}$          &                       &  &                 &
\\[5pt]	
 0.1   & -0.464\,605\,37           & 0.812\,36 	        & 10.0 & 17.199\,03  & 1.188
\\
  	   & -0.464\,605\,38 $^{a}$    & 0.812\,32 $^{{a}}$    &  & 17.198\,97 $^{a}$ & 1.186 $^{{a}}$
\\
  	   & -0.464\,6 $^{b}$          &                       &  & 17.2 $^{b}$     &
\\[5pt]
 0.5   & -0.022\,213\,4            & 2.045\,6              & 100.0 & 194.148\,9 & 0.539
\\
	   & -0.022\,213\,9 $^{a}$     & 2.045\,7 $^{{a}}$     &  & 194.148\,3$^{a}$ & 0.535 $^{{a}}$
\\
       &                           &                       &  & 194.14 $^{b}$     &
\\
	   &                           &                      &  & 194.177\,4 $^{c}$ &                      \\[5pt]
  $\gamma_c$ & 0.000\,001          & 2.054\,24            & 500.0 & 990.698    & 0.319
\\
		     & 0.000\,000 $^{{a}}$ & 2.054\,15 $^{{a}}$   &     & 990.695 $^{a}$ & 0.314 $^{{a}}$  \\[5pt]
 1.0         & 0.719\,204          & 2.073\,0             & 1\,000 & 1\,988.801 & 0.258
\\
			 & 0.719\,202 $^{a}$   & 2.072\,7 $^{{a}}$    &  & 1\,988.796 $^{a}$ & 0.253$^{{a}}$  \\
			 & 0.7192 $^{b}$       &                      &  &                   &                       \\[5pt]
 2.0         & 2.380\,622          & 1.870\,7             & 10\,000.0 & 19\,980.5 & 0.13
\\
			 & 2.380\,615 $^{a}$   & 1.870\,4 $^{{a}}$    &  & 19\,980.6 $^{a}$ & 0.11 $^{{a}}$     \\[5pt]
\hline
\hline
\end{tabular}}
\begin{tablenotes}
\small
\item $\gamma_c=0.516\,302\,965$
\item  $^{a}$\,LMM (present calculation),\, $^{b}$\,\cite{PsShertzer},\, $^{c}$\,\cite{PsWunner}
\end{tablenotes}
\end{threeparttable}
\end{table}

\vspace{-5mm}
\section*{Conclusions}
\vspace{-4mm}

A simple uniform locally accurate approximation for the ground state, nodeless function is constructed for a {\it neutral} system of two Coulomb charges of different masses at rest in a constant uniform magnetic field of positive and negative parity, ${(1s_0)}$ and ${(2p_0)}$ states, respectively. It is shown that by keeping the mass and charge of one body fixed all systems with different second body masses are related. This allows us to consider the second body as infinitely-massive and to take such a system as basic, which simplifies consideration. Three physical systems are considered: the Hydrogen atom with (in)-finitely massive proton (deuteron, triton) and positronium.

Concretely, 10-parameter approximation for the ground state functions of different parities for the hydrogen atom with infinitely-massive proton (the so-called one-center case) in a constant uniform magnetic field in the interval $\gamma \in [0\,,\,10^4]$ a.u. is proposed. If taken as a variational trial function it allows us to calculate with accuracy of not less than 6 s.d. ($\leq 10^{-6}$ in relative deviation) in the whole domain of the considered magnetic fields the total energy and not less than 3 s.d. for the quadrupole moment $Q_{zz}$. For the quadrupole moment such accuracies are reached for the first time. As for the energy at small magnetic fields $\gamma \leq 1$\,a.u. the relative deviation ranges from $\sim 10^{-11}$ at $\gamma=0.01$\,a.u. to $\sim 10^{-8}$ for $\gamma=1$\,a.u. with the increase of a magnetic field. Benchmark results used for comparison are established using the Lagrange mesh method with 16K mesh points. For both ground states of positive/negative parities the critical magnetic fields $\gamma^{(1s_0)}_c=2.065211858$ a.u. and $\gamma^{(2p_0)}_c=0.436\,663\,244$, where the Schr\"odinger operator has the zero mode, are calculated for the first time. The presented approximation remains the same functionally for an {\it arbitrary} two-body neutral system, it depends effectively on the reduced mass of the system only. This allows us to study the effects of finite proton (deuteron, triton) mass in Hydrogen atom as well as in the positronium - the system of electron and positron. It manifests an approximate solution of the problem of two Coulomb charges of opposite signs in a constant uniform magnetic field for the two lowest energy states of different parities.

Remarkably, for $c=e=m_e=1$, the perturbation series for energy appears in powers of $\gamma \hbar^3$ with constant coefficients. This implies that the PT in powers of a magnetic field coincides with semiclassical expansion in powers of $\hbar^3$. A fundamental result of the present study, based on the exploration of the RB/GB equations in PT in powers of $\gamma$, is the novel semi-classical expansion of the ground state energy in powers of $\hbar^3$ for a true two-dimensional problem.

Due to the algebraic nature of the PT for the RB equation the first 100 corrections to the ground state energy, all are rational numbers in atomic units, $c=e=m_e=\hbar=1$, and the exponential phase (in the form of polynomials in variables $\rho, r$ with rational coefficients) are calculated for the first time. The use of a Pad\'e-Borel re-summation technique for energy leads to highly accurate results (not less than 11 s.d.) at small values of $\gamma \leq 1$\,a.u. but fails for larger magnetic fields. Similar results can be obtained for the ground state of negative parity.

The key element of the procedure is a construction for exponential phase $\Phi(\rho,\,r)$ (the logarithm of the wavefunction), as a simultaneous interpolation between (i) the asymptotic series in the weak $\gamma \ll 1$ and strong $\gamma \gg 1$ magnetic field regimes, and between (ii) the semi-classical and perturbation expansions at large and small distances, respectively. The dimensionless RB (II.16) and GB (II.26) equations for $\Phi(\rho,\,r)$ help us construct the analytic interpolation in the form of a 10-parametric trial wave function, the \emph{Approximant} $\Psi^{(t)}$, (II.66) and (II.82) for the states of positive and negative parity, respectively.

For both the $(1s_0)$ and $(2p_0)$ states the Phase Approximant $\Phi_t$ has a similar functional form
\begin{equation}
\label{PA:1,2}
\Phi_t(\rho,r)\ = \
       \frac{\al_0+\al_1\,r+\,\al_3\,r^2+a_3\,\gamma\,\rho^2+\al_4\,\gamma\,\rho^2\,r}
       {\sqrt{1+ \beta_0\,w_{\pm} + \beta_1\,r + \beta_2\,r^2+\beta_3\,\rho^2}}\ +\  q\,\log(1+ \beta_0\,w_{\pm} + \beta_1\,r + \beta_2\,r^2+\beta_3\,\rho^2) \ ,
\end{equation}
with the only difference in $w$: $w_+=\sqrt{1+\frac{\gamma^2\,\rho^2}{12}}$ for the $(1s_0)$ state
and $w_-=\sqrt{\frac{1}{4}+\frac{\gamma^2\,\rho^2}{12}}$ for the $(2p_0)$ state. It can be shown
that for an arbitrary state with quantum number $m$ and parity $p$ the Phase Approximant $\Phi_t$ (\ref{PA:1,2}) remains of the same functional form but with different $w$, $$w_{m,p}=\sqrt{\frac{1}{(|m|+p+1)^2}+\frac{\gamma^2\,\rho^2}{12}}\ ,$$
while the pre-factors, which define nodal surfaces, can be quite complicated and non-trivial.
Note that the leading term in the semi-classical expansion (which is an analogue of the classical action in the one-dimensional case) has a surprisingly simple, closed analytic form,
\begin{equation}
\label{firstgen-wmp}
  \phi_0(u,v)\ =\ w_{m,p}\,v \ +\ \log w_{m,p}\ +\ (|m|+p+1) \log(\frac{1}{(|m|+p+1)}\ +\ {w}_{m,p})\ .
\end{equation}
Excluding $\beta_0$ and $q$, all other variational parameters in $\Phi_t$ (\ref{PA:1,2}) are positive (except $\al_0$, which grows as $\gamma \rar 0$) and exhibit a monotonous growth as a function of the magnetic field $\gamma$. The parameter $\beta_0$ is extremely small for all studied magnetic fields, it influences far distant digits in the energy and can be set equal to zero without loosing much accuracy. The parameter $q$ has a pretty surprising behavior: it is close to 1 for $\gamma \lesssim 10$\,a.u., then it sharply changes to almost zero for larger magnetic fields. In spite of this fact the optimal 8-parametric function at $q=1,\,\beta_0=0$ provides a relative deviation from the {exact} numerical solution of order $\lesssim 10^{-5}$ in the whole domain $\gamma \in [0.01,\,10^4]$ a.u.

In general, variational results with the 10-parametric trial function agree with the ones based on the Lagrange Mesh Method with 16K mesh points with high accuracy for all studied magnetic fields. The comparison with other calculations was made in Tables I and II for $1s_0$ state. As for the ground state $2p_0$ of negative parity the results are presented in Table IV.

As for the less studied problem of the positronium atom Ps, the trial function $\Phi_t(\rho,r)$ with only $8$ variational parameters ($q=1,\,\beta_0=0$) provides a ground state energy that agrees systematically with the accurate numerical result in 10 s.d. for small $\gamma$ and up to 6 s.d. at large $\gamma$ in the whole domain $\gamma \in [0.01,\,10^4]$ a.u. In this case, the critical magnetic field $\gamma_c=0.436663$ a.u. turned out to be almost five times smaller than the one for the hydrogen atom. And, not surprisingly, an excellent agreement between the results obtained variationally with use of the Approximant and the Lagrange Mesh method occurs. This reflects the high quality of the trial function used.

All two-body neutral systems we studied are at rest, they are not moving, ${\bf P}={\bf 0}$. Dynamics is defined by relative coordinates, see (\ref{spsi}). The effects of cms motion will be studied elsewhere.

\hskip 1cm {\bf Acknowledgments}
%\vspace{-16mm}
%\section*{Acknowledgments}
%\vspace{-5mm}

The authors thank J.C.~Lopez Vieyra and H.~Olivares~Pil\'on for their interest to the work and useful discussions. J.C.~del\,V. thanks E.~Dominguez-Rosas for assistance with Mathematica package. J.C.~del\,V. is supported by CONACyT PhD Grant No.570617 (Mexico). This work is partially supported by CONACyT grant A1-S-17364 and DGAPA grant IN113819 (Mexico).

%%%%%%%%%%%%%%%%%%%%%%%%%%%%%%%%%%%%%%%%%%%%%%%%%%%%%%%%%%%%%%%%%%%%%%%%
%\newpage
%%%%%%%%%%%%%%%%%%%%%%%%%%%%%REFERENCES%%%%%%%%%%%%%%%%%%%%%%%%%%%%%%%%%

\newpage

\appendix
\section{Ground state $1s_0$: PT Corrections}
\label{appendixA}

We present here the explicit forms the first perturbative corrections  $\Phi_n$, $n=2,3,4$
in the expansion (\ref{ptrb}) in addition to $\Phi_1$, see (\ref{Phi_1}),
\begin{align}
   -\Phi_2(s,t)\ =\ &\ \quad\frac{1}{1152}s^4 t\ +\ \frac{1}{1440}s^2 t^3
   \non\\
&\ +\ \frac{11}{4608}s^4\ +\ \frac{13 }{1440} s^2 t^2\ +\ \frac{1}{2880} t^4
   \non\\
&\ +\ \frac{193}{5760}s^2 t\ +\ \frac{1}{120} t^3
 \non\\
&\ +\ \frac{193 }{3840}s^2\ +\ \frac{337}{5760}t^2 \ ,
\\
 \non\\
%%%%%%%%%%%%%%%%%%%%%%%%%%%%%%%%%%%%%%%%%%%%%%%%%%%%%%%%%%%%%%%%%
 \Phi_3(s,t)\ =\ &\ \quad\frac{1}{27648}s^6 t\ +\ \frac{1}{11520}s^4 t^3\ +\ \frac{1}{60480}s^2 t^5
 \non\\
&\ +\ \frac{7}{55296}s^6\ +\ \frac{163}{138240}s^4 t^2\ +\ \frac{131}{241920}s^2 t^4\ +\
      \frac{1}{181440}t^6
 \non\\
&\ +\ \frac{61 }{11520}s^4 t\ +\ \frac{8063 }{1209600}s^2 t^3\ +\ \frac{53 }{201600}t^5
 \non\\
&\ +\ \frac{803 }{92160}s^4+\frac{33311 }{806400}s^2 t^2+\frac{2927 }{604800}t^4
 \non\\
&\ +\ \frac{90877}{691200} s^2 t\ +\ \frac{2027 }{43200}t^3
 \non\\
&\ +\ \frac{90877}{460800} s^2\ +\ \frac{188173}{691200} t^2\ ,
\\
 \non\\
%%%%%%%%%%%%%%%%%%%%%%%%%%%%%%%%%%%%%%%%%%%%%%%%%%%%%%%%%%%%%%%%%
%
 -\Phi_4(s,t)\ =\ &\ \quad\frac{5}{2654208} s^8 t\ +\ \frac{}{110592}s^6 t^3\ +\
                   \frac{163}{29030400}s^4 t^5\ +\ \frac{1}{2419200}s^2 t^7
 \non\\
&\ +\ \frac{163}{21233664}s^8\ +\ \frac{293 }{2211840}s^6 t^2\ +\
       \frac{9833}{58060800}s^4 t^4\ +\ \frac{727 }{29030400}s^2 t^6\ +\ \frac{1}{9676800}t^8
 \non\\
&\ +\ \frac{8819}{13271040}s^6 t\ +\ \frac{1663979}{812851200}s^4 t^3\ +\
       \frac{24733}{40642560}s^2 t^5\ +\ \frac{167 }{20321280}t^7
 \non\\
&\ +\ \frac{10577}{8847360}s^6\ +\ \frac{13945163}{1083801600}s^4 t^2\ +\
       \frac{22721}{2822400}s^2 t^4\ +\ \frac{5989}{22579200} t^6
 \non\\
&\ +\ \frac{27927329}{650280960}s^4 t\ +\ \frac{29335139 }{451584000}s^2 t^3\ +\
      \frac{4828099 }{1016064000}t^5
 \non\\
&\ +\ \frac{816005783}{13005619200}s^4\ +\ \frac{1349713153}{4064256000} s^2 t^2\ +\
      \frac{146213807}{2709504000}t^4
 \non\\
&\ +\ \frac{16222576613}{16257024000} s^2 t\ +\ \frac{141801871}{338688000} t^3
 \non\\
&\ +\ \frac{16222576613}{10838016000} s^2\ +\ \frac{36642046037}{16257024000} t^2 \ .
\end{align}
Besides that in Tables \ref{Table:upto10} - \ref{Table:upto100}
the higher order energy corrections $\veps_n$ are shown.

\begin{table}[ht]
\caption{Ground state $1s_0$: First ten perturbative coefficients $\veps_n$
         written in form of ratios for the perturbation series for $\veps$ calculated in the Non-Linearization Procedure, see (\ref{ptrb}). {  These dimensionless coefficients are universal, they do not depend on the concrete two-body system in hand.}}
\label{Table:upto10}
	{\setlength{\tabcolsep}{1.0cm}
\begin{tabular}{cl}
\hline \hline
			\rule{0pt}{4ex}			
			$n$&\hskip4.cm $(-1)^{n+1}\varepsilon_n$ \\[4pt]
			\hline	
			\rule{-4pt}{4ex}		
			0 & 1 \\[4pt]
			1 & 1/2 \\[4pt]
			2 & 53/96 \\[4pt]
			3 & 5581/2304 \\[4pt]
			4 & 21577397/1105920 \\
			5 & 31283298283/132710400 \\[4pt]
			6 & 13867513160861/3538944000 \\[4pt]
			7 & 5337333446078164463/62426972160000 \\[4pt]
			8 & 995860667291594211123017/419509252915200000 \\[4pt]
			9 & 86629463423865975592742047423/1057163317346304000000 \\[4pt]
			10& 6127873544613551793091647103033033/1776034373141790720000000 \\[4pt]
\hline\hline
\end{tabular}}
\end{table}

%\begin{table}[H]
%	\centering
%	\caption{Coefficients $\varepsilon_{10n}$, $n=1,2,...,10$,  of the series expansion for $\varepsilon$ calculated in Non-Linearization Procedure, see (\ref{ptrb}). Coefficients rounded to 3 displayed decimal digits. }
%	\label{upto100}
%	{\setlength{\tabcolsep}{1.0cm}
%		\begin{tabular}{clcl}
%			\hline
%			\hline
%			\rule{0pt}{4ex}			
%			$n$&$-\varepsilon_{10n}$  &$n$&$-\varepsilon_{10n}$\\[4pt]
%			\hline	
%			\rule{-2pt}{4ex}		
%			1&$3.450\times10^9$     &6  &$5.655\times10^{140}$ \\[4pt]
%			2&$2.160\times10^{29}$  &7  &$1.410\times10^{173}$ \\[4pt]
%			3&$3.215\times10^{53}$  &8  &$6.046\times10^{206}$ \\[4pt]
%			4&$3.720\times10^{80}$  &9  &$3.127\times10^{241}$ \\[4pt]
%			5&$6.263\times10^{109}$ &10 &$1.479\times{10}^{277}$      \\[4pt]
%			
%			\hline
%			\hline
%	\end{tabular}}
%\end{table}

\begin{table}[ht]
\centering
\caption{Ground state $(1s_0)$: Exact PT coefficients $\veps_{n=10k}$, $k=1,2,...,10$
         of the series expansion for $\veps$ calculated in Non-Linearization Procedure, see (\ref{ptrb}). The results marked by  $\veps_{10k}^{(asymp)}$ obtained in the $1/n$-expansion (\ref{Avron}) at leading order. Coefficients rounded to 4~s.d.
         {  These dimensionless coefficients are universal, they do not depend on the concrete two-body system considered.}}
\label{Table:upto100}
	{\setlength{\tabcolsep}{0.5cm}
		\begin{tabular}{cllcll}
			\hline
			\hline
			\rule{0pt}{4ex}			
		$k$\ &\ $-\veps_{10k}$
         &\ $-\veps_{10k}^{(asymp)}$\ &\ $k$\ &\ $-\veps_{10k}$\ &\ $-\veps_{10k}^{(asymp)}$\\[4pt]
\hline	
			\rule{-2pt}{4ex}		
		1\ &\ $3.450 \times 10^9$   \ &\ $4.623 \times 10^9$\ &\ 6\ &\ $5.655 \times 10^{140}$\ &
            \ $5.911 \times 10^{140}$
\\[4pt]
		2\ &\ $2.160 \times 10^{29}$ \ &\ $2.478 \times 10^{29}$\ &\ 7 \ &\ $1.410 \times 10^{173}$
       \ &\ $1.464 \times 10^{173}$
\\[4pt]
		3\ &\ $3.215 \times 10^{53}$ \ &\ $3.518 \times 10^{53}$\ &\ 8 \ &\ $6.046 \times 10^{206}$
       \ &\ $6.250 \times 10^{206}$
\\[4pt]
		4\ &\ $3.720 \times 10^{80}$ \ &\ $3.978 \times 10^{80}$\ &\ 9 \ &\ $3.127 \times 10^{241}$
       \ &\ $3.220 \times 10^{241}$
\\[4pt]
		5\ &\ $6.263 \times 10^{109}$ \ &\ $6.606 \times 10^{109}$\ &\ 10 \ &\ $1.479 \times {10}^{277}$ \ &\ $1.519 \times 10^{277}$
\\[4pt]
			\hline
			\hline
	\end{tabular}}
\end{table}

\section{Optimal Variational Parameters of (\ref{approximant1s0}) for $(1s_0)$ state}
\label{appendixB}

Plots of the optimal variational parameters for the 8-parametric Approximant (\ref{approximant1s0}) at $(q=1, \beta_0=0)$ are shown below.
\begin{figure}[H]
	\centering
	\begin{subfigure}[t]{0.35\textwidth}
		\centering
		\includegraphics[width=\linewidth]{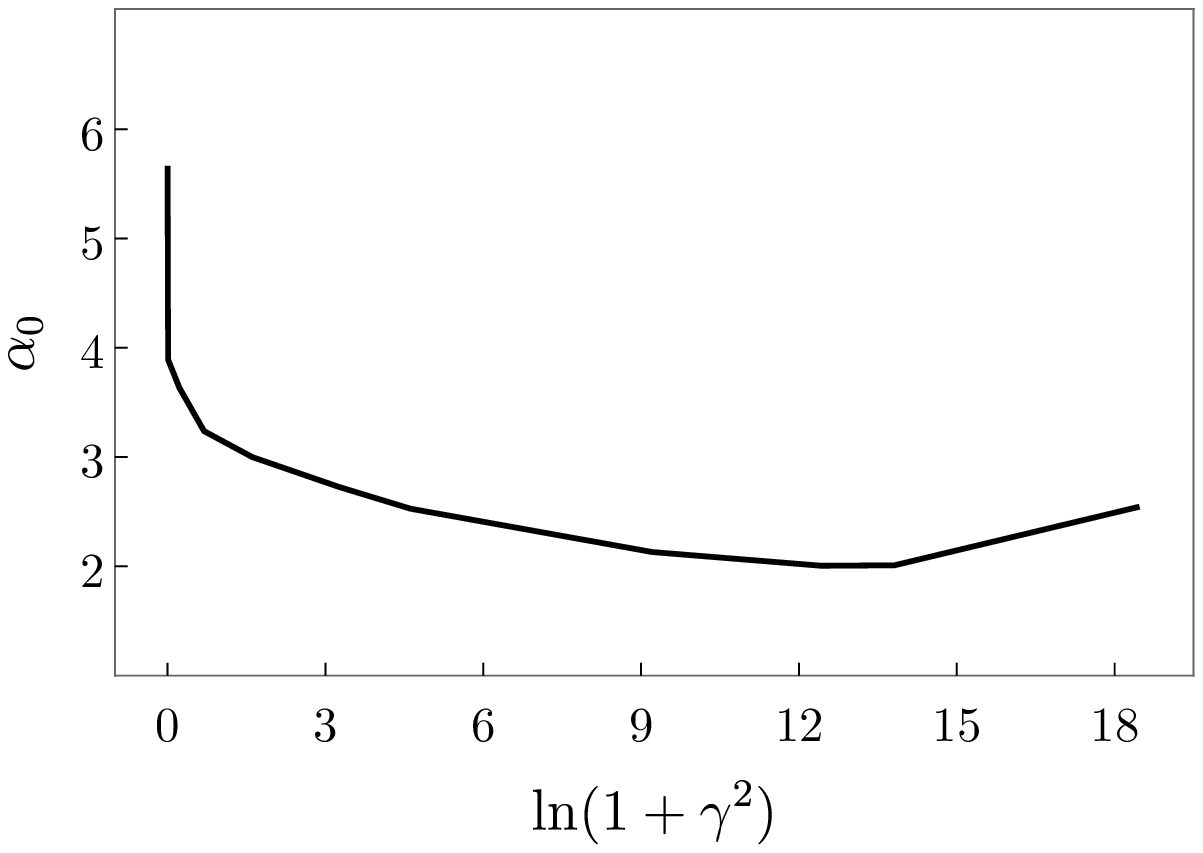}
		\label{fig:a0_1s0}
	\end{subfigure}
	\hspace{2cm}
	\begin{subfigure}[t]{0.35\textwidth}
		\centering
		\includegraphics[width=\linewidth]{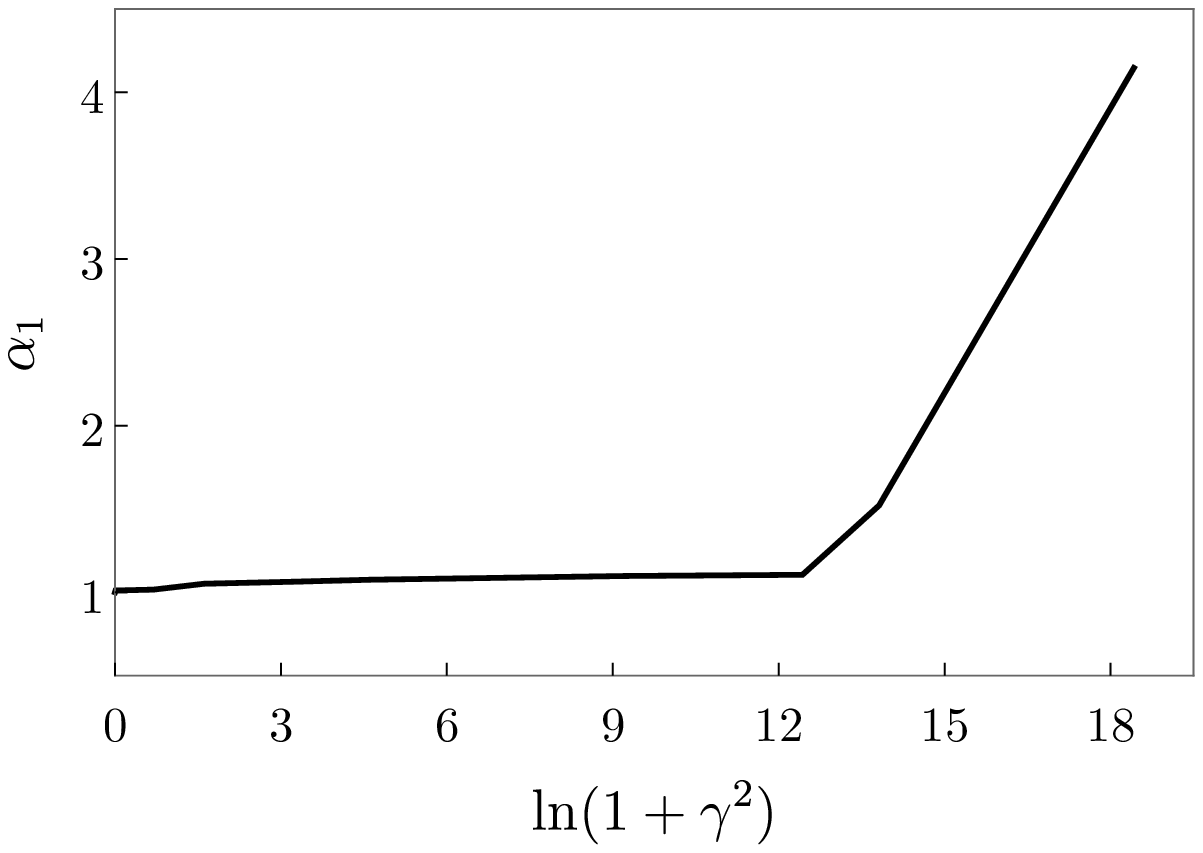}
		\label{fig:a1_1s0}
	\end{subfigure}
	\vspace{0.1cm}
	\begin{subfigure}[t]{0.35\textwidth}
		\centering
		\includegraphics[width=\linewidth]{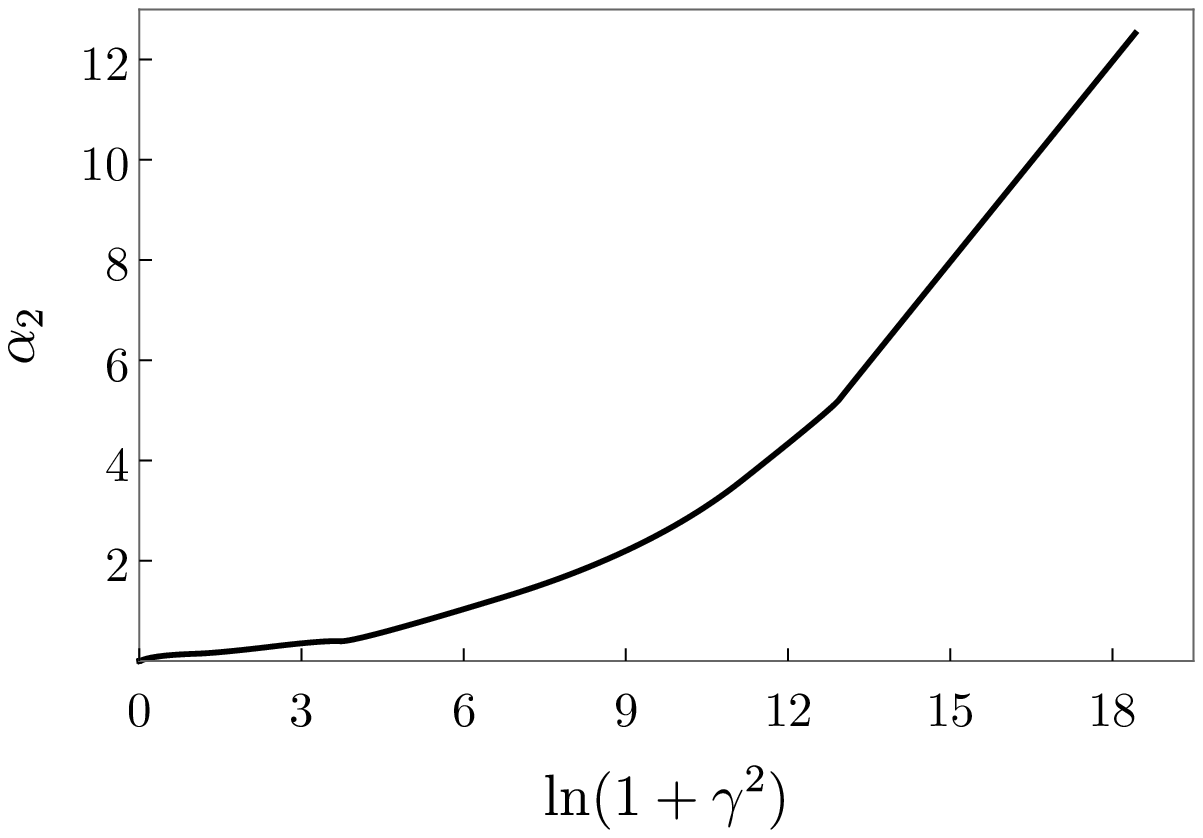}
		\label{fig:a2_1s0}
	\end{subfigure}
	\hspace{2cm}
	\begin{subfigure}[t]{0.35\textwidth}
		\centering
		\includegraphics[width=\linewidth]{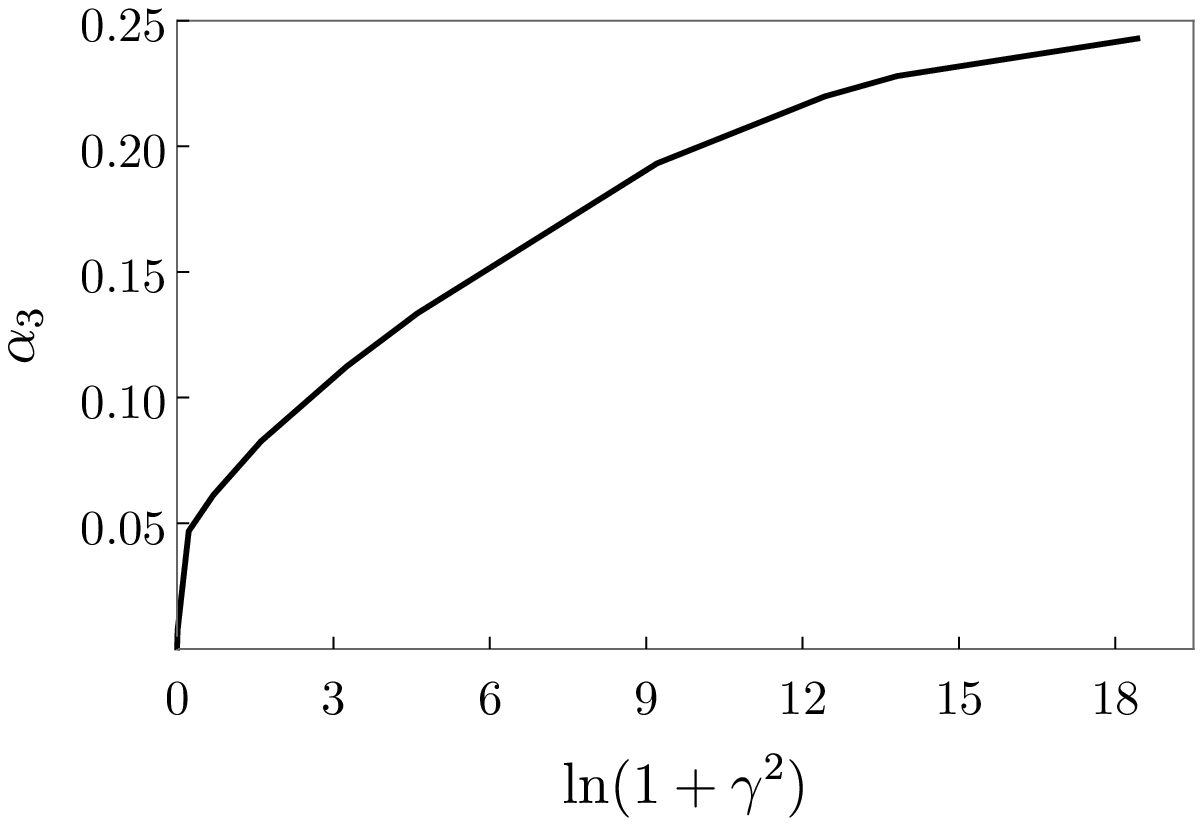}
		\label{fig:a3_1s0}
	\end{subfigure}
	\vspace{0.1cm}
	\begin{subfigure}[t]{0.35\textwidth}
		\centering
		\includegraphics[width=\linewidth]{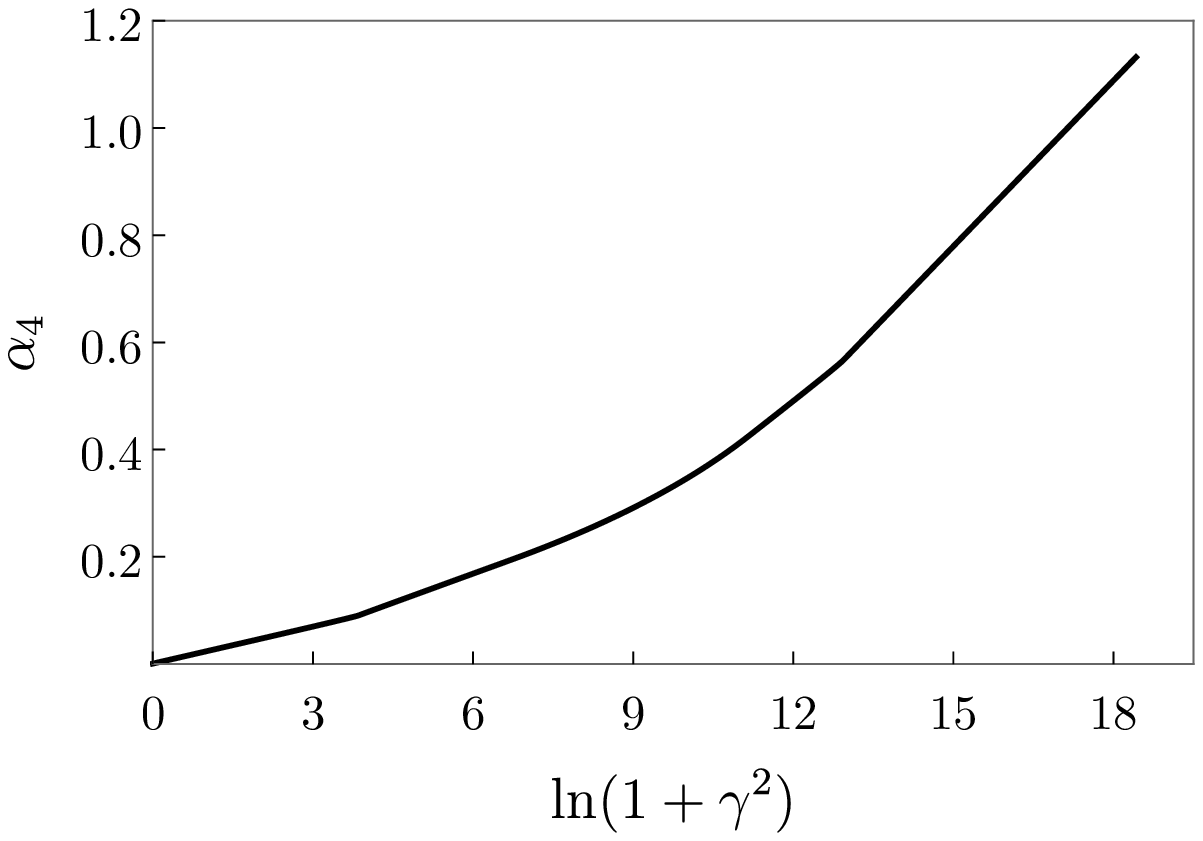}
		\label{fig:a4_1s0}
	\end{subfigure}
	\hspace{2cm}
	\begin{subfigure}[t]{0.35\textwidth}
		\centering
		\includegraphics[width=\linewidth]{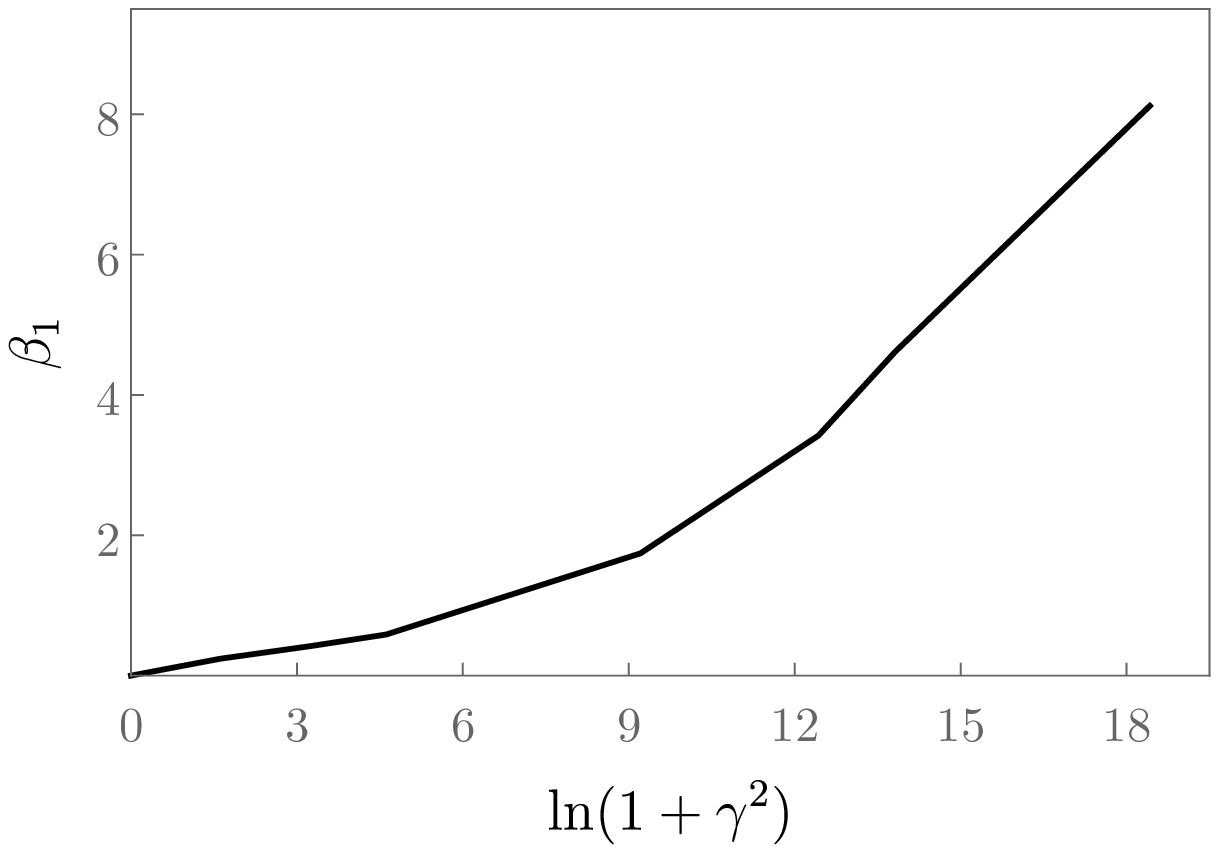}
		\label{fig:b1_1s0}
	\end{subfigure}
	\begin{subfigure}[t]{0.35\textwidth}
		\centering
		\includegraphics[width=\linewidth]{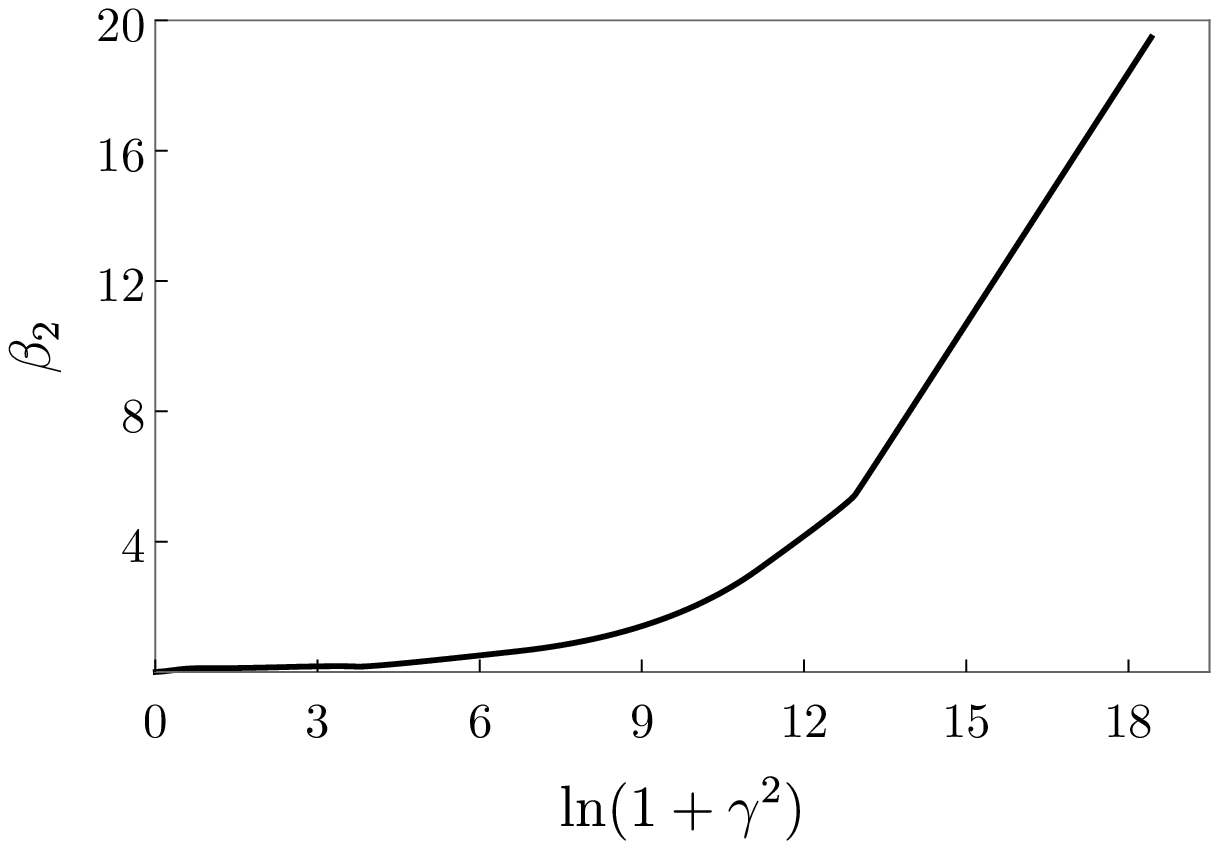}
		\label{fig:b2_1s0}
	\end{subfigure}
	\hspace{2cm}
	\begin{subfigure}[t]{0.35\textwidth}
		\centering
		\includegraphics[width=\linewidth]{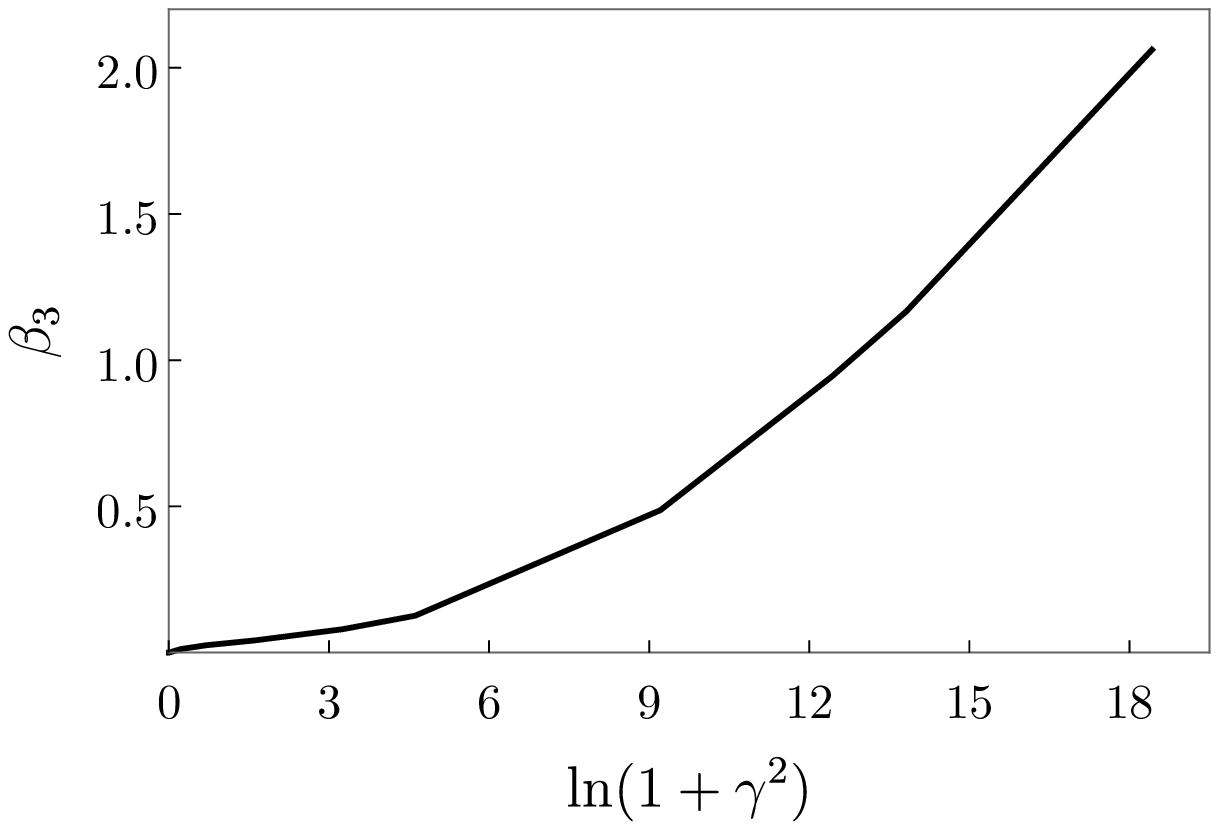}
		\label{fig:b3_1s0}
	\end{subfigure}
\caption{Optimal variational parameters
         $\{\al_0,\al_1,\al_2,\al_3,\al_4,\beta_1,\beta_2,\beta_3\}$
         of the 8-parametric Approximant (\ref{approximant1s0}) at $q=1, \beta_0=0$
         as functions of $\ln(1+\gamma^2)$.}
\label{fig:varpar}
\end{figure}

\section{Optimal Variational Parameters of (\ref{approximant2p0}) for $(2p_0)$ state}
\label{appendixC}
% \textcolor{red}{Plots of the optimal variational parameters for the Approximant % (\ref{approximant2p0}) are shown below.}
\begin{figure}[H]
	\centering
	\begin{subfigure}[t]{0.35\textwidth}
		\centering
		\includegraphics[width=\linewidth]{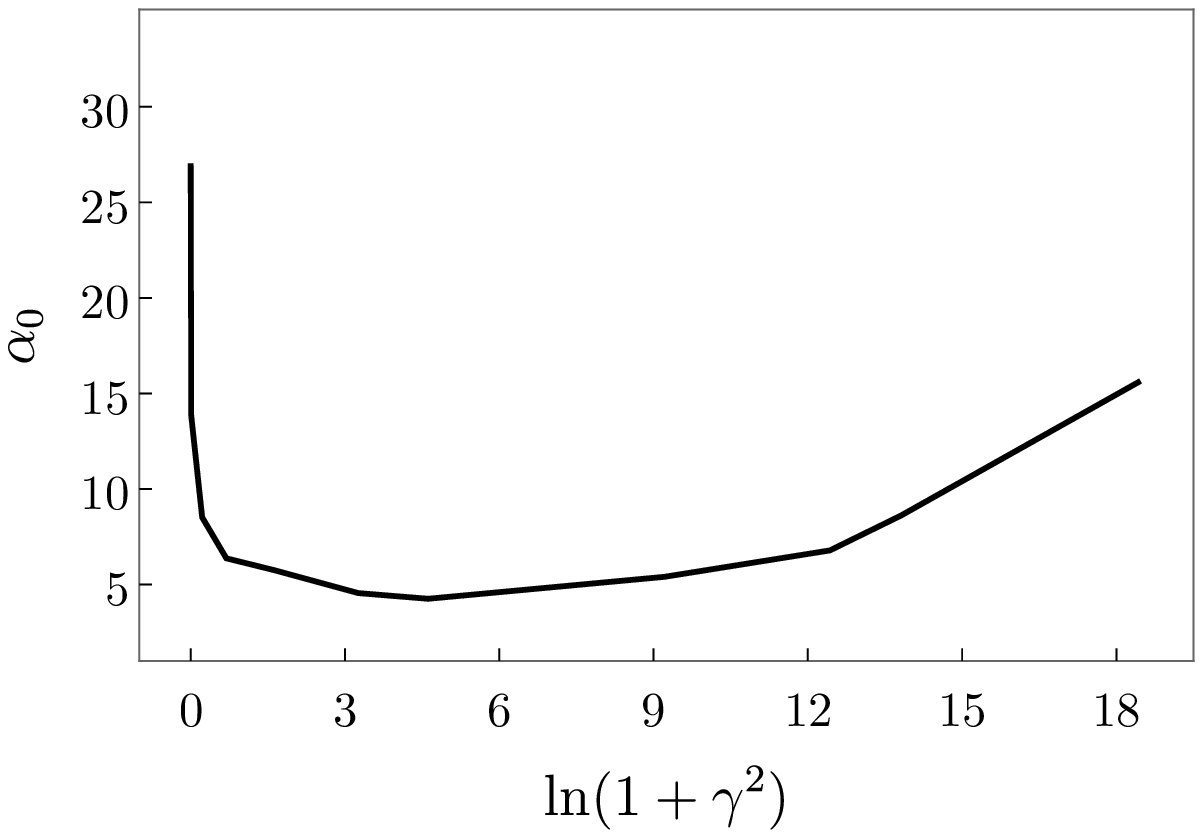}
		\label{fig:a0_2p0}
	\end{subfigure}
	\hspace{2cm}
	\begin{subfigure}[t]{0.35\textwidth}
		\centering
		\includegraphics[width=\linewidth]{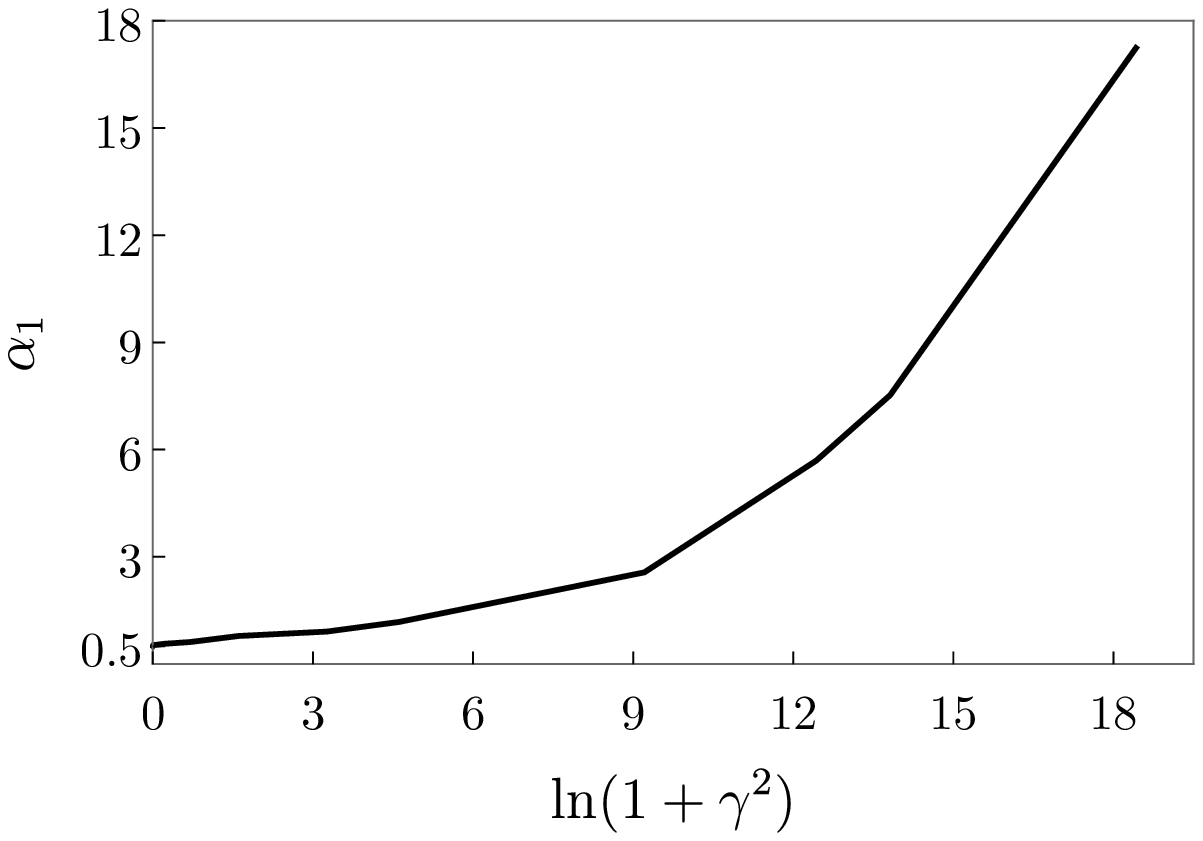}
		\label{fig:a1_2p0}
	\end{subfigure}
	\vspace{0.1cm}
	\begin{subfigure}[t]{0.35\textwidth}
		\centering
		\includegraphics[width=\linewidth]{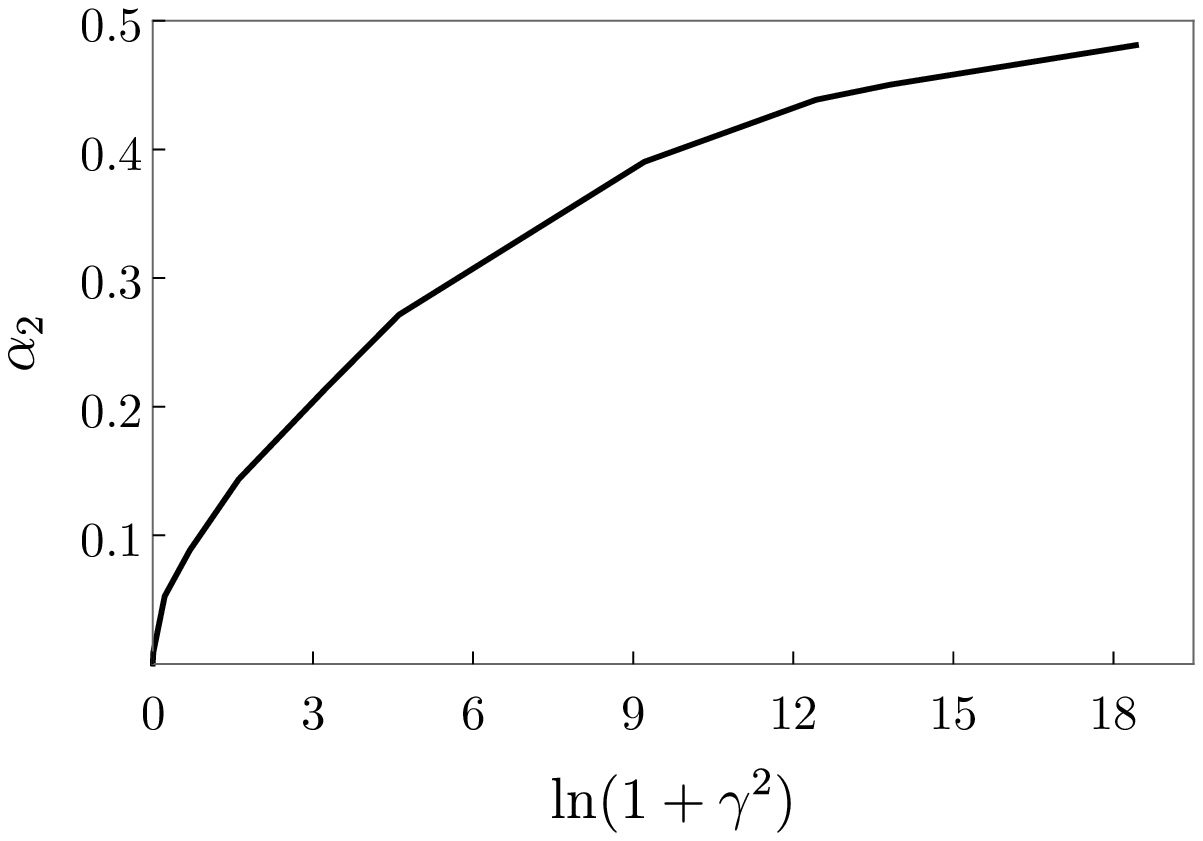}
		\label{fig:a2_2p0}
	\end{subfigure}
	\hspace{2cm}
	\begin{subfigure}[t]{0.35\textwidth}
		\centering
		\includegraphics[width=\linewidth]{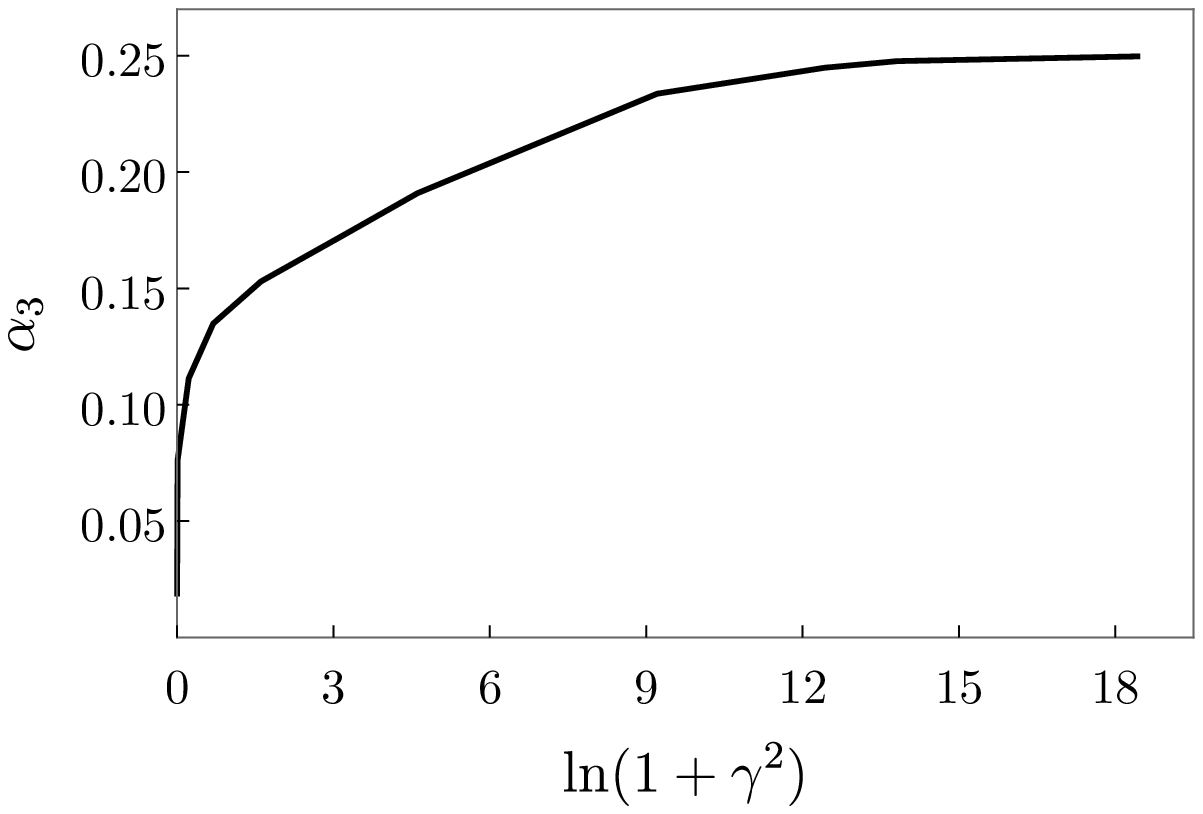}
		\label{fig:a3_2p0}
	\end{subfigure}
	\vspace{0.1cm}
	\begin{subfigure}[t]{0.35\textwidth}
		\centering
		\includegraphics[width=\linewidth]{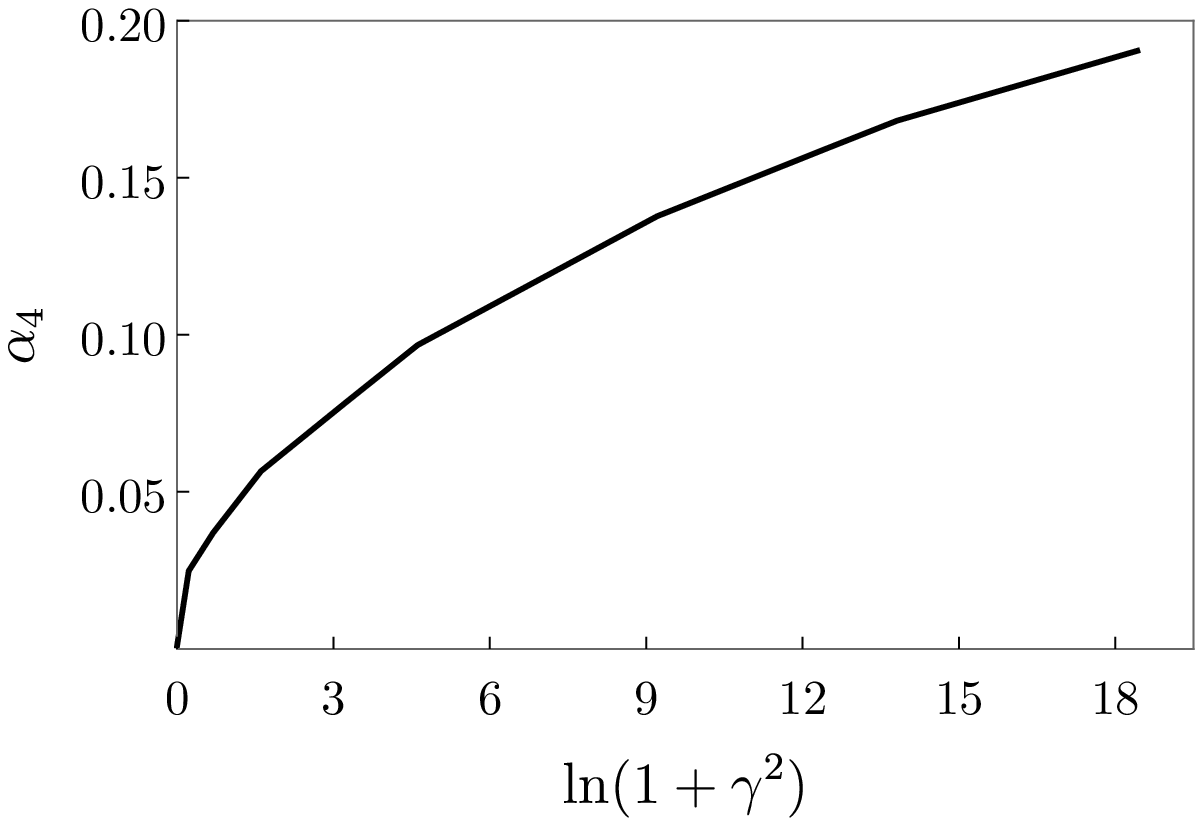}
		\label{fig:a4_2p0}
	\end{subfigure}
	\hspace{2cm}
	\begin{subfigure}[t]{0.35\textwidth}
		\centering
		\includegraphics[width=\linewidth]{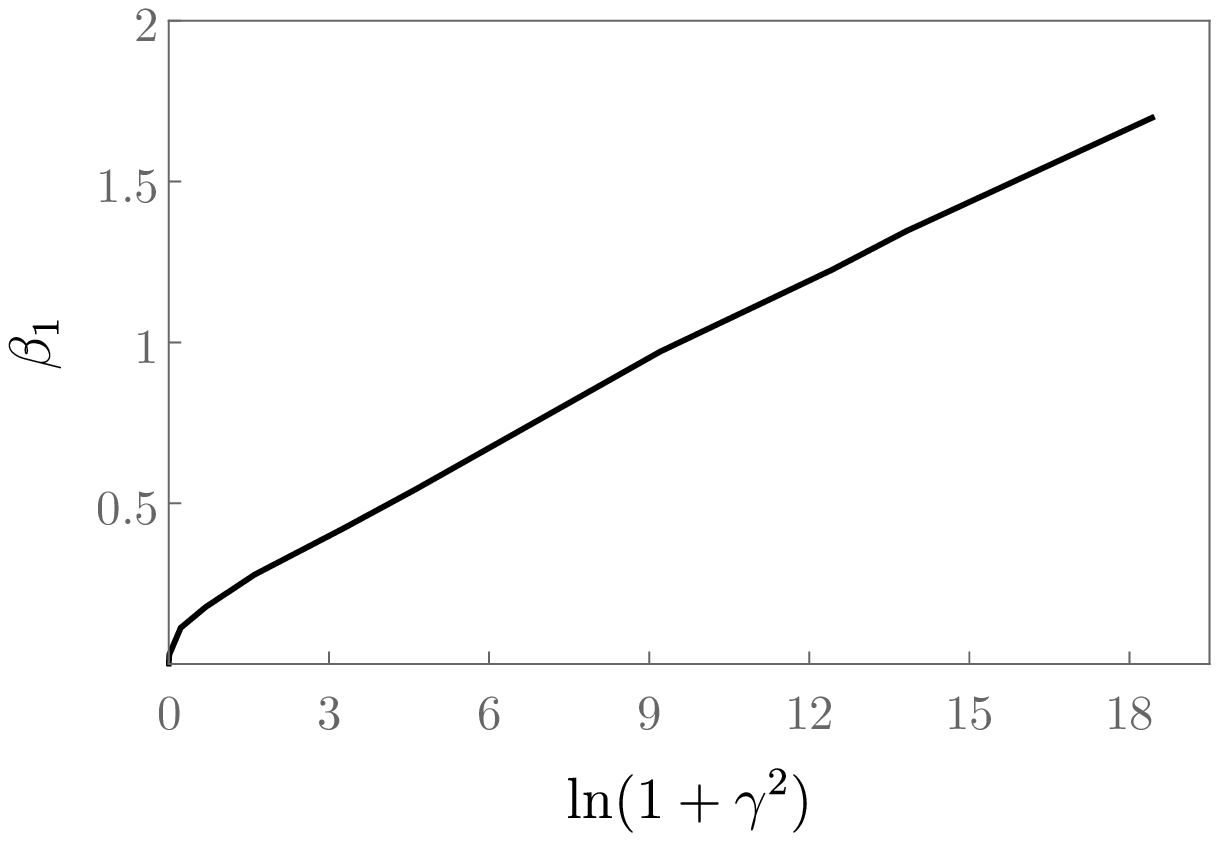}
		\label{fig:b1_2p0}
	\end{subfigure}
	\begin{subfigure}[t]{0.35\textwidth}
		\centering
		\includegraphics[width=\linewidth]{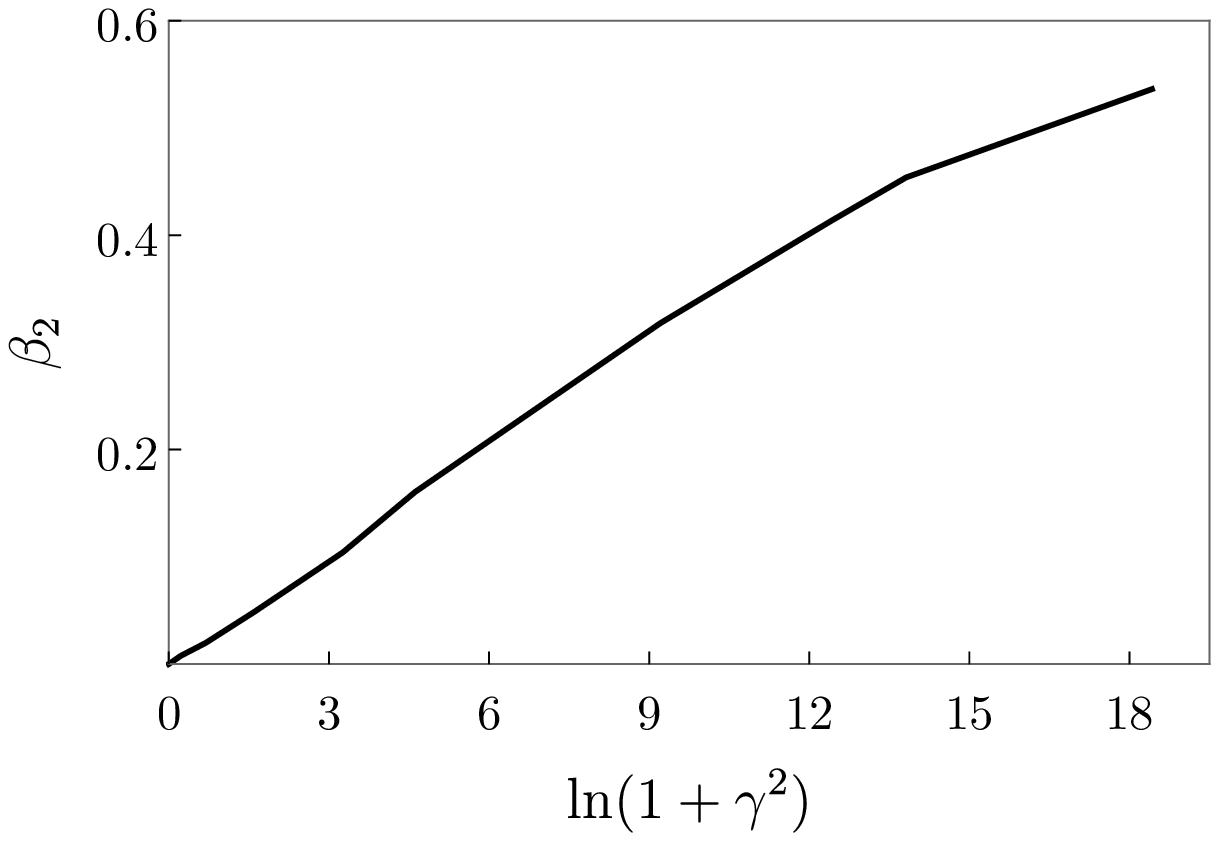}
		\label{fig:b2_2p0}
	\end{subfigure}
	\hspace{2cm}
	\begin{subfigure}[t]{0.35\textwidth}
		\centering
		\includegraphics[width=\linewidth]{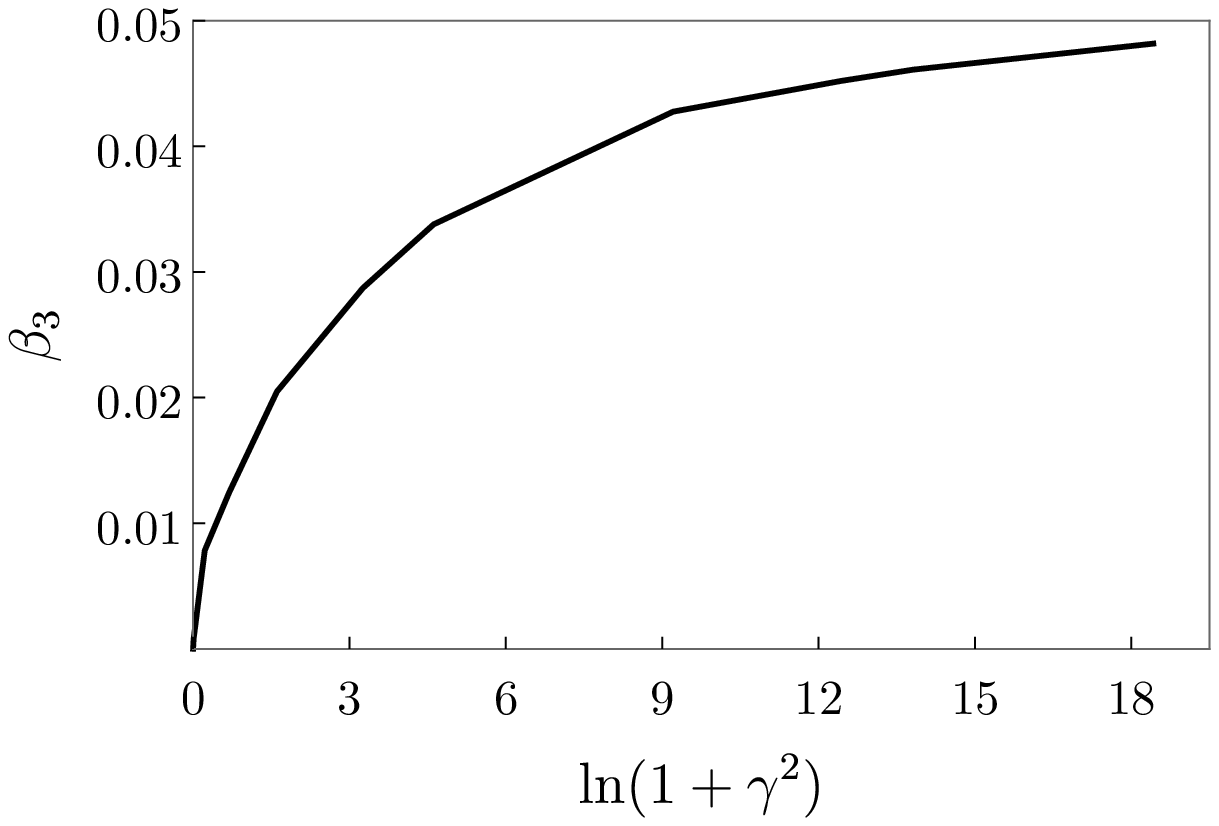}
		\label{fig:b3_2p0}
	\end{subfigure}
	\caption{Optimal variational parameters
		$\{\al_0,\al_1,\al_2,\al_3,\al_4,\beta_1,\beta_2,\beta_3\}$
		of the 8-parametric Approximant (\ref{approximant2p0}) with $(q=1,\beta_0=0)$
        as functions of $\ln(1+\gamma^2)$, cf. Figs.\ref{fig:varpar}.}
\label{fig:varpar2s0}
\end{figure}

\section{Function(\ref{approximant0})}

Parameter-less function (\ref{approximant0}) leads to very accurate energies at $\ga \leq 1$\, a.u., see below Table VIII, cf. Table I, 
but fails for larger $\gamma$.

\begin{table*}[h]
	
	{\setlength{\tabcolsep}{0.3cm}
		\caption{Variational energy $E^{(\infty)}$ through trial function (5.4). Energies are in Ry.}
		\label{Auxiliar}
		\begin{tabular}{cc}
			\hline
			\hline\rule{0pt}{3ex}
			$\gamma$ (a.u.)&$E^{(\infty)}$    \\[3pt]
			\hline\rule{0pt}{4ex}
			0&-1.000\,000\,000\,000   \\
			0.01&-0.999\,950\,005\,31   \\
			0.1&-0.995\,051\,191\,1   \\
			0.5	&-0.89\,421\,726    \\
			1.0&-0.659\,657\,07  \\
			2.0&-0.002\,271\,9   \\
			5.0& 2.950\,728  \\
			10.0&10.650\,295   \\[5pt]
			\hline
	\end{tabular}}
\end{table*}


\begin{thebibliography}{99}

\bibitem{Elliott:1960}
     R.J.~Elliott, R.~Loudon,
     {\it J. Phys. Chem. Solids \bf 15}, 196 (1960)

\bibitem{Turbiner:2006}
     A.V.~Turbiner and J.C.~Lopez Vieyra,
     {\it Physics Reports \bf 424}, 309-396 (2006)

\bibitem{LL:1977}
     L.D.~Landau and E.M.~Lifshitz,\\
     {\it Quantum Mechanics, Non-relativistic Theory} {\rm (Course of Theoretical Physics, vol.3)},\\
     {3rd edn (Oxford:Pergamon Press)}, 1977

\bibitem{Garstang:1977}
	 R.~Garstang,
     {\it Rep. Prog. Phys. \bf 40}, 105 (1977)
	%     PT

\bibitem{Gorkov:1968}
	L.~Gorkov and I.~Dzyaloshinskii,
    {\it Sov. Phys. JETP \bf 26}, 449 (1968)

\bibitem{Pavlov-Verevkin:1980}
	V.~Pavlov-Verevkin and B.~Zhilinskii,
    {\it Phys. Lett. \bf A 78}, 244 (1980)

\bibitem{Burkova:1976}
	L.~Burkova, I.~Dzyaloshinskii, S.~Drukarev, and B.~Monozon,
    {\it Sov. Phys. JETP \bf 44}, 276 (1976)

\bibitem{Kravchenko:1996}
	Y.P.~Kravchenko, M.A.~Liberman and B.~Johansson,
    {\it Phys. Rev. \bf A 54}, 287 (1996)

\bibitem{Baye:2015}
    D.~Baye,
    {\it Phys. Rep. \bf 565}, 1-108 (2015)

\bibitem{Stubbins:2004}
     C.~Stubbins, K.~Das, Y.~Shiferaw,
     {\it J. Phys. \bf B 37}, 2201 (2004) 	

\bibitem{Yafet:1956}
	Y.~Yafet, R.~Keyes and E.~Adams,
    {\it J. Phys. Chem. Solids \bf 1}, 137 (1956)

\bibitem{Larsen:1968}
	D.~Larsen,
    {\it J. Phys. Chem. Solids \bf 29}, 271 (1968)
	
\bibitem{Brandi:1975}
    H.S.~Brandi,
    {\it Phys. Rev. \bf A 11}, 1835 (1975)

\bibitem{Larsen:1982}
	 D.~Larsen,
     {\it Phys. Rev. \bf B 25}, 1126-1132 (1982)

\bibitem{Turbiner:1984}
	A.V.~Turbiner,
    {\it J. Phys. \bf A 17}, 859 (1984)

\bibitem{Turbiner:1987}
    A.V.~Turbiner,
    {\it Soviet Phys. - Yad. Fiz. \bf 46}, 204-218 (1987),\\
    {\it Sov. Journ. of Nucl. Phys. \bf 46}, 125-134 (1987)
           (English Translation)

\bibitem{Potekhin:2001}
	A.Y.~Potekhin and A.V.~Turbiner,
    {\it Phys. Rev. \bf A 63}, 065402 (2001)

\bibitem{Turbiner:2007}
      A.V.~Turbiner,\\
      Plenary Talk given at {\it The Neutron Star: Crust and Surface},\\ Institute for Nuclear Theory, University of Washington, Seattle, USA (2007)
      	
\bibitem{Baye:2008}
    D.~Baye, M.~Vincke and M.~Hesse,
    {\it J. Phys. \bf B 41}, 055005 (2008)

\bibitem{Schmelcher:1988}
	P.~Schmelcher and L.S.~Cederbaum,
    {\it Phys. Rev. \bf A 37}, 672-681 (1988)

\bibitem{Pop-Kar:2014}
    V.S.~Popov, B.~Karnakov,
    {\it Phys.-Usp. \bf 57}, 257-279 (2014)

\bibitem{Potekhin:1998}
    A.Y.~Potekhin,
    {\it J. Phys. \bf B 31}, 49-63 (1998)

\bibitem{Escobar-Turbiner:2014}
    A.M.~Escobar-Ruiz, A. V. Turbiner,
    {\it Annals of Physics \bf 359}, 405-418 (2014)

%%%%%%%%%%%%%
%%%%%%%%%%%%%
%%%%%%%%%%%%%
	
\bibitem{DELVALLE:2019-20}
    J.C.~del Valle and  A.V.~Turbiner,
    {\it Int. J. Mod. Phys. \bf A34} (2019) 1950143;\\
    {\it ibid \bf A35} (2020) 2050005

\bibitem{twe}
   A.~V.~Turbiner, W.~Miller Jr and M.A.~Escobar Ruiz,\\
   {\it Journal of Physics \bf A50} (2017) 215201;\\
   {\it Journ of Math Physics \bf A59} (2018) 022108;\\
   {\it Journal of Physics \bf A51} (2018) 205201;\\
   {\it Journ of Math Physics \bf A60} (2019) 062101

\bibitem{ESCOBAR:2016}
    M. A. Escobar-Ruiz, E. Shuryak, and A. V. Turbiner,
    {\it Phys. Rev. \bf D93}, 105039 (2016)

\bibitem{Turbiner:1981}
    A.V.~Turbiner,
    {\it J. Phys. \bf A14}, 1641 (1981)

\bibitem{ESCOBAR:2017}
    M.A.~Escobar-Ruiz, E.~Shuryak, and A.V.~Turbiner,
    {\it Phys. Rev. \bf D 96}, 045005 (2017)

\bibitem{Sh-Tur:1999}
     M.A.~Shifman, A.V.~Turbiner,
     {\it Phys.Rev. \bf A59}, 1791 (1999)
	
\bibitem{delValle:2016}
     J.C.~del Valle,\\
     Master Thesis (UNAM, July, 2016), pp.1-80

\bibitem{TURBINER:2005}
     A.V.~Turbiner,
     {\it Lett.Math.Phys. \bf 74}, 169-180 (2005)

\bibitem{TURBINER:2010}
     A.V.~Turbiner,
     {\it Int.Journ.Mod.Phys. \bf A25}, 647-658 (2010)

\bibitem{Dyson:1952}
     F.J.~Dyson,
     {\it Phys. Rev. \bf 85}, 631 (1952)
	
\bibitem{Avron:1979}
	J.~Avron, B.~Adams, J.~ {\v{C}}{\'\i}{\v{z}}ek, M.~Clay, M.~Glasser, P.~Otto, J.~Paldus
    and E.~Vrscay,\\
    {\it Phys.Rev. \bf 43}, 691 (1979)
	
\bibitem{Avron:1981}
	J.E.~Avron,
    {\it Ann. Phys. \bf 131}, 73 (1981)

\bibitem{Bender-O}
    C.M.~Bender, S.A.~Orszag,\\
    \textit{Advanced Mathematical Methods for Scientists and Engineers I}\\
    \textit{(Asymptotic Methods and Perturbation Theory)},\\
    Springer-Verlag New York, 1978, pp.XIV, 593
	
\bibitem{Hasegawa-Howard:1961}
    H.~Hasegawa, R.E.~Howard,
%    \textit{Optical Absorption Spectrum Of Hydrogenic Atoms In A Strong Magnetic Field},\\
    {\it J. Phys. Chem. Solids \bf 21}, 179 - 198 (1961)
		
\bibitem{Praddaude:1972}
	H.C.~Praddaude,
    {\it Phys. Rev. \bf A 6}, 1321 (1972)
	
\bibitem{Wang:1995}
    J.H.~Wang and C.S.~Hsue,
    {\it Phys. Rev. \bf A 52}, 4508 (1995)

\bibitem{PsShertzer}
    J.~Shertzer, J.~Ackermann and P.~Schmelcher,
    {\it Phys. Rev. \bf A 58}, 1129 (1998)

\bibitem{PsWunner}
    G.~Wunner, H.~Ruder and H.~Herold,
    {\it J. Phys. \bf B 14}, 765 (1981)

%%%%%%%%%%%%%%%%%%%%%%%%%%%
%%%%%%%%%%%%%%%%%%%%%%%%%%%


%    \bibitem{SCHMELCHER1988}
%	P. Schmelcher and L. S. Cederbaum,\\
%    {\it Phys. Rev. \bf A 37}, 672-681 (1988)
%	
%%	\bibitem{AVRON1978}
%%	J. E. Avron, I. W. Herbst and B. Simon, Ann. Phys. 114, 431-451 (1978).
%	
%	\bibitem{BEHNKE1976}
%	G.~Behnke, H.~B\"uttner, and J.~Pollmann,\\
%    {\it Solid State Commun.\bf 20}, 873 (1976).
%	
%	\bibitem{BURKOVA1976}
%	L. Burkova, I. Dzyaloshinskii, S. Drukarev, and B. Monozon, Sov. Phys. JETP 44, 276 (1976).
%	
%	\bibitem{HEROLD1981}
%	H. Herold, H. Ruder, and G. Wunner, J. Phys. B 14, 751 (1981).
%	
%    \bibitem{POTEKHIN1998}
%	A.Y.~Potekhin,\\
%    {\it J. Phys. \bf B 31}, 49 (1998)
	
\end{thebibliography}
\end{document}